\documentclass[twocolumn, twocolappendix, trackchanges]{aastex631}

\newcommand{\cc}{\rm cm^{-3}}	% per cubic cm
\newcommand{\nh}{n_{\text{H}}} % hydrogen number density
\newcommand{\dtsn}{\Delta t_{\rm SN}}
\newcommand{\Myr}{{\rm Myr}}

\newcommand{\pcmq}{{\rm cm^{-3}}}
\newcommand{\psec}{{\rm s^{-1}}}
\newcommand{\erg}{{\rm erg}}
\newcommand{\kelvin}{{\rm K}}
\newcommand{\Msun}{{\rm  M_{\odot}}}
\newcommand{\kB}{{k_{\rm B}}}

\newcommand{\Zsun}{Z_{\odot}}

\newcommand{\himpc}{{h^{-1}{\rm Mpc}}}

\newcommand{\kms}{{\rm km\,s^{-1}}}
\defcitealias{2019MNRAS.484.2632S}{S19}

\usepackage{amsmath}	% Advanced maths commands
\usepackage{cases}
\usepackage{amssymb}	% Extra maths symbols
\usepackage{bm}
\usepackage{tabularx}
\usepackage{tikz}

\shorttitle{Modeling SN Feedback}
\shortauthors{Oku et al.}
\graphicspath{{./}{figures/}}
\begin{document}

\title{Osaka Feedback Model II: Modeling Supernova Feedback Based on High-Resolution Simulations}

\correspondingauthor{Yuri Oku}
\email{oku@astro-osaka.jp}

\author[0000-0002-5712-6865]{Yuri Oku}
\affiliation{Theoretical Astrophysics, Department of Earth \& Space Science, Graduate School of Science, Osaka University, 
1-1 Machikaneyama, Toyonaka, Osaka 560-0043, Japan}

\author[0000-0001-8105-8113]{Kengo Tomida}
\affiliation{Astronomical Institute, Tohoku University, Aoba, Sendai, Miyagi 980-8578, Japan}

\author[0000-0001-7457-8487]{Kentaro Nagamine}
\affiliation{Theoretical Astrophysics, Department of Earth \& Space Science, Graduate School of Science, Osaka University, 
1-1 Machikaneyama, Toyonaka, Osaka 560-0043, Japan}
\affiliation{Kavli IPMU (WPI), The University of Tokyo, 5-1-5 Kashiwanoha, Kashiwa, Chiba 277-8583, Japan}
\affiliation{Department of Physics \& Astronomy, University of Nevada, Las Vegas, 4505 S. Maryland Pkwy, Las Vegas, NV 89154-4002, USA}

\author{Ikkoh Shimizu}
\affiliation{Shikoku Gakuin University, 3-2-1 Bunkyocho, Zentsuji, Kagawa, 765-8505, Japan}

\author[0000-0001-8531-9536]{Renyue Cen}
\affiliation{Department of Astrophysical Sciences, Princeton University, Peyton Hall, Princeton, NJ 08544, USA}
\affiliation{Theoretical Astrophysics, Department of Earth \& Space Science, Graduate School of Science, Osaka University, 
1-1 Machikaneyama, Toyonaka, Osaka 560-0043, Japan}

%% Note that the \and command from previous versions of AASTeX is now
%% depreciated in this version as it is no longer necessary. AASTeX 
%% automatically takes care of all commas and "and"s between authors names.

%% AASTeX 6.31 has the new \collaboration and \nocollaboration commands to
%% provide the collaboration status of a group of authors. These commands 
%% can be used either before or after the list of corresponding authors. The
%% argument for \collaboration is the collaboration identifier. Authors are
%% encouraged to surround collaboration identifiers with ()s. The 
%% \nocollaboration command takes no argument and exists to indicate that
%% the nearby authors are not part of surrounding collaborations.

%% Mark off the abstract in the ``abstract'' environment. 
\begin{abstract}

% Supernova\,(SN) feedback has been recognized as an essential process for self-regulating the growth of galaxies, and a better model of SN feedback is still needed in galaxy formation simulations. 
Feedback from supernovae\,(SNe) is an essential mechanism that self-regulates the growth of galaxies, and a better model of SN feedback is still needed in galaxy formation simulations. 
In the first part of this paper, using an Eulerian hydrodynamic code {\sc Athena++}, we find universal scaling relations for the time evolution of momentum and radius for a superbubble, when the momentum and time are scaled by those at the shell-formation time.  
In the second part of this paper, we develop an SN feedback model based on the {\sc Athena++} simulation results utilizing 
%the anisotropic SPH particle distribution and 
Voronoi tessellation around each star particle, and implement it into the {\sc GADGET3-Osaka} smoothed particle hydrodynamic code. 
% Our feedback model is guaranteed to be isotropic, energy- and momentum-conserving. 
Our feedback model was demonstrated to be isotropic and conservative in terms of energy and momentum. 
We examined the mass/energy/metal loading factors and find that our stochastic thermal feedback model produced galactic outflow that carries metals high above the galactic plane but with weak suppression of star formation. 
Additional mechanical feedback further suppressed star formation and brought the simulation results in better agreement with the observations of the Kennicutt--Schmidt relation,
% although all model results are within the uncertainties of observed data. 
with all the results being within the uncertainties of observed data.
We argue that both thermal and mechanical feedback are necessary for the SN feedback model of galaxy evolution when an individual SN bubble is unresolved. 

\end{abstract}

%% Keywords should appear after the \end{abstract} command. 
%% The AAS Journals now uses Unified Astronomy Thesaurus concepts:
%% https://astrothesaurus.org
%% You will be asked to selected these concepts during the submission process
%% but this old "keyword" functionality is maintained in case authors want
%% to include these concepts in their preprints.
%\keywords{***KEYWORD***}


%% From the front matter, we move on to the body of the paper.
%% Sections are demarcated by \section and \subsection, respectively.
%% Observe the use of the LaTeX \label
%% command after the \subsection to give a symbolic KEY to the
%% subsection for cross-referencing in a \ref command.
%% You can use LaTeX's \ref and \label commands to keep track of
%% cross-references to sections, equations, tables, and figures.
%% That way, if you change the order of any elements, LaTeX will
%% automatically renumber them.
%%
%% We recommend that authors also use the natbib \citep
%% and \citet commands to identify citations.  The citations are
%% tied to the reference list via symbolic KEYs. The KEY corresponds
%% to the KEY in the \bibitem in the reference list below. 

\section{Introduction} \label{sec:introduction}

It is widely recognized that baryonic feedback processes play vital roles in the formation and evolution of galaxies in the $\Lambda$ cold dark matter\,($\Lambda$CDM) universe \citep[e.g.,][]{1974MNRAS.169..229L, 1984Natur.311..517B, 1986ApJ...303...39D}.
Supernova\,(SN) feedback is one of the most dominant processes; it impacts the interstellar medium\,(ISM) and circumgalactic medium\,(CGM) by generating hot bubbles \citep[][]{1977ApJ...218..148M, 1990ARA&A..28...71S}, driving turblence \citep[][]{1974ApJ...188..501C, 2011ApJ...731...41O}, and launching galactic winds \citep[][]{1985Natur.317...44C, 2005ARA&A..43..769V, 2009ApJ...697.2030S}.
Hence, the study of SN feedback becomes necessary to better understand galaxy-formation processes.

It is challenging to spatially and temporally resolve SN explosions in galaxy formation simulations due to a lack of resolution. 
In the early days of cosmological galaxy simulation (i.e., the 1990s), SN explosions were modeled as thermal energy injections \citep{1992Cen, 1996ApJS..105...19K,1999ApJ...514....1C}. 
However, due to low resolution, the thermal energy was dissipated and radiated away as soon as it was injected \citep[i.e., the overcooling problem,][]{1992ApJ...391..502K}.

% Many attempts have been made to model the SN feedback effect in large-scale simulations.
% Some previous studies shut down cooling after the SN event to avoid the overcooling problem \citep[``delayed-cooling model'',][]{2001ApJ...555L..17T,2006MNRAS.373.1074S}, and others introduce the multiphase ISM model \citep{2003MNRAS.339..289S, 2014MNRAS.442.3013K}.
% Another strategy is to model the SN feedback effect as kinetic energy injection \citep{1993MNRAS.265..271N}.
% Previous studies compare different feedback models \citep{2017MNRAS.470.3167V, 2017MNRAS.466...11R, 2018MNRAS.478..302S}, and showed that different models predict different galactic properties.

To model SN feedback, we must consider what to model. 
As described above, SN feedback affects galaxy evolution via both kinetic feedback\,(driving turbulence) and thermal feedback\,(generating hot bubbles and outflows).
When the resolution is high enough to resolve the Sedov--Taylor phase, its kinetic and thermal energies get converted to one another, and the difference between the thermal and kinetic form, or a combination of both forms, is negligible \citep{2012MNRAS.419..465D}.
According to \citet{2015ApJ...802...99K}, the spatial resolution of $\Delta x = R_{\rm sf}/3$, where $R_{\rm sf}$ is the shell-formation radius\,(see equation~\ref{eq:SB:Rsfsingle} below), is necessary to resolve the Sedov--Taylor phase.
In most cases of cosmological galaxy simulations, such small scales cannot be resolved.
Hence, we need to include both kinetic and thermal feedbacks in the SN feedback model.
We note that simply distributing thermal energy is ineffective \citep[][]{2019MNRAS.483.3363H} because it leaves the mass of the SN ejecta unresolved in simulations as long as we use simple stellar population\,(SSP) approximation and the particle masses of stars and gas are comparable \citep{2012MNRAS.426..140D}.
Therefore, the effective modeling of thermal feedback is necessary. 

There are four key techniques to overcome the overcooling problem: i) ignoring some physical process, e.g., cooling or hydrodynamical interactions, ii) skipping unresolved physics, iii) scaling up the energy dynamics to a resolvable scale, and iv) modeling the unresolved scale.

The first method is to ignore some physical processes to make the thermal or kinetic SN feedback effective.
\citet[][]{2001ApJ...555L..17T, 2006MNRAS.373.1074S} developed a model to shut down cooling after the SN event to avoid overcooling and make thermal feedback more effective; this is known as the ``delayed-cooling model''.
\citet[][]{2003MNRAS.339..289S} modeled galactic winds driven by SN feedback and  
stochastically chose gas particles to attain a constant wind velocity.
These ``wind particles'' are briefly decoupled from hydrodynamical interactions to avoid their thermalization and subsequent radiative cooling.
The advantage of this method is that it is easier to control the SN feedback effect by tuning the time to cut-off; the disadvantage is that it is unphysical.

The second method is to predict the outcome of kinetic SN feedback and inject the resultant energy and momentum into the simulation.
\citet[][]{1993MNRAS.265..271N} developed a kinetic SN feedback model under the assumption that some fraction of the SN energy couples with the ISM as kinetic energy.
The ``mechanical feedback model'' developed by \citet[][]{2014ApJ...788..121K, 2018MNRAS.477.1578H} estimate the terminal momentum of supernova remnant (SNR) using scaling relations obtained from high-resolution simulations \citep[][]{1974ApJ...188..501C, 1988ApJ...334..252C, 1998ApJ...500..342B} and inject momentum to neighboring fluid elements.
Similar models were also developed for kpc-scale ISM simulations \citep{2014A&A...570A..81H, 2015MNRAS.449.1057G, 2017ApJ...846..133K}.
The galactic wind model by \citet[][]{2003MNRAS.339..289S} could be considered as a combination of the ``ignore'' and ``skip'' methods as they skip wind acceleration and kick wind particles at a velocity consistent with observations.
The advantage of this method depends on how one estimates the effect of SN feedback; the model can be realistic if the estimation is based on high-resolution numerical simulations, and can be helpful to study the baryon cycle if based on observational results.
Its disadvantage, on the other hand, is that it necessitates a good estimator applicable to the wide spectrum of variations in the ISM environment, and we may miss some feedback processes by skipping the unresolved aspects of the physics.

The third method is to scale up the thermal energy of SN feedback to resolvable levels.
\citet[][]{2003MNRAS.343..608K,2012MNRAS.426..140D} developed a thermal feedback model 
that stochastically increases the temperature of neighboring fluid elements by a constant $\Delta T$.
The advantage offered by this method is that we can solve the growth of hot bubbles by thermal feedback using a hydro solver without any ad hoc treatments.
However, its disadvantage is that while there remains the freedom of choosing $\Delta T$, its physical interpretation of $\Delta T$ is insufficient.

The fourth method is to model the unresolved hot phase formed by thermal SN feedback.
\citet[][]{1997MNRAS.284..235Y, 2003MNRAS.339..289S} developed a two-phase ISM model considering thermal energy injection from SN feedback within the sub-resolution scale.
Another two-phase ISM model is developed by \citet[][]{2014MNRAS.442.3013K} who decomposed the hot and cold phases to store thermal energy injected from stellar particles in neighboring gas particles.
The benefit of this method is that the two-phase model enables us to handle the hot phase, which is small in mass and otherwise challenging to resolve.
It does, however, have the drawback of the simple two-phase model being unable to solve the kinetic effect of SN feedback on the sub-resolution scale.

Previous studies have compared different feedback models \citep{2017MNRAS.470.3167V, 2017MNRAS.466...11R, 2018MNRAS.478..302S} and demonstrated that different models predict different galactic properties.
In \citet[][hereafter S19]{2019MNRAS.484.2632S}, we combined the kinetic feedback with the delayed-cooling model using the Sedov--Taylor self-similar solution. 
In the Fiducial model, 70\% of the SN energy was injected as thermal energy, while 30\% was as kinetic energy.
We also temporarily shut off cooling for the neighboring gas particles after the thermal energy injection to increase the effectiveness of thermal feedback.
While the ``Osaka feedback model'' by \citetalias{2019MNRAS.484.2632S} succeeded in providing an understanding of the self-regulation of star formation and naturally produced galactic outflow, it did contain some unphysical treatment, such as turning off cooling for the thermal feedback. 
%For the thermal feedback, we turned off cooling, which is an unphysical treatment.
For kinetic feedback, \citetalias{2019MNRAS.484.2632S} injected the kinetic energy into scales larger than that of the radius at the Sedov--Taylor phase, which neglects radiative energy dissipation.
% the momentum-conserving phase due to radiative cooling.

% One strategy to account for the evolutionary stage of SNRs in the kinetic SN feedback model is to use the terminal momentum of the SNR obtained by numerical simulations \citep[``mechanical feedback model'',][]{2014ApJ...788..121K, 2018MNRAS.477.1578H}. The advantage of this model is that the momentum injected to the surrounding ISM has little dependence on the resolution by construction \citep{2018MNRAS.477.1578H, 2020MNRAS.492.1243G}.
% By utilizing the results of high-resolution simulation of SNR, the physics occurring at sub-grid scale is naturally included.

% need for high-reso multiple SNe simulation
% To construct a mechanical feedback model, we need a scaling relations of momentum injection by SN feedback based on high-resolution simulations 

To build a good subgrid SN feedback model, we need to go back to the evolution of the single SNR and superbubble. 
The time evolution of the SNR and superbubble have been analytically and numerically investigated \citep[e.g.,][]{1974ApJ...188..501C, 1977ApJ...218..377W, 1986PASJ...38..697T, 1988RvMP...60....1O, 2015MNRAS.450..504M, 2015ApJ...802...99K, 2017ApJ...834...25K}. 

The SNR formed by a single SN explosion grows through four steps: 
1) free expansion phase, 2) Sedov--Taylor phase, 3) pressure-driven snowplow phase, and 4) momentum-conserving snowplow phase. 
Amongst these four stages of evolution, about 77 \% of the terminal momentum of SNR is acquired in the Sedov--Taylor phase \citep{2015ApJ...802...99K}.
% Thus, the duration of the Sedov--Taylor phase, determined by density and metallicity, sets the terminal momentum of the SNR in the case of a single SN.
Thus, the terminal momentum of the SNR is set by the duration of the Sedov--Taylor phase, i.e., its cooling time, which depends on the density and metallicity.
\citet{2020ApJ...896...66K} studied the effects of metallicity on the momentum of SNR using 1D simulation.
They showed that terminal momentum saturates at low metallicities ($Z < 10^{-2}\,Z_\odot$), and that the power-law fit used in previous works \citep{1998ApJ...500...95T, 2015MNRAS.450..504M} overestimates the terminal momentum;
this can be attributed to the contribution of metal-line cooling being smaller %from metals to the cooling function becomes small compared 
than that of hydrogen and helium recombination at low metallicities.
They derived a three-parameter fitting function to the terminal momentum of SNR and demonstrated it to decrease the mass-loading by a factor of two over the power-law fit from 3D simulations of vertically stratified ISM patches.

In the case of the superbubble formed by multiple SNe, there are two classical theory of its time evolution after the adiabatic phase; pressure-driven and momentum-driven models (Appendix~\ref{sec:SBtheory}). 
The pressure-driven model is applicable to the case where the gas inside the shell is overpressured with respect to the ambient medium, and the momentum-driven model otherwise.
Since multiple SN explosions occur repeatedly, the thermodynamic state of the hot bubble fluctuates significantly, which makes it difficult to solve the time evolution of a superbubble with analytical methods \citep{2019MNRAS.488.4753D}, necessitating a full numerical simulation. 

\citet{2017ApJ...834...25K} performed 3D hydrodynamical simulations with the \textsc{Athena} code and found that superbubbles follow the momentum-driven model.
They also found that the momentum per SN event is insensitive to the ambient density, but more sensitive to the interval of SN explosions. 

Studying the dynamics of a bubble created by stellar wind can also give significant insight for the evolution of an SN superbubble. 
Recently, the effect of thermal transfer due to mixing between the shell and its interior has been studied in the context of stellar wind bubbles by 
\citet{2021ApJ...914...89L, 2021ApJ...914...90L} %studied bubbles formed by stellar winds 
using analytical and numerical methods. 
They showed that the bubbles cool efficiently due to mixing at the fractal interface of the hot bubble surface and that its growth is described by the momentum-driven model.
Therefore, 1D simulations neglecting the mixing may overestimate the momentum as was shown by  \citet{2019MNRAS.490.1961E} and \citet{2019MNRAS.483.3647G} in the context of SN superbubbles. 
\citet[][]{2020ApJ...894L..24F} have also studied turbulent,
fractal mixing/cooling interfaces (Kelvin--Helmholtz instability) between hot and cold gas with
resolved simulations. 
%, although their simulation did not model the structure of entire superbubble.}

% ---------------------

In this study, we investigate the time evolution of the superbubble momentum via 3D hydrodynamical simulations with a focus on its dependence on metallicity, to supplement the 
% information revealed in the work of 
results of \citet[][]{2017ApJ...834...25K},
% which featured
who focused on its dependency on ISM density and time interval of SN explosions. 
%\citet{2017ApJ...834...25K} studied dependency of superbubble momentum on density and time interval of SNe. %In this work, we further investigate the dependence on metallicity. 
Shell formation is delayed when the metallicity is small due to weaker metal-line cooling, which results in longer cooling times. 
Since the energy loss is negligible, the superbubble efficiently gains momentum in the adiabatic phase.
Thus, one can expect the momentum input from a superbubble to the ISM to be more significant in a low-metallicity environment. 
Since the metal becomes more abundant during galaxy evolution, it is vital to study the effect of metallicity on SN feedback. % on the galaxy formation in a cosmic time.

% \citet{2019MNRAS.488.4753D} classified superbubbles into seven groups by interval time of SNe and duration time. They predicted the time evolution of superbubble in each group.

% recent works eg. SMAUG project tries to construct physically motivated subgrid model in galaxy simulation. 
Recent zoom-in simulations of dwarf galaxies reach the pc-scale resolution and resolve the Sedov--Taylor phase of a single SNR \citep[][]{2015Kimm, 2016Kimm, 2019MNRAS.483.3363H, 2020MNRAS.491.1656A, 2021MNRAS.501.5597G}, and exascale computing systems are expected to enable star-by-star simulations up to the scale of the Milky Way galaxy within the next decade \citep[][]{2021PASJ...73.1036H}.
While subgrid SN feedback models become less necessary in high-resolution zoom-in simulations, there is a growing demand for large-scale cosmological simulations for large observational projects on $>100\,\himpc$ such as the Subaru Prime Focus Spectrograph \citep[PFS,][]{2014PASJ...66R...1T}, the Dark Energy Spectroscopic Instrument \citep[DESI,][]{2013arXiv1308.0847L}, the WEAVE \citep[][]{2012SPIE.8446E..0PD}, the Euclid \citep[][]{2011arXiv1110.3193L}, and the Nancy Grace Roman Space Telescope \citep[][]{2015arXiv150303757S}.
Therefore, a better subgrid SN feedback model is still needed.

In this work, we use high-resolution simulation with \textsc{Athena++} to investigate the momentum input from the superbubble and used it, along with previous high-resolution small-box simulations of SN-driven outflow, to develop an SN feedback model for the SPH cosmological simulation code \textsc{GADGET3-Osaka}.
Our strategy to develop a physically motivated model based on high-resolution numerical simulations is similar to the Simulating Multiscale Astrophysics to Understand Galaxies (SMAUG) project\footnote{\url{https://www.simonsfoundation.org/flatiron/center-for-computational-astrophysics/galaxy-formation/smaug/}} \citep[][]{2020ApJ...900...61K}, where a galactic wind model of the multiphase outflow was developed with TIGRESS simulations \citep[][]{2020ApJ...903L..34K} and  
probability density functions of mass, momentum, energy, and metal loading were obtained in terms of the outflow velocity and sound speed.
%Whereas their work constructs a model that gives the galactic wind with respect to the star formation surface density, 
Our goal here is to develop a model of SN feedback from individual stellar particles, which can be applied in the future in both zoom-in and large-scale cosmological simulations for high-redshift galaxies that do not necessarily 
have fully-developed disks yet.   

The structure of this paper is as follows.  We discuss our superbubble simulation with Athena++ in Section~\ref{sec:SB}, supplementing it with derivations of scaling relations for terminal momentum and radius of a superbubble shell for multiple SN explosions.
We then construct our SN feedback model (both mechanical and thermal) in Section~\ref{sec:FBmodel}, and apply it to an isolated galaxy simulation in Section~\ref{sec:Isogal}. 
We then compare our results with those of previous numerical works and discuss its caveats concerning its applications in cosmological simulations in Section~\ref{sec:Discussion}.
We then conclude in Section~\ref{sec:Conclusions}.
In Appendix~\ref{sec:SNRtheory} we briefly review the analytic theory of single SNR and provide equations that are necessary for the formulation of our model, while in Appendix~\ref{sec:SBtheory}, we review the analytic theory of a superbubble. 
Appendix~\ref{sec:ResolutionRequirement} discusses the resolution requirement for thermal and kinetic feedback, as the final part,
Appendix~\ref{sec:ResolutionTest}, depicts the resolution test results of the SN feedback model in an isolated galaxy simulation.

\section{Superbubble simulation with {\sc Athena++}}
\label{sec:SB}
In this section, we study the momentum input from the superbubble and its dependence on the ISM environment. We first use analytical calculations to study the dependence of the shell-formation time (i.e., when the adiabatic phase finishes and the cold, dense shell forms) on metallicity. 
Next, we perform high-resolution simulations using \textsc{Athena++} to investigate the time evolution of superbubbles. Finally, we derive fitting functions of typical momentum input by a superbubble per SN and a typical shock radius affected by the superbubble.

\subsection{Analytic theory}
When multiple SN explosions occur clustered in space and time in stellar clusters, superbubbles are formed with the shell-formation time being a critical timescale for their growth. 
In Section \ref{sec:SB:shellformationtime}, we analytically calculate the shell-formation time and investigate its dependence on metallicity.
% In Section \ref{sec:SB:pressuredrivenmodel} and \ref{sec:SB:momentumdrivenmodel}, we overview classical theories of the time evolution of superbubble after shell formation.

\subsubsection{Shell-formation time}
\label{sec:SB:shellformationtime}
We first consider the shell-formation time of an SNR formed by a single SN and extend it to the case of a superbubble formed by multiple SNe. 

\paragraph{Single SN}
After SN explosion occurs, heated gas adiabatically expands and forms an SNR. 
The adiabatic phase is called the Sedov--Taylor phase, during which the SNR time evolution is described by Sedov--Taylor solutions. 
When the cooling effect is no longer negligible, the SNR can no longer be considered adiabatic, terminating the Sedov--Taylor phase.

SNR cooling is most effective at the shock front, where its density is its highest. The cooling time immediately after the shock wave is estimated to be 
\begin{multline}
    t_{\rm c} = \frac{e}{de/dt}
    = \frac{n_t k_{\rm B}T_{\rm ST}}{(\gamma+1)\nh n_e \Lambda(T_{\rm ST})}\\
    = \frac{2.3}{1.2}\frac{k_{\rm B}T_{\rm ST}}{(\gamma+1)\nh\Lambda(T_{\rm ST})},
    \label{eq:SB:t_c}
\end{multline}
where $\nh$, $n_e$, and $n_t$ are the number densities of hydrogen, electrons, and total ions and electrons, 
% {\bf $n_e$ and $n_t$ do not appear in the equation above} 
respectively, while $\Lambda$ represents the cooling function. 
Here, we assume the ratio of the number densities of hydrogen to helium atoms to be 10:1.
$T_{\rm ST}$ is the temperature at the post-shock layer of the shock wave, obtained from the Sedov--Taylor solution and the Rankine--Hugoniot relation \citep[][see also Appendix~\ref{sec:SNRtheory}]{2015ApJ...802...99K}:
\begin{equation}
    T_{\rm ST} = 5.3\times10^7\,\kelvin~E_{51}^{2/5}\,n_0^{-2/5}\,t_3^{-6/5},
    \label{eq:SB:T_ST}
\end{equation}
where $E_{51} = E/(10^{51}\,\erg),\,n_0 = \nh/(1\,\pcmq),\,t_3 = t/(1\,{\rm kyr})$ denote the energy injected by the SN explosion, the number density of hydrogen atoms in the ISM, and the time from the SN explosion. 

Cooling via metal-line emission is most effective at $T\sim10^5\,\kelvin$, and the cooling function at $10^5\,\kelvin < T < 10^7\,\kelvin$ is well-approximated as
% a straight line with a slope of $-0.7$.
% To analytically determine the cooling time, we approximate the cooling function as
\begin{equation}
    \Lambda(T) = 
    10^{-22}\,\erg\,{\rm cm}^3\,\psec~\Lambda_{6, -22} \left(\frac{T}{10^6\,\kelvin}\right)^{-0.7},
    \label{eq:SB:Lambda}
\end{equation}
where $\Lambda_{6, -22} = \Lambda(10^6\,\kelvin)/(10^{-22}\,\erg\,{\rm cm}^3\,\psec)$ depicts the value of the cooling function at $T=10^6\,\kelvin$ (see Fig.~\ref{fig:SB:coolingcurve}).
From equations (\ref{eq:SB:t_c}), (\ref{eq:SB:T_ST}), and (\ref{eq:SB:Lambda}), the cooling time of the gas at the shock front at time $t_{\rm s}$ is 
\begin{equation}
    t_{\rm c}(t_{\rm s}) = 2.6\times10^7\,{\rm yr}~\left(\frac{t_{\rm s}}{1\,{\rm kyr}}\right)^{-2.04}\,E_{51}^{0.68}\,n_0^{-1.68}\,\Lambda_{6, -22}^{-1}
    \label{eq:SB:tc_1}.
\end{equation}
\citet{2015ApJ...802...99K} found that when the cooling time is set to $t_c = 0.6\,e/(de/dt)$, the shell-formation time obtained from Eq.~(\ref{eq:SB:tsfsingle}) is in good agreement with the numerical results \citep{2017ApJ...834...25K}. 

The gas that is stirred up at time $t_{\rm s}$ cools down to form a low-temperature ($T<10^5\,\kelvin$) dense shell at 
\begin{equation}
    t_{\rm sf}(t_{\rm s}) = t_{\rm s} + t_{\rm c}(t_{\rm s}).
\end{equation}
The shell-formation time is determined as the minimum value of $t_{\rm sf}(t_{\rm s})$ following the method of \citet{1982ApJ...253..268C, 1986ApJ...304..771C}:
\begin{equation}
    t_{\rm sf} = 4.5\times10^4\,{\rm yr}~E_{51}^{0.22}\,n_0^{-0.55}\,\Lambda_{6,-22}^{-0.33}.
    \label{eq:SB:tsfsingle}
\end{equation}
Combining Eq.~(\ref{eq:SB:tsfsingle}) with the Sedov--Taylor self-similar solution (Eq.~\ref{eq:sedovtaylor}), we obtain the radius, mass, and momentum at shell formation as:
\begin{eqnarray}
    R_{\rm sf} &=& 
    \xi_0\left(\frac{E}{\rho}\right)t_{\rm sf}^{2/5} = 
    23\,{\rm pc}~E_{51}^{0.29}\,n_0^{-0.42}\,\Lambda_{6, -22}^{-0.13},
    \label{eq:SB:Rsfsingle}\\
    M_{\rm sf} &=&
    \frac{4}{3}\pi R_{\rm sf}^3 \rho =
    1.8\times10^3\,\Msun~E_{51}^{0.87}\,n_0^{-0.26}\,\Lambda_{6,-22}^{-0.39},
    \label{eq:SB:Msfsingle}\\
    % \begin{multline}
    % p_{\rm sf} &=& M_{\rm sf}\,\frac{3}{4} \dot{R}_{\rm sf} \nonumber \\
    p_{\rm sf} &=& 2.69 \frac{3}{4\pi}M_{\rm sf}\,\dot{R}_{\rm sf} \nonumber \\
    &=& 2.2\times10^5\,\Msun\,\kms~E_{51}^{0.93}\,n_0^{-0.13}\,\Lambda_{6,-22}^{-0.19},
    % \end{multline}
    \label{eq:SB:psfsingle}    
\end{eqnarray}
where the coefficient 2.69 is the value obtained when integrating the profiles of the Sedov--Taylor self-similar solution \citep[see Eq.~(16) of][]{2015ApJ...802...99K}.
The $\Lambda_{6,-22}$ value depends on metallicity. Since the cooling rate due to metal-line emission is proportional to the metal abundance, it proportionally varies with metallicity for a fixed metal abundance ratio.
%
%The cooling function used in this study by \citet{1993ApJS...88..253S} is 
In this study, we use the cooling function of \citet{1993ApJS...88..253S}. They assumed a solar abundance ratio \citep{1989GeCoA..53..197A} for $\log (Z/\Zsun) > 0.0$,  
and a primordial abundance ratio \citep{1989ARA&A..27..279W, 1991ApJ...383L..71B} for $\log (Z/\Zsun) \leq -1.0$,
where $Z_\odot = 0.0194$ is the solar metallicity.
Here, we fit the cooling function at $T=10^6\,\kelvin$ as follows: 
% \begin{subnumcases}{\Lambda_{6,-22} = \label{eq:L_6}}
%         \left( 1.9-0.85\frac{Z}{\Zsun} \right) \frac{Z}{\Zsun} + 10^{-1.33} & ($Z < \Zsun$)\\
%         1.05\frac{Z}{\Zsun} + 10^{-1.33} & ($Z > \Zsun$)    
% \end{subnumcases}
\begin{equation} \label{eq:L_6}
    \Lambda_{6, -22}(Z) = \max\left(1.9-0.85\frac{Z}{\Zsun}, 1.05\right)\frac{Z}{\Zsun} + 10^{-1.33}
\end{equation}
Combining Eqs. (\ref{eq:SB:tsfsingle}) and (\ref{eq:L_6}), we obtained the metallicity-dependent shell-formation time. 
% When shell formation begins at shell-formation time, the Sedov--Taylor phase ends, and the pressure-driven snowplow phase begins.

\paragraph{Multiple SNe}
One can estimate the average time interval between the SN explosions in a stellar cluster by multiplying the stellar initial mass function with stellar age,  %relative to its mass, 
which is approximately evenly spaced \citep{2010A&A...510A.101F}.
The temporal evolution of the superbubble can be thought of in the same way as that of an SNR formed by a single SN explosion, by assuming SN explosions to occur at regular time intervals with a continuous injection of energy.

The superbubble evolves adiabatically until the cooling effect manifests and a shell forms. 
Its time evolution is represented by a self-similar solution, where the dimensionless quantity $\xi$ is a combination of the radius $r$, the energy released in a single SN explosion $E_{\rm SN}$, the SN explosion interval $\dtsn$, the density of the ISM $\rho$, and the time $t$:
\begin{equation}
    \xi = \left(\frac{\rho}{E_{\rm SN}/\dtsn}\right)^{1/5}\frac{r}{t^{3/5}}
    \label{eq:xi_multi}.
\end{equation}
Compared to the case of a single SN explosion, the energy $E$ in the dimensionless quantity of the Sedov--Taylor solution (Eq.~\ref{eq:sedovtaylor}) is replaced by the energy injection rate $E/\dtsn$, resulting in a difference in the time dependence from $t^{2/5}$ to $t^{3/5}$. The value of $\xi$ %of the dimensionless quantity 
at the shock is $\xi_0 = 0.88$, and the ratio of kinetic and thermal energies is $E_{\rm k}:E_{\rm th} = 22:78$ \citep{1977ApJ...218..377W}.

The shell-formation time can be estimated by replacing the energy $E$ in Eq.~(\ref{eq:SB:tsfsingle}) with the total energy injected until time $t$, $E_{\rm SN}(t/\dtsn)$, and solving for $t$: 
\begin{equation}
    t_{\rm sf, m} = 1.7\times10^4\,{\rm yr}~E_{51}^{0.28}\,\Delta t_{\rm SN, 6}^{-0.28}\,n_0^{-0.71}\,\Lambda_{6,-22}^{-0.42}.
    \label{eq:tsfmulti} 
\end{equation}

%Due to the different values of the dimensionless quantities of the self-similar solution in the shock wavefront, the cooling time of the super bubble is about 0.15 times longer and the shell-formation time is about 0.53 times longer in the estimation by Eq. (\ref{eq:tc_line3}). Considering the different time powers of the self-similar solution, the shell-formation time is about 1.05 times longer. However, since the power of each parameter does not change from the case where the shell-formation time is obtained starting from the self-similar solution (\ref{eq:selfsimilarmulti}) of the adiabatic period of the super bubble, the equation (\ref{eq:tsfmulti}) is used as the estimate of the shell-formation time in this study.  The power of each parameter should not change because the only difference is whether the substitution of $E = Et/\dtsn$ is done first or second.

The radius, velocity, mass, and momentum at this point in time are 
\begin{eqnarray}
    R_{\rm sf, m} &=& 5.3\,{\rm pc }~E_{51}^{0.37}\,\Delta t_{\rm SN, 6}^{-0.37}\,n_0^{-0.63}\,\Lambda_{6,-22}^{-0.25}, \label{eq:Rsfmulti}\\ 
    V_{\rm sf, m} &=& \nonumber \\
    1.8 &\times&10^2\,\kms~E_{51}^{0.088}\,\Delta t_{\rm SN, 6}^{-0.088}\,n_0^{0.084}\,\Lambda_{6, -22}^{0.17}, \label{eq:Vsfmulti}\\
    M_{\rm sf, m} &=& 22\,\Msun~E_{51}^{1.11}\,\Delta t_{\rm SN, 6}^{-1.11}\,n_0^{-0.89}\,\Lambda_{6,-22}^{-0.75},
    \label {eq:Msfmulti} \\
    p_{\rm sf, m} &=& \nonumber \\ 3.3&\times&10^3\,\Msun\,\kms~E_{51}^{1.2}\,\Delta t_{\rm SN, 6}^{-1.2}\,n_0^{-0.80}\,\Lambda_{6,-22}^{-0.59}. 
    \label{eq:psfmulti} 
\end{eqnarray}
We use Eqns.~(\ref{eq:tsfmulti}), (\ref{eq:Rsfmulti}), and (\ref{eq:psfmulti}) to normalize each physical quantity while examining the time evolution in Fig.~\ref{fig:SB:radius} \& \ref{fig:SB:momentum}.

\subsection{Numerical Method}
\begin{figure}
    \centering
    \includegraphics[width=\columnwidth]{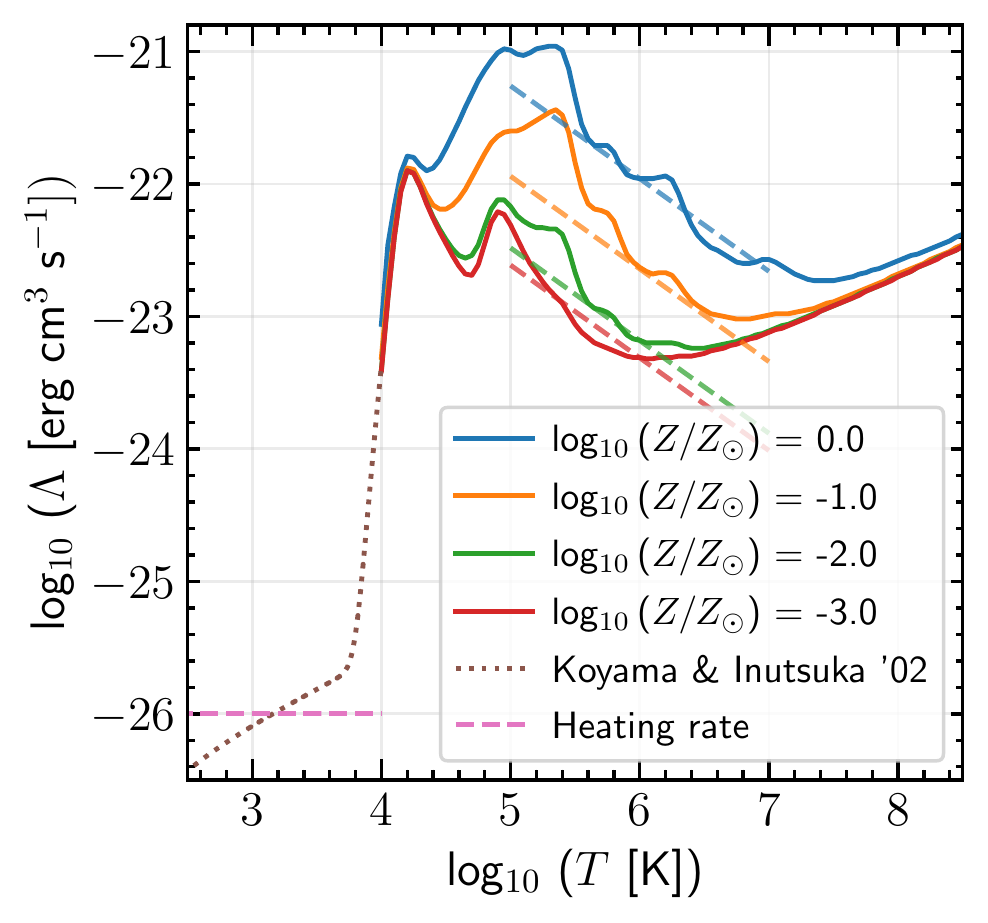}
    \caption{Cooling curves used in this work at different metallicities from \citet{1993ApJS...88..253S} combined with those of \citet{2002ApJ...564L..97K}.
    The dashed colored lines show approximations given in Eq.~(\ref{eq:SB:Lambda}) for the same metallicities given in the legend.
    The horizontal line indicates the constant photoelectric heating function $\Gamma = 1\times 10^{-26} (n_H / 1\,{\rm cm}^{-3})\,\erg\,\psec$ at $T<10^4$\,K. }
    \label{fig:SB:coolingcurve}
\end{figure}

We carried out 3D hydrodynamical simulations with the \textsc{Athena++} code\footnote{\url{https://www.athena-astro.app}} \citep{2020ApJS..249....4S} with the second-order accurate van Leer time integrator, the HLLC solver, and second-order spatial reconstruction.
We solved for the equations of continuity, momentum conservation, and energy while incorporating cooling, heating, and thermal conduction:
\begin{eqnarray}
    \frac{\partial \rho}{\partial t} + \nabla \cdot (\rho \bm{v}) &=& 0\\
    \frac{\partial (\rho \bm{v})}{\partial t} + \nabla \cdot (\rho \bm{v}\bm{v} + P) &=& 0 \\
    \frac{\partial E}{\partial t} + \nabla \{ (E + P)\cdot \bm{v} &\}& \nonumber \\
    = \nabla \cdot (\kappa \nabla T) &+& \nh\Gamma(T) - \nh^2 \Lambda(T),
\end{eqnarray}
where $E$ is the energy density.
We did not consider self-gravity and magnetic fields in our simulation.
Assuming a mean molecular weight of $\mu = 1.4$, the gas temperature is taken at $T = P/(1.1 \nh k_{\rm B})$, where the hydrogen number density is $\nh = \rho / (1.4 m_{\rm H})$.
We followed \citet{2015ApJ...802...99K} for the implementation of radiative cooling and heating.
For the cooling function $\Lambda(T)$ at low ($T < 10^4$ K) and high ($T > 10^4$ K) temperatures of gas, we adopted the fitting formulae from \citet{2002ApJ...564L..97K}\footnote{The cooling function given by Eq.\,(4) of \citet{2002ApJ...564L..97K} contains two typographical errors. See Eq.\,(4) of  \citet{2007ApJ...657..870V} for the corrected functional form.} and piecewise power-law fits to the cooling function of \citet{1993ApJS...88..253S} with collisional ionization equilibrium\footnote{\url{https://www.mso.anu.edu.au/~ralph/data/cool/}}, respectively, as shown in Fig.~\ref{fig:SB:coolingcurve}. 
Heating was applied, with the following heating function, only at $T < 10^4$ K to model the photoelectric heating of warm/cold ISM:
\begin{subnumcases}{\Gamma(T) =}
    10^{-26} \left(\frac{\nh}{1\,\pcmq}\right) ~\erg\,{\rm s}^{-1}  & ($T < 10^4\,\kelvin$)\\
    0 ~~\erg\,{\rm s}^{-1} & ($T > 10^4\,\kelvin$).
\end{subnumcases}
We used different cooling functions at $T > 10^4$ K for different metallicities, but we did not take into account the metallicity dependence at $T < 10^4$ K.
We did not consider the metallicity dependence of the heating function. 
At $Z\ll1$, the low-temperature cooling and heating rates would be quite different due to reduced photoelectric heating and cooling by fine-structure C and O lines when the metal abundance is low. 
However, these differences do not affect the results of our simulations, because the shell formation time is determined by the cooling time of shock-heated gas at high temperatures.
We imposed a temperature floor of $10^3$\,K.

We followed \citet{2019MNRAS.490.1961E} to implement thermal conduction.
For thermal conductivity $\kappa$ at low ($T \leq 6.6 \times 10^4 {\rm K}$) and high $(T > 6.6 \times 10^4 {\rm K})$ temperatures of gas, we applied, respectively, the thermal conductivities due to the neutral atomic collision \citep{1963idp..book.....P}:
\begin{equation}
    \kappa = 2.5 \times 10^5\,T_4^{1/2}\,{\rm erg}\,{\rm s}^{-1}\,{\rm cm}^{-1}\,{\rm K}^{-1},
\end{equation}
where $T_4 = T/10^4\,{\rm K}$,
and hydrogen plasma \citep{1962pfig.book.....S}:
\begin{equation}
    \kappa = \frac{1.7 \times 10^{11} T_7^{5/2}}{1 + 0.029 \ln (T_7 n_{e, -2}^{-1/2})}\,{\rm erg}\,{\rm s}^{-1}\,{\rm cm}^{-1}\,{\rm K}^{-1},
\end{equation}
where $T_7 = T/10^7\,{\rm K},\ n_{e, -2} = n_e/10^{-2} \cc$\ and $n_e$ denotes the electron number density.
The conductive heat flux $q = \kappa \nabla T$ is limited by the energy flux that can actually be transported by the electrons $q_{\rm max} = \frac{3}{2} \rho c_{\rm s, iso}^3$ \citep{1963idp..book.....P, 1977ApJ...211..135C}.
The effective thermal conductivity $\kappa_{\rm eff}$, which includes the saturation resulting from limiting the heat flux, becomes
\begin{equation}
 \kappa_{\rm eff}^{-1} = \kappa^{-1} + \frac{|\nabla T|}{q_{\rm max}}.
\end{equation}
To save calculation time, we imposed a ceiling on the thermal conductivity at the following value:  
\begin{equation}
    \kappa_{\rm ceiling} = 1.8\times10^{12}\left(\frac{\nh}{1\,{\rm cm}^{-3}}\right)\,{\rm erg}\,{\rm s}^{-1}\,{\rm cm}^{-1}\,{\rm K}^{-1}.
\end{equation}
This value is sufficiently high and has negligible effects on our results.
We note that our resolution is not high enough to resolve the conductive interface.

\begin{deluxetable}{cl}
\label{tab:SB:initialcondition}
% \tablenum{1}
\tablecaption{Initial conditions of the \textsc{Athena++} simulations.}
% \tablewidth{0pt}
\tablehead{Parameters & Their Values}
% \decimalcolnumbers
\startdata
Number density $\nh$\,[$\pcmq$] & 0.1, 1, 10 \\
Metallicity $Z$\,[$\Zsun$] & 10$^{-3}$, 10$^{-2}$, 10$^{-1}$, 1\\
Time interval of SN explosions $\dtsn$\,[Myr] & 0.01, 0.1, 1\\
\enddata
\end{deluxetable}

We ran 36 simulations by varying the initial values of the parameters as listed in Table~\ref{tab:SB:initialcondition}.
The spatial resolution was fixed in time and uniform over the simulation box with periodic boundary conditions (we did not use static mesh refinement (SMR) or adaptive mesh refinement (AMR)).
The spatial resolutions were $\Delta x$ = 6, 3, and 0.75\,pc for number densities of $\nh=$ 0.1, 1, and 10 $\pcmq$, respectively, to satisfy the convergence condition by \citet[][]{2015ApJ...802...99K}, $\Delta x < R_{\rm sf}/10$. 
To set up a turbulent initial condition, we start from a uniform density field with pressure of  $2.0\times10^3\,(\nh/1\,\pcmq)\,\kB\,\kelvin\,\pcmq$.
We then generate the decaying turbulence with only initial forcing and evolve to 1\,Myr. 
The initial forcing field was normalized to a \citet{1941DoSSR..30..301K} power spectrum $E(k) \propto k^{-5/3}$, while the range of driving was set to $2 \leq kL/2\pi \leq 20$. The velocity dispersion in the box was set to $\sigma = 5$\,km\,s$^{-1}$ 
adopted based on Larson's law \citep{1981MNRAS.194..809L, 2019MNRAS.485.3887O}.
We note that this velocity dispersion of 5\,$\kms$ is lower than observed values at the scale of the simulation box, and that the initial evolution of 1\,Myr is not long enough for multiphase ISM to develop via thermal instability. As a result, the inhomogeneity in the background is very weak, and the bubble expansion can be much more spherical than in realistic cases, which in turn affects the development of the Kelvin--Helmholtz instabilities that are normally driven by the shear at interfaces.

% how to inject thermal energy of 10^51 erg
Thermal energy of $10^{51}$ erg per SN was injected into cells whose centers were at a distance $< r_{\rm init}$ from the site of the SN explosion, where $r_{\rm init}$ is the radius within which the mass is 1 $\Msun$.
When thermal energy was injected, a mass of 1 $\Msun$ was also simultaneously injected.
We repeated this ten times, at intervals of $\dtsn$, setting $t = 0$ at the time of the initial energy injection.
% definition of SNR and comparrison with other definitions.
% We define shock radius by radial velocity $v_{\rm r} > 3$ km\,s$^{-1}$ and time scale of density change $\rho/\dot{\rho} > 0.25 \Delta x/c_s$, where $\rho$ is density, $\Delta x$ is simulation mesh size and $c_s$ is sound speed. SNR is region inside shock radius.
% 

To study the time evolution of the physical quantities of the superbubble, we define the bubble radius $R_{\rm bub}$ as the largest radius which satisfies following two criteria: (i) $v_r > 3\,\kms$, and (ii) $t_{\rm dens} < 4t_{\rm sc}$, where $v_r$ is the radial velocity averaged over a spherical shell of radius $r$,
$t_{\rm dens} = \rho/\dot{\rho}$ indicates the timescale of the density change, and $t_{\rm sc} = \Delta x/c_s$ depicts the sound-crossing time. In practice, we compute $v_r$, $t_{\rm dens}$, and $t_{\rm sc}$ for each cell, draw averaged radial profiles of these quantities, and then apply the above criteria.
We have described the time evolution of total mass, energy, and momentum within $R_{\rm bub}$ in Section~\ref{sec:SB:result}.

%In the $R_{\rm bub}$ definition, the superbubble is defined to include the region with large radial velocity, so when the superbubble grows non-uniformly, the region of the superbubble is overestimated due to the fast growth.
% Since the energy density of the bubble is larger than that of the ISM, there is no problem in discussing the energy and momentum, but it is more convenient to define the radius using the volume of the hot gas region inside the shell to discuss the time evolution of the radius compared to the 1D analytical model. 

%We compare this definition to $v_{\rm r} > 1$ km\,s$^{-1}$ used in \citet{2015ApJ...802...99K} and find difference of 10 \%.

To discuss the time evolution of the hot bubble, 
we also define the hot-gas radius as $R_{\rm hot} = (V_{\rm hot}/(4/3)\pi)^{1/3}$, where $V_{\rm hot}$ is the region at which $T > 10^4$\,K. 
%In Figure~\ref{fig:SB:slice plot}, we see that $R_{\rm hot}$ nicely captures the size of hot bubble inside the shell, and  $R_{\rm bub}$ also encloses the shell iself.  
The time evolution of $R_{\rm hot}$ is discussed in Section~\ref{sec:SB:fitting}.

\subsection{Results from Athena++ simulations}

\subsubsection{Results for $\nh = 1\,\pcmq$, $Z = 1\,\Zsun$, $\dtsn = 0.1\,\Myr$}
\label{sec:SB:result}
%  shock wave propagation in turbulent medium 
% A snapshot of n1\_Z1\_dt01 run at $t = 1.53$ Myr, about 30 kyr after fifth energy injection, is illustrated in Figure \ref{fig:slice plot}. The blast wave propagates in turbulent ISM and forms of a dense perturbed shell. The gas inside the shell is heated up to $\sim 10^7$ K by thermal energy injection. On the other hand, the dense shell cools rapidly, and the temperature is $\sim 10^4$ K. These gases are in contact at the internal boundary of the shell. 

\begin{figure}
\begin{tabular}{c}
\begin{minipage}{\columnwidth}
	\includegraphics[width=\columnwidth]{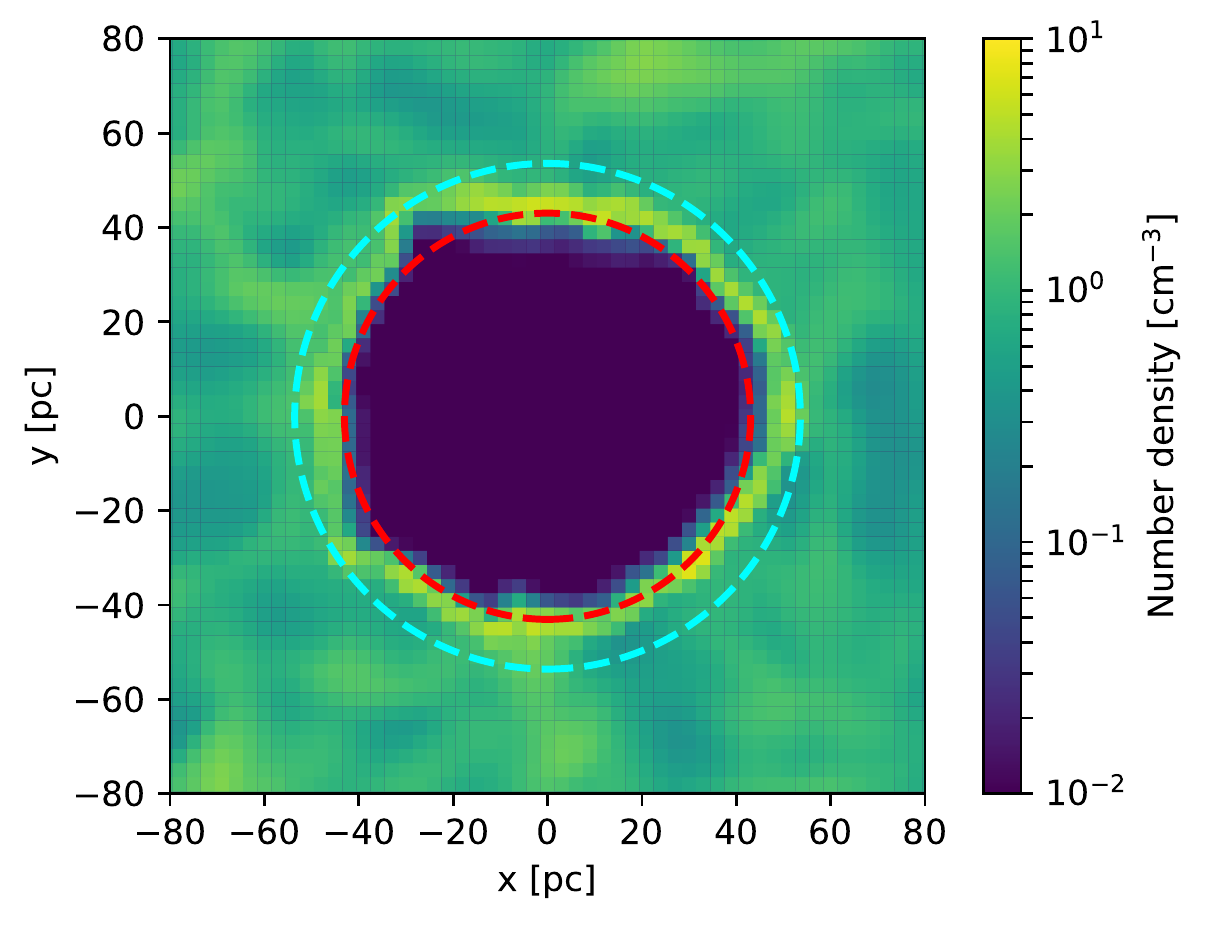}
\end{minipage}\\
\begin{minipage}{\columnwidth}
	\includegraphics[width=\columnwidth]{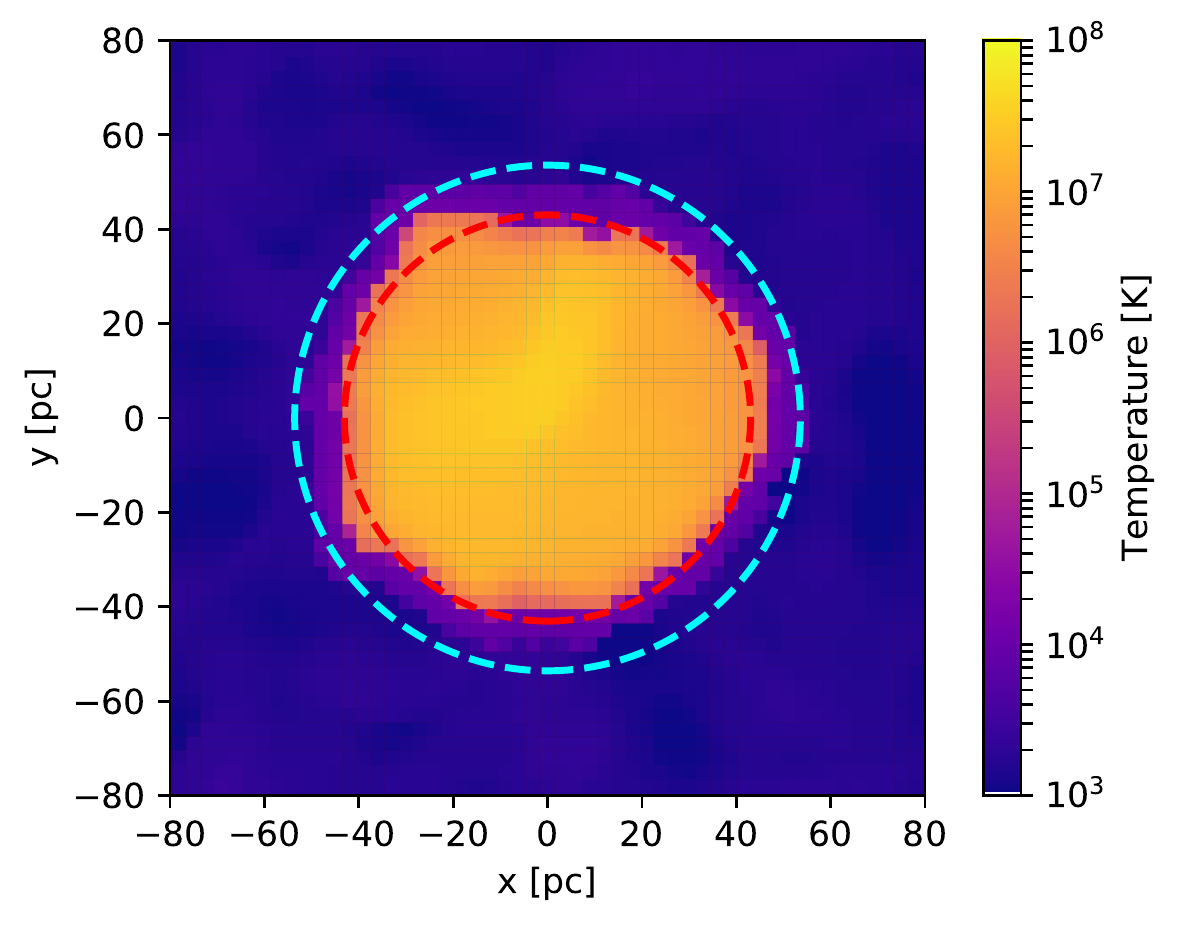}
\end{minipage}\\
\end{tabular}
    \caption{Slice plot of gas number density and temperature distribution %for n1\_Z1\_dt01 
    at $t = 0.53\,\text{Myr}$ for the case of 
    $\nh = 1\,\pcmq$, $Z =\Zsun$, $\dtsn = 0.1$ Myr.
    The inner red circle shows $R_{\rm hot}$, while the outer cyan circle shows $R_{\rm bub}$.
    }
    \label{fig:SB:slice plot}
\end{figure}

%%Calculation Setup

Figure \ref{fig:SB:slice plot} shows a slice plot of density and temperature at 0.03\,Myr after the fifth energy injection for the case of $\nh = 1\,\pcmq,\,Z = \Zsun,\,\Delta t_{\rm SN} = 0.1\,\Myr$.
The collected gas forms a low-temperature, high-density shell, inside which exists a high-temperature, low-density gas.
The shell grows almost spherically, but with a slight distortion due to the density fluctuations caused by turbulence.
We see that $R_{\rm hot}$ adequately captures the size of the hot bubble inside the shell, with $R_{\rm bub}$ enclosing the shell itself.  

\begin{figure}
    \centering
    \includegraphics[width=\columnwidth]{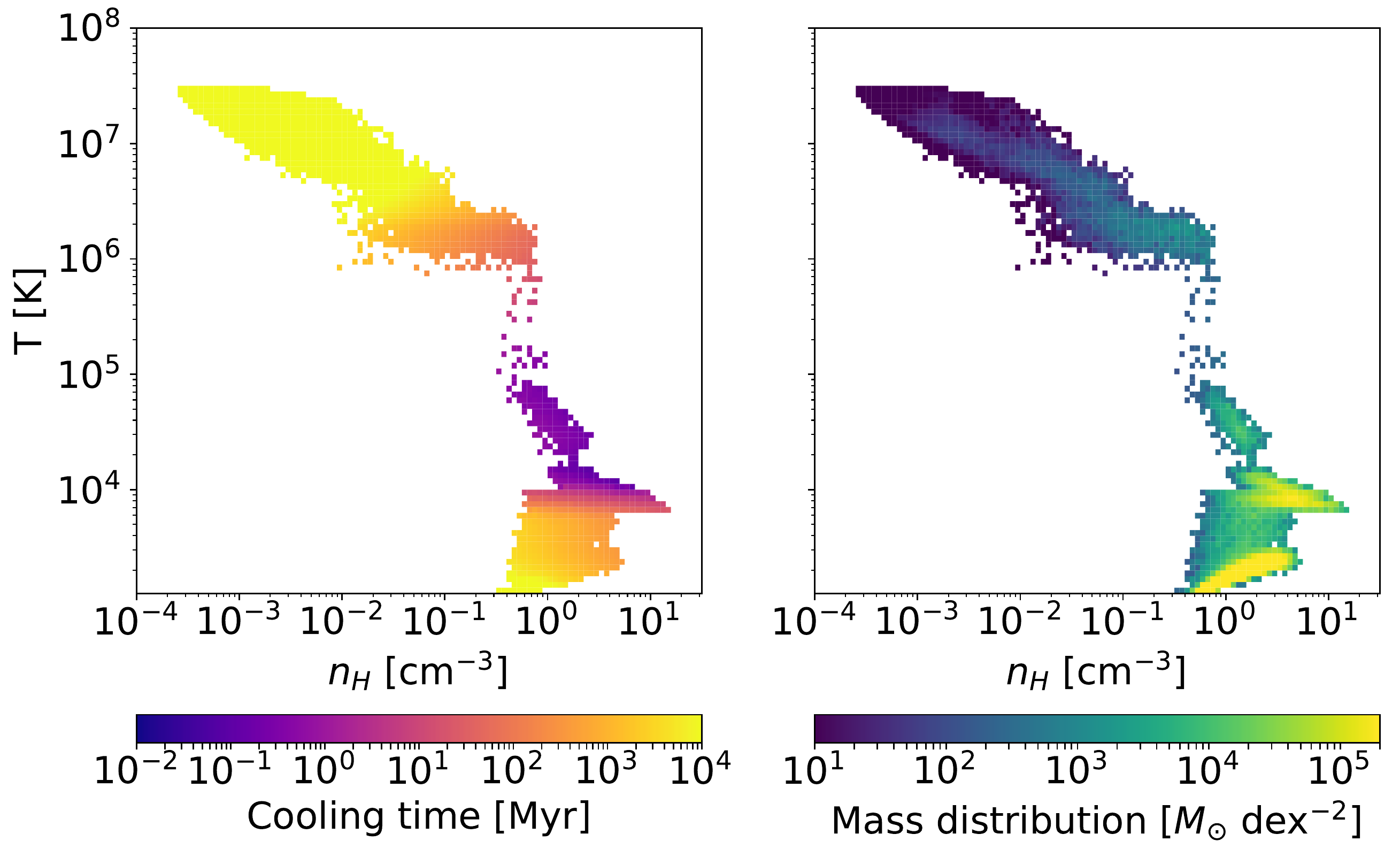}
    \caption{Phase diagram of gas in computational box, weighted for its cooling time (left panel) and mass (right panel), at the same time as in Fig.~\ref{fig:SB:slice plot}. 
    % Various components of superbubble structure are indicated in the figure. 
    In the left panel, one can see that the cooling time is short in the shell of $T \sim 10^4\,\kelvin$ and $\nh \sim 10\,\pcmq$.}
    \label{fig:SB:phasediagram}
\end{figure}

Figure \ref{fig:SB:phasediagram} shows the phase diagram at this time. 
The low-temperature ($T < 2\times10^3\,\kelvin$) gas at the bottom right corner is the ISM gas.
The gas swept up at front of the shell is heated by the shock, but as the temperature rises to $T > 10^4\,\kelvin$, it loses energy due to hydrogen recombination.
Since the shock compresses the gas to a high density, its cooling time is shorter than the dynamical time, and a shell of $T \sim 10^4\,\kelvin,\,\nh \sim 10\,\pcmq$ is formed.
At the post-shock layer of the shell, 
% \deleted{the gas is heated by the hot, low-density gas inside the shell.}
the gas can be heated by compression (if a shock propagates through the layer), or by mixing (if hot gas is advected into a cell containing cool gas), or by thermal conduction (if conduction is resolved). 
In our simulation, the upper left plume of hot gas in Figure~\ref{fig:SB:phasediagram} is heated by compression and thermal conduction.
% \deleted{If the effect of heating is to raise the temperature to $T < 10^6\,\kelvin$, the energy is immediately dissipated by cooling due to metal-line emission at $T < 10^6\,\kelvin$, however, if heated to $T > 10^6\,\kelvin$, adiabatic expansion occurs within the cooling time, causing the mass to flow from the shell to the interior of the superbubble.}

\begin{figure}
    \centering
    \includegraphics[width=\columnwidth]{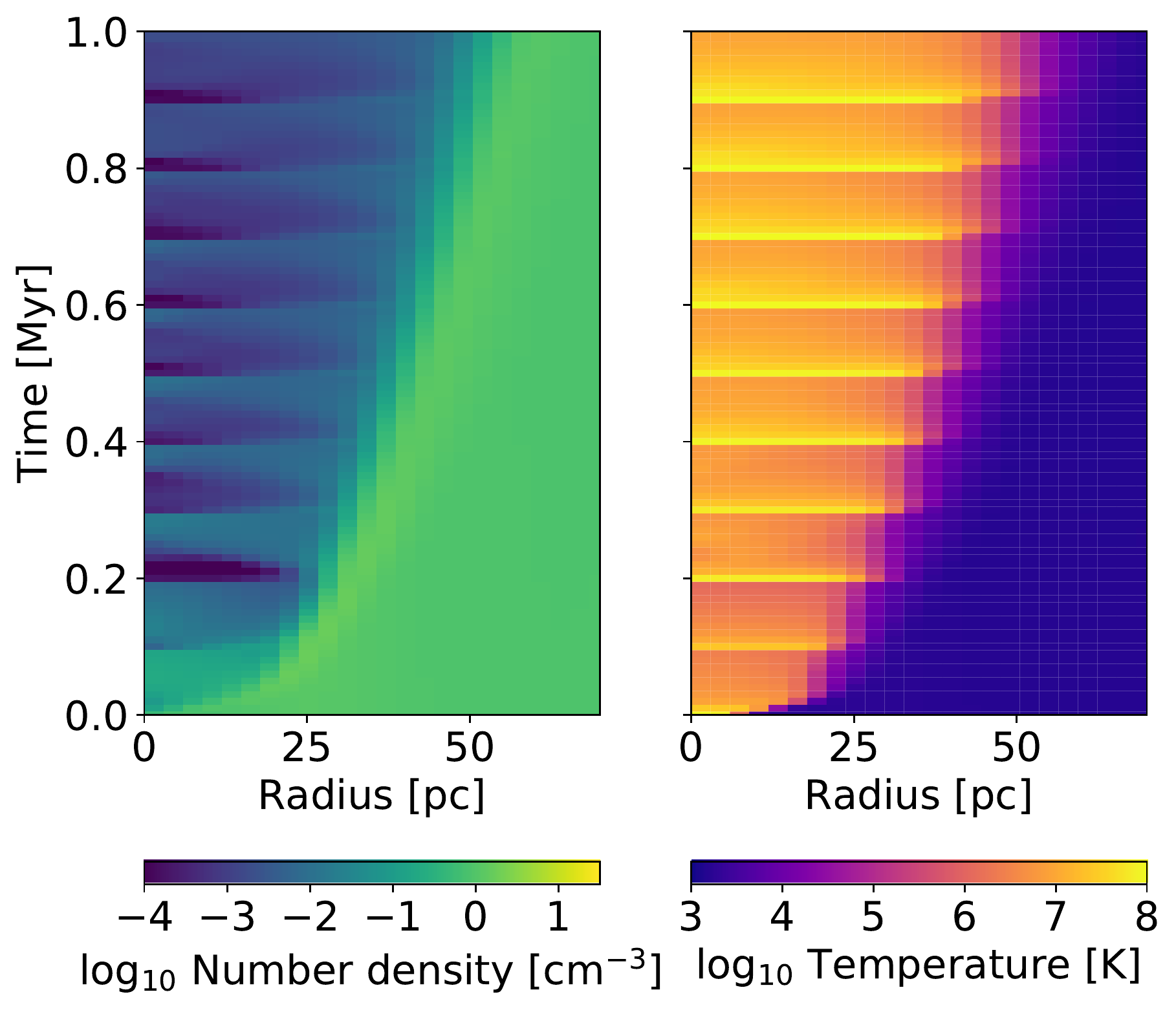}
    \caption{Evolution of superbubble as a function of radius, weighted for gas number density (left panel) and gas temperature (right panel), for the case of 
    $\nh = 1\,\pcmq$, $Z =\Zsun$, $\dtsn = 0.1$ Myr. One can see the power-law expansion of the hot bubble. The horizontal features indicate the intermittent SN injection.}
    \label{fig:radius-time}
\end{figure}

Figure \ref{fig:radius-time} shows the time evolution of the superbubble.
The shell forms when it grows to a radius of about 20 pc, after which its growth slows down from the adiabatic phase.
The subsequent energy injection stirs up the gas inside the shell and reduces its density.
However, the density increases again due to the evaporation from the shell to the interior, and the mass is transferred from the cold shell to the hot interior.

\subsubsection{Results for $\dtsn = 0.01\,\Myr$}
\begin{figure*}
    \centering
    \includegraphics[width=\textwidth]{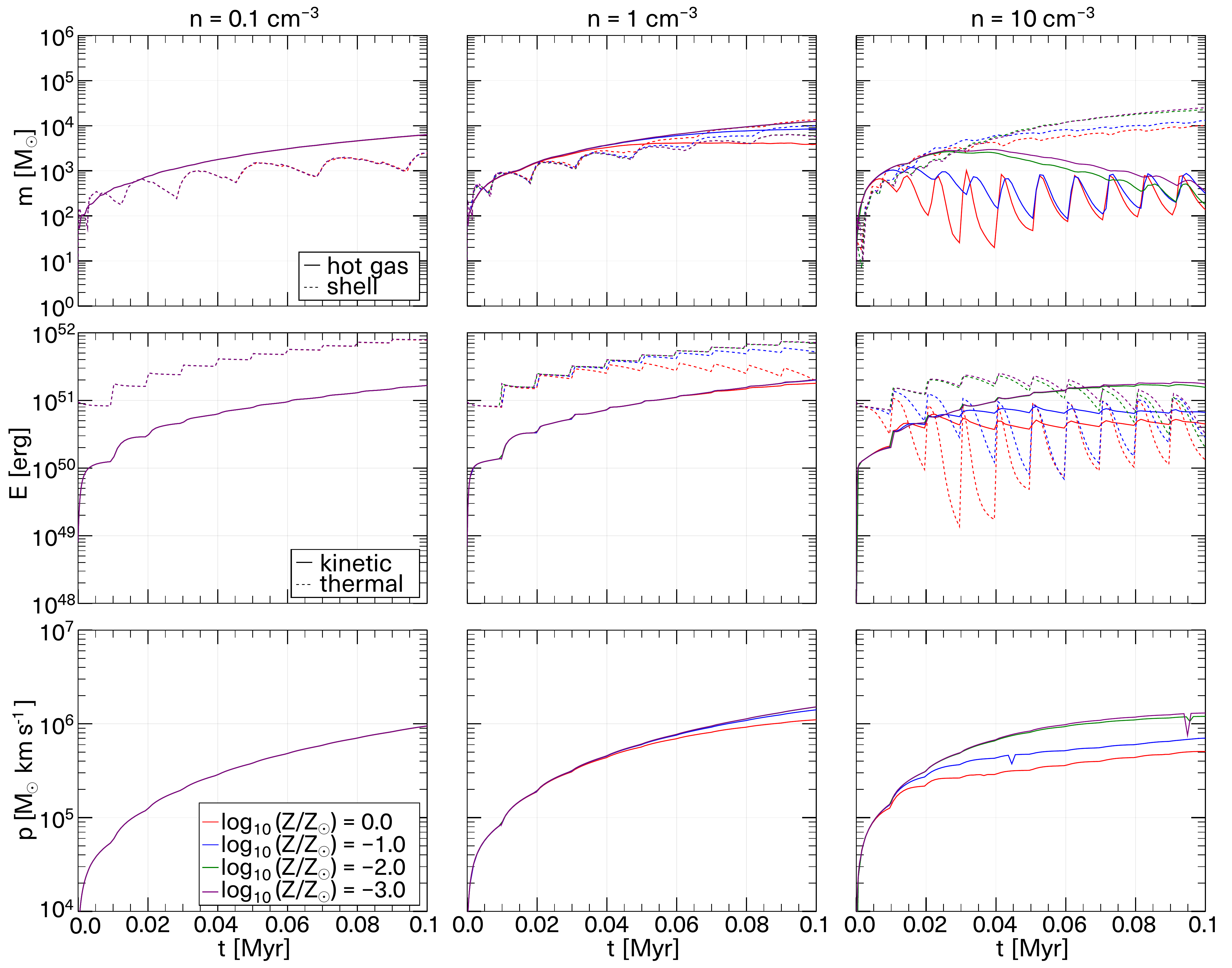}
    \caption{Time evolution of mass (top row), energy (middle row) and momentum (bottom row) of the superbubble for SN injection time intervals of $\dtsn$ = 0.01 Myr. For each component, we also show four different metallicities as indicated in the legend. Left column: $\nh = 0.1\,\cc$. Middle column: $\nh = 1\,\cc$. Right column: $\nh = 10\,\cc$.
    In the top panels for mass, the solid lines are for the hot gas inside the superbubble while dotted lines are for the shell.    
    In the middle panels for energy, the solid lines are for kinetic energy while the dotted lines are for the thermal energy of superbubble gas including the shell. 
    In the left column, the lines for the four metallicities overlap because of low density and long shell-formation time, which prolongs the adiabatic phase with little metal-cooling. In the right column, the effect of intermittent SN injection is visible. }
    \label{fig:SB:timeevolution 0.01}
\end{figure*}

The time evolution of mass, energy, and momentum for an SN explosion with a time interval of 0.01 Myr is shown in Figure~\ref{fig:SB:timeevolution 0.01}.

For $\nh = 0.1\,\pcmq$, the system evolves adiabatically because the shell-formation time is longer than 0.1 Myr for any metallicity.
Since the metallicity affects the shell-formation time but not the system's dynamics, no metallicity dependence is observed in this case.
Since the system evolves adiabatically, its total energy after 10 SN explosions is $10^{52}\,\erg$.
In the case of $\nh = 1\,\pcmq$, the cooling time is shorter due to the higher density, and this cooling effect is seen for $\log (Z/\Zsun) = 0, -1$.
The cooling effect is seen earlier for $\log (Z/\Zsun) = 0$ because the larger the metallicity is, the shorter the cooling time is.
Shell formation does not immediately get completed and, in the case of $\log (Z/\Zsun) = 0$, the cooling effect begins to appear at $t\sim 0.04\,\Myr$, while shell formation completes at $t\sim 0.1\,\Myr$.
During shell formation, the mass of the shell increases while that of the hot gas remains constant because it proceeds from the dense and short cooling time gas gathered in front of the superbubble.
% While there is a decrease in thermal energy due to cooling, there is little metallicity dependence in the time evolution of the kinetic energy because the thermal energy is converted to kinetic energy as in the adiabatic case until the effect of cooling appears in the hot gas inside the shell that is accelerating the shell.

In the case of $\nh = 10\,\pcmq$, the shell-formation time is shorter than 0.1 Myr at all metallicities, and a low-temperature, high-density shell is formed.
After shell formation, most of the system's mass is in the shell, and at $t = 0.1\,\Myr$, the mass of the hot gas is about 2\% of the total.
Comparing the cases of $\log (Z/\Zsun) = -2.0, -3.0$, there is little difference in the time evolution of the physical quantities.
In the low-metallicity environment, bremsstrahlung is a more dominant cooling mechanism than metal line emission at $T > 10^6\,\kelvin$, and hence the difference in metallicity becomes less apparent.

% Looking at the cooling curve (Figure \ref{fig:SB:coolingcurve}), the temperature at which the cooling rate increases as the temperature decreases is $10^6\,\kelvin$ for both $\log (Z/\Zsun) = -2.0, -3.0$.
% When the gas temperature decreases below $10^6\,\kelvin$, the cooling rate increases as the temperature decreases; therefore, the gas cools rapidly to form a shell.
% For $\log (Z/\Zsun) = -2.0, -3.0$, the temperature at which the shell formation starts is almost equal. 
% Hence, the shell-formation time is also almost equal, and there is almost no difference in the time evolution of physical quantities.

\subsubsection{Results for $\dtsn = 0.1\,\Myr$}
\begin{figure*}
    \centering
    \includegraphics[width=\textwidth]{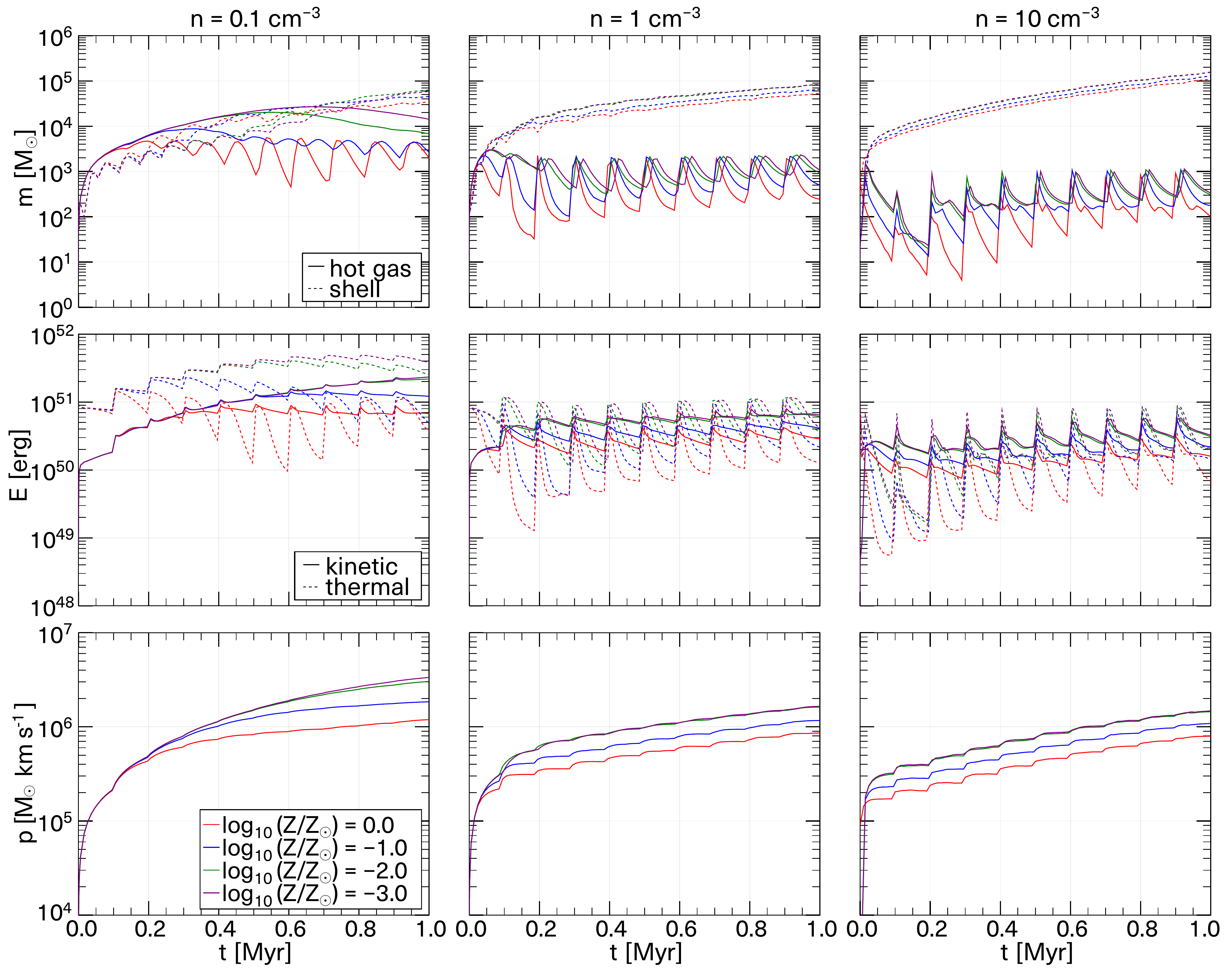}
    \caption{Same as Figure~\ref{fig:SB:timeevolution 0.01}, but for SN injection time interval $\dtsn$ = 0.1 Myr. Note that the time range in the abscissa is different from that in Figure~\ref{fig:SB:timeevolution 0.01}. }
    \label{fig:SB:timeevolution 0.1}
\end{figure*}

The time evolution of mass, energy, and momentum of a superbubble in the cases of $\dtsn = 0.1\,\Myr$ is shown in Fig.~\ref{fig:SB:timeevolution 0.1}.
In this case, the shell-formation time is shorter due to the lower energy injection rate than $\dtsn = 0.01\,\Myr$.

In cases of $\nh = 0.1\,\pcmq$ and $\log (Z/\Zsun) = 0.0, -1.0$, the shell formation is completed within the simulation time, after which the kinetic energy becomes roughly constant as shown by the red solid line in the left middle panel of the Fig.~\ref{fig:SB:timeevolution 0.1}.
% As shown in the right column of Figure \ref{fig:time evolution 0.01}, the shell formation completes and the kinetic energy becomes constant when the thermal energy decreases due to cooling and becomes almost equal to the kinetic energy.
% This is because when the SNR is in the adiabatic phase, about 70\% of the total energy is thermal energy, and half of it is in the shell, so when the thermal energy in the shell is dissipated by cooling, the thermal energy and kinetic energy are approximately equal.

In cases of $\nh = 1\,\pcmq$, the shell formation completes before the second SN explosion.
% The time evolution of mass shows that each SN explosion heats a part of the shell to form hot gas after the shell is formed, which then cools down again. (maybe caused by shock heating and subsequen t radiative cooling at shock front)
Each SN explosion injects $10^{51}\,\erg$ of thermal energy;
some of it is converted to kinetic energy by $PdV$ work on the shell, and the rest is dissipated by cooling so that most of the thermal energy is lost within 0.1\,Myr following energy injection.
From the phase diagram at $t = 0.53\,\Myr$ for $\log (Z/\Zsun) = 0.0$ (Fig.~\ref{fig:SB:phasediagram}), the cooling time of the gas inside the shell is longer than 1\,Myr.
However, heat is transported to the shell by gas mixing at its rear surface and then radiated away \citep{2019MNRAS.490.1961E, 2020ApJ...894L..24F, 2021ApJ...914...89L, 2021ApJ...914...90L}, resulting in the loss of thermal energy within a duration shorter than 1\,Myr.
We note that our resolution is not high enough to resolve the conductive interfaces, and some of the thermal energy could be lost due to numerical diffusion.
% While the pressure inside the shell is high, the shell is driven by the pressure. 
% \deleted{As the pressure decreases, the shell grows while conserving momentum; as the mass increases, kinetic energy decreases.}
As shown in the top middle panel of Fig.~\ref{fig:SB:timeevolution 0.1}, the shell grows in mass while conserving momentum after each SN injection. 
At the same time, the kinetic energy decreases as shown in the central panel of Fig.~\ref{fig:SB:timeevolution 0.1}.

\subsubsection{Results for $\dtsn = 1\,\Myr$}

\begin{figure*}
    \centering
    \includegraphics[width=\textwidth]{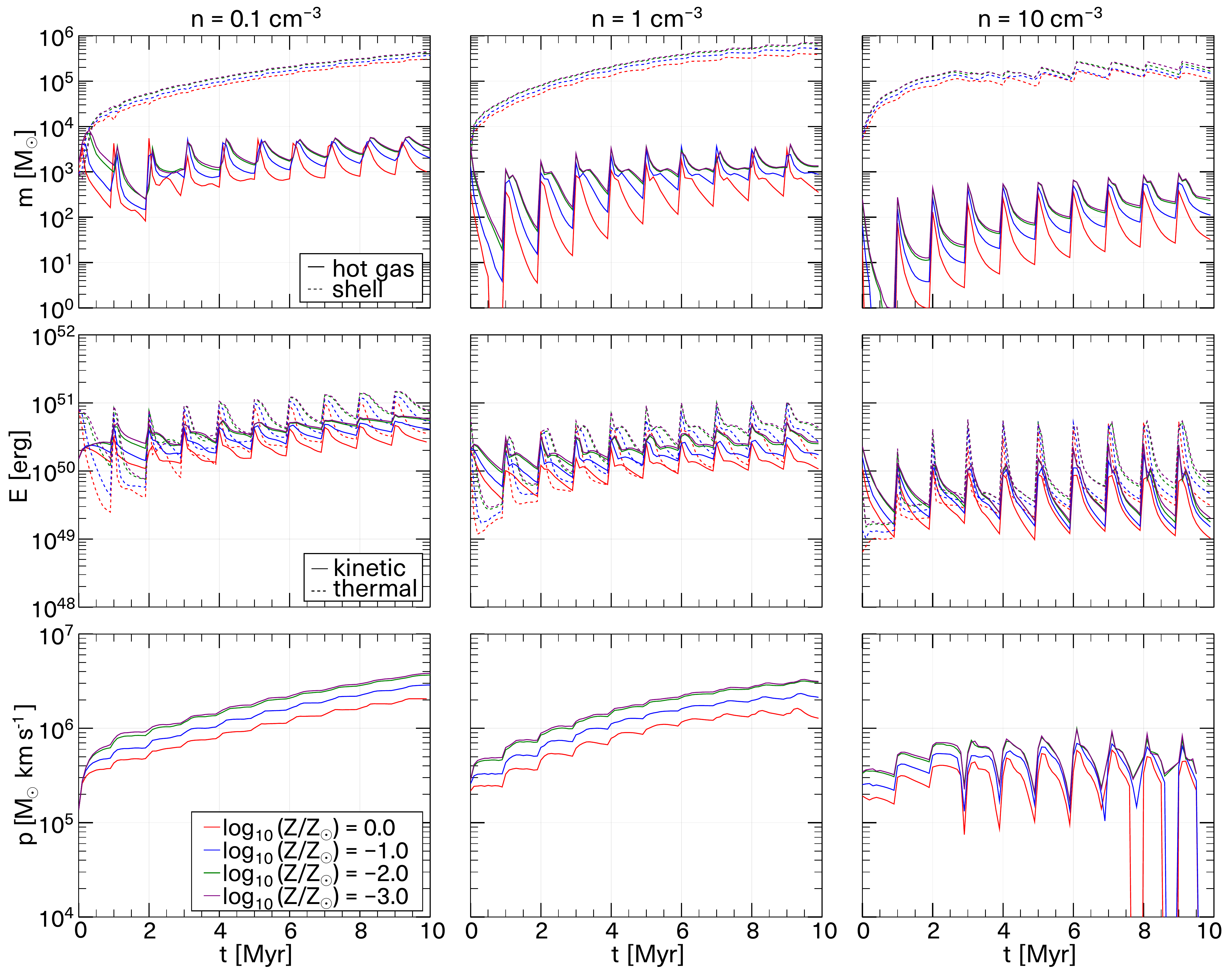}
    \caption{Same as Figure~\ref{fig:SB:timeevolution 0.01}, but for SN injection time interval $\dtsn$ = 1 Myr. Note that the time range in the abscissa is different from that in Figure~\ref{fig:SB:timeevolution 0.01}. }
    \label{fig:SB:timeevolution 1}
\end{figure*}

The time evolution of superbubble mass, energy, and momentum in the case of $\dtsn = 1\,\Myr$ is shown in Fig.~\ref{fig:SB:timeevolution 1}.
In this case, the shell is formed before the second SN explosion occurs in all cases; therefore the time evolution is similar to that for $\dtsn = 0.1\,\Myr,\,\nh = 1,\,10\,\pcmq$. 
At $t > 4\,\Myr$, thermal energy exceeds kinetic energy.
At this time, most of the mass is in the shell with its interior being cooled. Therefore, the gas in the shell contains most of the thermal energy of the superbubble.
As shell mass increases while conserving momentum, its kinetic energy decreases.
Let the velocity, temperature, and density of the shell be $v_{\rm shell}$, $T_{\rm shell}$, and $\rho_{\rm shell}$, respectively. 
When the kinetic energy decreases to $(1/2)\rho_{\rm shell}v_{\rm shell}^2 < (3/2)(\rho_{\rm shell}/\mu m_{\rm H})k_{\rm B} T_{\rm shell}$, the thermal energy becomes larger. 
If the temperature of the shell is around the ISM temperature $T_{\rm ISM}\sim 2\times 10^3\,\kelvin$, then the thermal energy exceeds kinetic energy when $v_{\rm shell} < 7\,\kms$.
When $\nh = 10\,\pcmq$, the mass and momentum appear to be decreasing when $t$ is large.
This is because the shell velocity has fallen below 3$\,\kms$, whereby we can no longer follow its time evolution.
In such a case, the shell is expected to gradually mix with the ISM.

\subsection{Application to SN feedback model in galaxy simulations}
\label{sec:SB:fitting}
\begin{figure*}
    \centering
    \includegraphics[width=.8\textwidth]{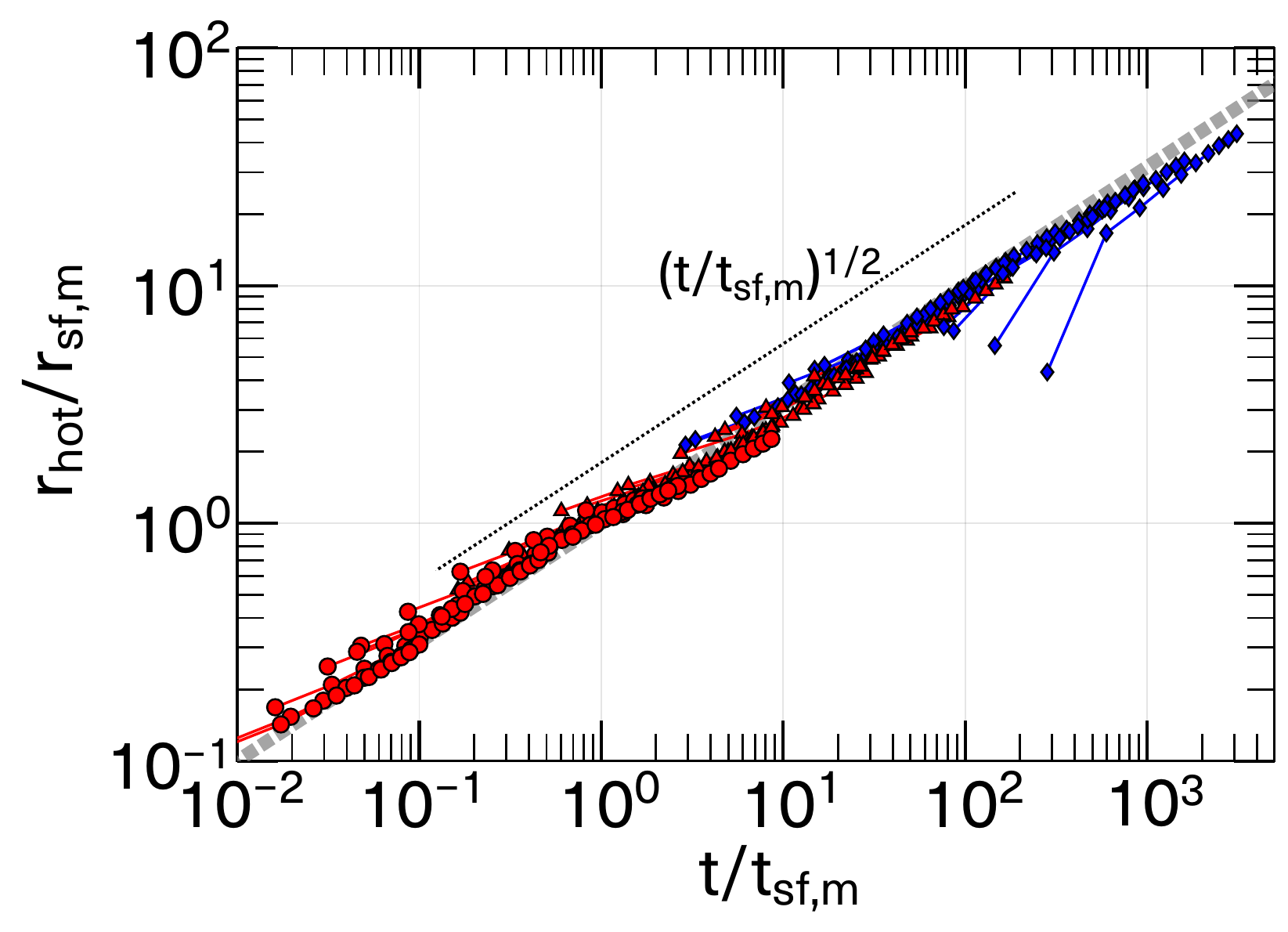}
    \caption{Evolution of $R_{\rm hot}$ normalized by $R_{\rm sf, m}$, versus time normalized by $t_{\rm sf, m}$.
    Red and blue points depict the runs with $\dtsn < 0.1 t_{\rm PDS}$ and  $\dtsn > 0.1 t_{\rm PDS}$, respectively.
    % Different symbols connected by line correspond to runs at different densities: $\nh = 0.1\,\pcmq$ (circle), $\nh = 1\,\pcmq$ (triangle), $\nh = 10\,\pcmq$ (diamond).
    Different symbols connected by line correspond to runs at different SN intervals: $\dtsn = 0.01$\,Myr (circle), $\dtsn = 0.1$\,Myr (triangle), $\dtsn = 1$\,Myr (diamond).
    % The black dotted line indicates the $(R_{\rm hot}/R_{\rm sf,m}) \propto (t/t_{\rm sf})^{1/2}$ power-law.
    The thick black dotted line shows the power-law of Eq.~(\ref{eq:SB:radiusfit}). %($R_{\rm hot}/r_{\rm sf, m}) = (t/t_{\rm sf, m})^{0.5}$.
    }
    \label{fig:SB:radius}
\end{figure*}

\begin{figure*}
    \centering
    \includegraphics[width=.8\textwidth]{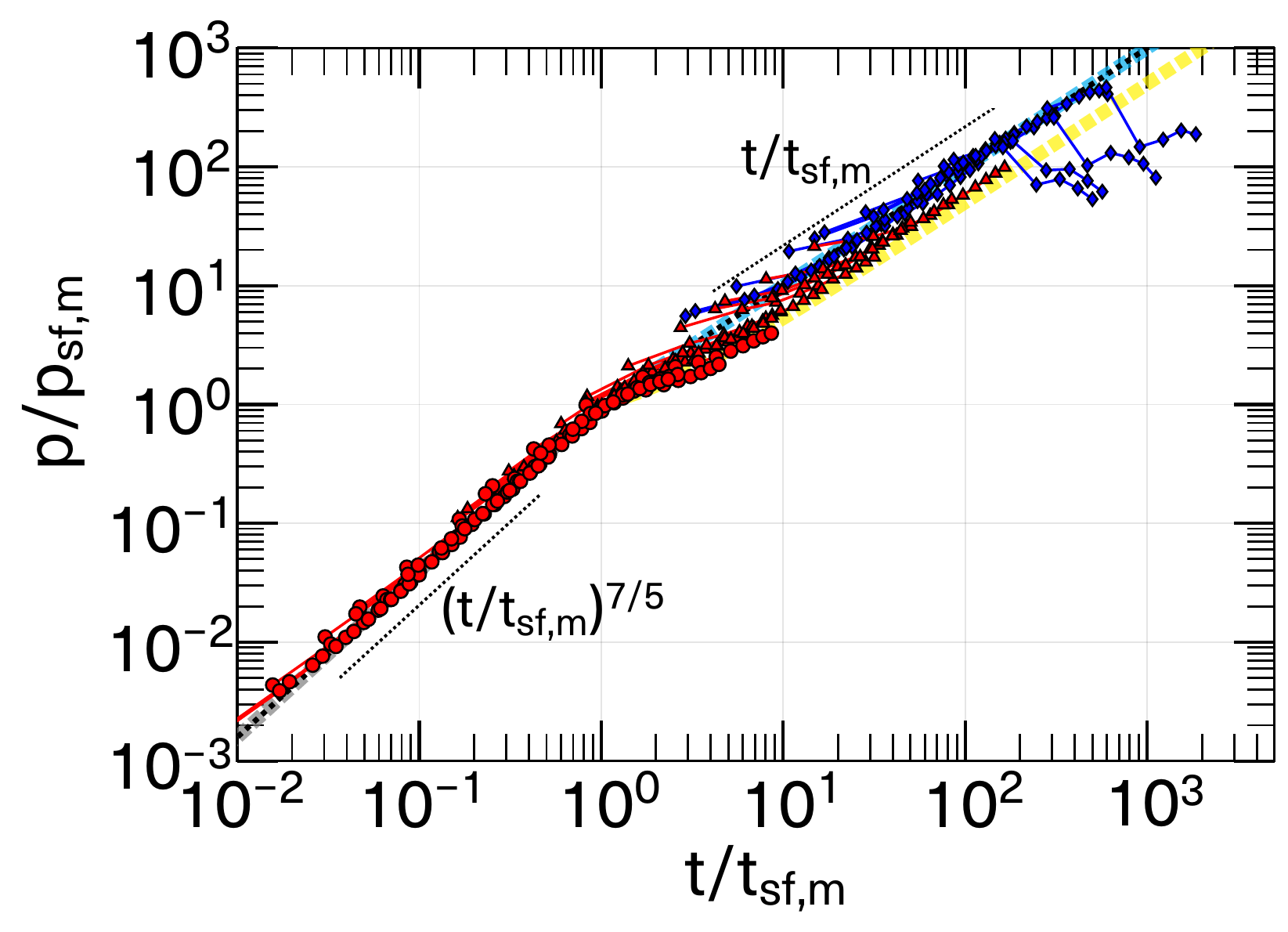}
    \caption{Evolution of momentum $p$ normalized by $p_{\rm sf, m}$, versus time normalized by $t_{\rm sf, m}$.
    Red and blue points depict the runs with $\dtsn < 0.1 t_{\rm PDS}$ and $\dtsn > 0.1 t_{\rm PDS}$, respectively.
    % Different symbols connected by line correspond to runs at different densities: $\nh = 0.1\,\pcmq$ (circle), $\nh = 1\,\pcmq$ (triangle), $\nh = 10\,\pcmq$ (diamond).
    Different symbols connected by line correspond to runs at different SN intervals: $\dtsn = 0.01$\,Myr (circle), $\dtsn = 0.1$\,Myr (triangle), $\dtsn = 1$\,Myr (diamond).
    The black dotted lines indicates the $(p/p_{\rm sf,m}) \propto (t/t_{\rm sf, m})^{7/5}$ and $(p/p_{\rm sf,m}) \propto (t/t_{\rm sf, m})$ power-law, respectively.
    The thick gray, yellow and cyan dotted lines indicate the fitting functions of  Eq.~(\ref{eq:SB:fit_momentum_a}), (\ref{eq:SB:fit_momentum_b}), and (\ref{eq:SB:fit_momentum_c}), respectively.
    %shows the power-law ($p/p_{\rm sf, m}) = (t/t_{\rm sf})^{7/5}$ at $t<t_{\rm sf, m}$
    }
    \label{fig:SB:momentum}
\end{figure*}
In this section, we derive the scaling relations for the time evolution of superbubble momentum and radius for the application to the SN feedback model in galaxy simulations.

% \subsubsection{Fit to simulation results}
% \label{sec:SB:fitting}
% In this section, we discuss the time evolution of superbubble radius and momentum. 

To discuss the environmental dependence of the superbubble in a unified manner, we normalize each physical quantity by its value at the shell-formation time.
The same was done for single SN explosion simulations by \citet{2015ApJ...802...99K}, and also by \citet{2017ApJ...834...25K} for only the time variable for multiple SN explosion simulations.

Figure \ref{fig:SB:radius} shows the time evolution of the radius, normalized by its value at the shell-formation time.
The time evolution of the radius shows some deviations, but in all cases, we can see that the radius evolves according to
\begin{equation}
    \frac{R_{\rm hot}}{R_{\rm sf,m}} = \left( \frac{t}{t_{\rm sf,m}} \right)^{0.5}\,, 
    \label{eq:SB:radiusfit}
\end{equation}
as was also found by \citet{2017ApJ...834...25K}.
When $t < t_{\rm sf, m}$, the superbubble is in an adiabatic period and expected to evolve with $R\propto t^{3/5}$ (Eq.~\ref{eq:xi_multi}).
The good fit of the straight line with a slope of 0.5 can be attributed to the fact that the energy injection is discrete rather than continuous, resulting in an intermediate time evolution with the Sedov--Taylor solution $R\propto t^{2/5}$ (Eq.~\ref{eq:sedovtaylor}).
The asymptote of the fitting line is due to the transition from the Sedov--Taylor solution to the adiabatic solution of the superbubble.
Time evolution after shell-formation can also be approximated by a straight line with a slope of 0.5.
The asymptotic behavior of the blue lines from the bottom is because the hot-gas radius $R_{\rm hot}$ could not correctly follow the radius of the inside of the shell when it was cooled down to $T<2\times10^4\,\kelvin$.
The scaling relation $R \propto t^{1/2}$ agrees with the prediction from the momentum-driven model and not with that from the pressure-driven model (Appendix~\ref{sec:SBtheory}).
% The time evolution model of the superbubble after shell formation is $R\propto t^{7/5}$ for the pressure-driven model and $R\propto t^{1/2}$ for the momentum-driven model (Appendix~\ref{sec:SBtheory}),
% Therefore, we believe that the momentum-driven model can explain the time evolution of the radius.
Substituting the values of $R_{\rm sf, m},\,t_{\rm sf, m}$ (Eqs. (\ref{eq:Rsfmulti}), (\ref{eq:tsfmulti})), the time evolution of radius and velocity are obtained as
\begin{eqnarray}
    R &=& 40\,{\rm pc}~t_6^{1/2}\,E_{51}^{0.23}\,\Delta t_{\rm SN, 6}^{-0.23}\,n_0^{-0.28}\,\Lambda_{6,-22}^{-0.040}, \label{eq:SB:Rbubble}\\
    V &=& 20\,\kms~t_6^{-1/2}\,E_{51}^{0.23}\,\Delta t_{\rm SN, 6}^{-0.23}\,n_0^{-0.28}\,\Lambda_{6,-22}^{-0.040}.
\end{eqnarray}

Next, we show the time evolution of momentum in  Figure~\ref{fig:SB:momentum}.  
It shows two different trends after the shell-formation time.
If the interval between SN explosions is longer than the duration of the pressure-driven snowplow (PDS) phase of a single SN explosion, $t_{\rm PDS}$\,(Eq.~\ref{eq:t_PDS}), energy injection is considered to be discrete and the superbubble is expected to display an evolution over time that is different from the continuous case.
Since thermal transfer to the shell was not considered in the derivation of $t_{\rm PDS}$, the actual duration of the PDS phase will be shorter than $t_{\rm PDS}$.

%Here we wish to derive a criteria to separate continuous vs. discrete limit of SN energy injection by comparing $\dtsn$ and $t_{\rm PDS}$, which is different from $t_{\rm subsonic}$ of \citet[][]{2019MNRAS.490.1961E}.
Here we find that the following fitting functions can describe the two separate regimes continuously as shown in Fig.~\ref{fig:SB:momentum}:
% \deleted{Assuming that the comparison of $\dtsn$ and $st_{\rm PDS}$ distinguish the two trends, 
% we searched for the value of the constant $s$ and found that the trends were distinguishable when $0.05 < s < 0.2$.
% Here, we adopt $s = 0.1$, and the time evolution of momentum is}
% \begin{equation}
%     \frac{p}{p_{\rm sf, m}} = 
%     \begin{cases}
%     \left(\frac{t}{t_{\rm sf, m}}\right)^{7/5} & (t < t_{\rm sf, m})\\
%     \frac{1}{2}\left(\frac{t}{t_{\rm sf, m}} + \left(\frac{t}{t_{\rm sf, m}}\right)^{-1/5}\right) & (t > t_{\rm sf, m}\,\&\, \dtsn < 0.1 t_{\rm PDS})\\
%     \frac{t}{t_{\rm sf, m}} & (t > t_{\rm sf, m}\,\&\,\dtsn > 0.1t_{\rm PDS})
%     \end{cases}.
%     \label{eq:SB:fit_momentum}
% \end{equation}
\begin{subnumcases}{\frac{p}{p_{\rm sf, m}} = \label{eq:SB:fit_momentum}}
    \tau^{7/5} & ($t < t_{\rm sf, m}$) \label{eq:SB:fit_momentum_a} \\
    \frac{1}{2}\left(\tau + \tau^{-1/5}\right) & ($t > t_{\rm sf, m}\,\&\, \frac{\dtsn}{t_{\rm PDS}}< 0.1$)  \label{eq:SB:fit_momentum_b} \\
    \tau & ($t > t_{\rm sf, m}\,\&\,\frac{\dtsn}{t_{\rm PDS}} > 0.1$),  \label{eq:SB:fit_momentum_c}
\end{subnumcases}
where $\tau = t/t_{\rm sf, m}$.
Since the momentum can be estimated as $p = (4/3)\pi R^3 \rho \dot{R}$, when the time evolution of the radius is expressed as $R \propto t^{\alpha}$, $p \propto t^{4\alpha - 1}$.
Before shell formation, the superbubble grows adiabatically, and the time evolution of its radius is $R\propto t^{3/5}$; thus, the time evolution of momentum is $p \propto t^{7/5}$.

When $t \gg t_{\rm sf, m}$, there occurs a deviation from the fitting line.
This was also seen when $\dtsn = 1\,{\rm Myr},\,\nh = 10\,\pcmq$ (Fig.~\ref{fig:SB:timeevolution 1}), because the shell velocity dropped below $3\,\kms$ and its time evolution could not be followed.

In passing, we note that \citet[][]{2019MNRAS.490.1961E} derived a criterion to distinguish discrete energy injection versus continuous limit in their Eq.\,(53).  They derived this criteria only by considering the blast waves from individual SNe before reaching the shell, and their simulation results asymptote to their modified energy-driven solution with cooling at $t > t_{\rm subsonic}$. 
%Our fitting function given in Eq.~(\ref{eq:SB:fit_momentum}) provides a more continuous and  broader perspective of evolution than Eq.\,(53) of \citet[][]{2019MNRAS.490.1961E}, although our usage of $0.1 t_{\rm PDS}$ in Eq.~(\ref{eq:SB:fit_momentum}) is somewhat arbitrary. 
We only simulated ten SN explosions, while \citet[][]{2019MNRAS.490.1961E} examined the evolution all the way to the continuous energy injection limit (although with 1-D simulation); therefore our time-scale shown in Fig.\,9 is somewhat of shorter time range than that of \citet[][]{2019MNRAS.490.1961E}'s work (see their Fig. 6).
The subsonic timescale $t_{\rm subsonic}$ in Eq.(53) of \citet[][]{2019MNRAS.490.1961E} is much longer than our $t_{\rm PDS}$, and therefore it is not straightforward to compare our results using their criteria based on $t_{\rm subsonic}$.

\section{Construction of SN feedback model}
\label{sec:FBmodel}
In this section, we describe the construction of our SN feedback model for galaxy simulations. 
Here we assume using the model in SPH simulation. 
Our model is based on the results from the previous section and on existing works on high-resolution small-box simulations.
% The weighting scheme described here is used for stellar wind feedback, Type II SN feedback, Type Ia SN feedback, and AGB feedback. Details of feedback model of these feedbacks are different as discribed in Section \ref{sec:Isogal:stellarfeedback}.

\subsection{Averaged momentum per SN for a stellar population}
From the fitting function in Sec.~\ref{sec:SB:fitting}, we derive the expression for momentum of a superbubble per SN, averaged over an initial stellar cluster mass function (ICMF). 
We assume the ICMF to be described by the power-law $dN/dM_\mathrm{c} \propto M_\mathrm{c}^{-2}$ with high-mass cutoff, where $M_\mathrm{c}$ is the mass of the stellar cluster \citep[][]{2019ARA&A..57..227K} and stellar initial mass function (IMF) is fixed. We also assume SNe occurs at a constant rate. The ICMF can be translated to an SN number function (probability distribution function of the number of SN explosions occurring in a stellar cluster)
\begin{equation}
    \frac{dN}{dN_\mathrm{SN}} = A N_\mathrm{SN}^{-2},
\end{equation}
where $N_\mathrm{SN}$ is the number of SN explosions occurring in a stellar cluster and $A$ is the normalization factor. 
The range of the number of SN explosions is assumed as $N_\mathrm{SN} = N_{\rm SN, min} - N_{\rm SN, max}$, where we adopt the range of ($N_{\rm SN, min}$, $N_{\rm SN, max}$) = (1, 500), and the normalization factor is $A = 1/\ln{500}$.
The value of $N_\mathrm{SN, max} = 500$ is the number of SN explosions expected to happen in the largest young massive star clusters (YMCs) in the Milky way with a mass of $M = 5\times10^4\,\Msun$ \citep{2010ARA&A..48..431P}.
The choice of high-mass cutoff has little effect on the result since such high-mass YMCs are rare objects. 
The SNe interval is obtained by dividing their duration by their number, $\Delta t_\mathrm{SN} = (t_\mathrm{SN, end} - t_\mathrm{SN, start})/N_\mathrm{SN}$, where $t_\mathrm{SN, start},\,t_\mathrm{SN, end}$ are the times at the first and the last Type II SN after the formation of the stellar cluster, i.e., the minimum and the maximum lifetimes of stars causes Type II SNe. 
The lifetimes of stars depend on their mass and metallicity. We calculate SNe duration using \textsc{CELib}\footnote{\url{https://bitbucket.org/tsaitoh/celib}} \citep{2017AJ....153...85S}, which compute it using the metallicity-dependent stellar lifetime table by \citet{1998A&A...334..505P}. 
We assume the mass range of the progenitors of Type II SNe to be 13--40 $\Msun$. 
In the range of $Z = 10^{-6}$--$10^{-2}$, SNe duration is %$t_\mathrm{SN,end} - t_\mathrm{SN,start} \sim 1.2 \times 10^7$ yr. 
\begin{equation}
t_\mathrm{SN,end} - t_\mathrm{SN,start} \sim 1.2 \times 10^7 {\rm yr}. 
\label{eq:tduration}
\end{equation}
Single SN explosion energy is set to $10^{51}$\,erg. 
The averaged momentum per SN is
\begin{equation}
    \hat{p}(n_0, Z) = \int_{N_{\rm SN, min}}^{N_{\rm SN, max}} p(n_0, Z, \Delta t_\mathrm{SN}(N'_\mathrm{SN}))\, \frac{dN}{dN'_\mathrm{SN}}\,dN'_\mathrm{SN}.
\end{equation}
The final integral is well-fitted by the following function:
\begin{equation}
    \hat{p}(n_0, Z) = 1.75 \times 10^5\,\Msun\,\kms~n_0^{-0.05} \Lambda_{6,-22}^{-0.17},
\end{equation}
and the momentum input by ${\cal N}_{\rm SN}$ SNe (superbubble momentum, $p_{\rm SB}$) is estimated as
\begin{equation}
    p_{\rm SB}({\cal N}_{\rm SN}, n_0, Z) = {\cal N}_{\rm SN}\,\hat{p}(n_0, Z).
    \label{eq:SB:terminalmomentum}
\end{equation}
Here we used a slightly different font of  ${\cal N}_{\rm SN}$ for the total number of SNe for a star particle in the simulation (i.e., a collection of stellar clusters, rather than a population of stars in a single stellar cluster).

\subsection{Shock radius for a star particle}

We similarly calculate the shock radius of SN feedback of a star particle. 
Assuming that stellar clusters follow the power-law ICMF, 
we determine the shock radius of the SN feedback as the radius of a sphere whose volume is the sum of those of the superbubbles. 
We first consider the averaged volume shocked by SN feedback {\em per SN} considering the variation of superbubbles created by a range of star clusters:
\begin{equation}\label{eq:FBmodel:vol}
    \hat{V}(n_0, Z) = \int_{N_{\rm SN, min}}^{N_{\rm SN, max}} \frac{4}{3}\pi R(n_0, Z, \Delta t_\mathrm{SN}(N'_\mathrm{SN}))^3\,\frac{dN}{dN'_\mathrm{SN}}\,dN'_\mathrm{SN}.
\end{equation}
Then the effective radius is obtained as
\begin{equation}
    \hat{R}(n_0, Z) = \left(\frac{3\hat{V}}{4\pi}\right)^{1/3} 
    = 61\,\mathrm{pc}~n_0^{-0.22}\,\Lambda_{6, -22}^{-0.04},
\end{equation}
and the shock radius by ${\cal N}_{\rm SN}$ SNe is calculated as
\begin{equation}
    R({\cal N}_{\rm SN}, n_0, Z) = {\cal N}_{\rm SN}^{1/3}\,\hat{R}(n_0, Z).
    \label{eq:SB:shockradius}
\end{equation}
Here, ${\cal N}_{\rm SN}$ is meant to be the number of SNe occurring in the stellar population of a star particle. 
%Since it is unphysical to assume that a star particle is one star cluster, we first derive a typical shocked volume of an SN, and then obtain the (effective) shock radius by multiplying by ${\cal N}_{\rm SN}^{1/3}$.
Since Eq.\,(\ref{eq:FBmodel:vol}) gives the averaged superbubble volume {\em per SN}, the effective volume for $\mathcal{N}_{\rm SN}$ SNe for a star particle would be $\mathcal{N}_{\rm SN} \hat{V}$ and the effective radius would be $\mathcal{N}_{\rm SN}^{1/3} \hat{R}$ as given in Eq.\,(\ref{eq:SB:shockradius}).

{Another way to determine the shock radius is to use the radius of a single superbubble described in Eq.~(\ref{eq:SB:Rbubble}) at %$t = N_{\bf SN}\dtsn$, 
\begin{equation}
t = N_{\rm SN}\dtsn, 
\label{eq:tNsn}
\end{equation}
assuming that the percolated superbubbles from multiple star clusters can be approximated well as a single superbubble. 
Then, inserting Eq.~(\ref{eq:tNsn}) into Eq.~(\ref{eq:SB:Rbubble}) and using Eq.~(\ref{eq:tduration}) with $E_{51}=1$ yields the shock radius %$R'({\cal N}_{\rm SN}, n_0, Z) = 78\,{\rm pc}\,{\cal N}_{\rm SN}^{0.23}\,n_0^{-0.28}\,\Lambda_{6, -22}^{-0.04}(Z)$.
\begin{equation}
R'({\cal N}_{\rm SN}, n_0, Z) = 78\,{\rm pc}\,{\cal N}_{\rm SN}^{0.23}\,n_0^{-0.28}\,\Lambda_{6, -22}^{-0.04}(Z).
\label{eq:refereeRsn}
\end{equation}

Similarly to the above, one can interpret  Eq.~(\ref{eq:SB:shockradius}) as the shock radius of a single superbubble formed by ${\cal N}_{\rm SN}$ SNe at the age of $t = 4.8\,\Myr\,{\cal N}_{\rm SN}^{0.38}\,n_0^{0.22}$, which can be obtained by equating Eq.~(\ref{eq:SB:Rbubble}) and (\ref{eq:SB:shockradius}). 
In other words, the above two different methods of computing the effective shock radius can be interpreted as single superbubble radius at different times. 

In the following, we use Eq.~(\ref{eq:SB:shockradius}) 
together with Eq.~(\ref{eq:L_6}) and  Eq.~(\ref{eq:SB:terminalmomentum}) 
for our SN feedback model described in the next section, 
%to determine the effective shock radius of a star particle in galaxy simulations, 
but the choice of either Eq.~(\ref{eq:SB:shockradius}) or (\ref{eq:refereeRsn})
is unlikely to affect our final results.} 

%In combination with equation (\ref{eq:L_6}), we use equations (\ref{eq:SB:terminalmomentum}) and (\ref{eq:SB:shockradius}) in our SN feedback model described in the next section. 

\subsection{Spherical superbubble model}
\label{sec:FBmodel:model}
\begin{figure*}[t]
    % \centering
    % \includegraphics[width=\textwidth]{figures/Isogal/SNFBmodel.pdf}
    \plotone{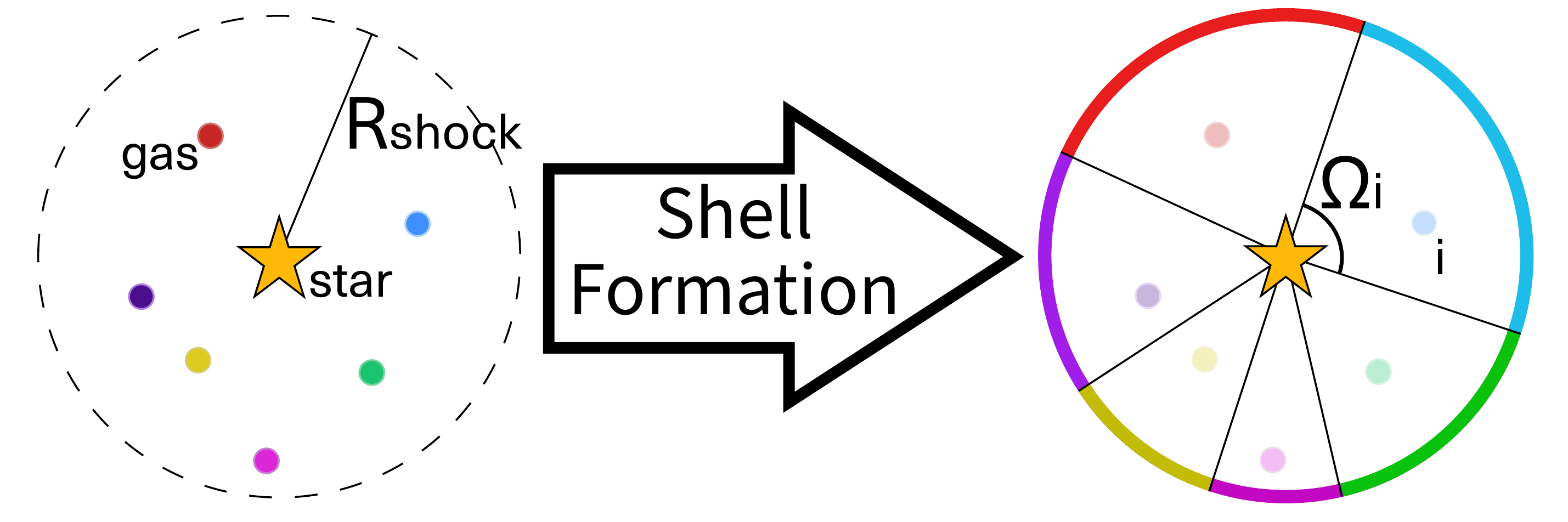}
    \caption{Schematic description of the spherical superbubble feedback model developed in this paper.
    The left hand side shows the particle distribution of gas and star particles when the supernova explosions begin, whereas the right hand side shows the superbubble shell formation, and how we split the shell using Voronoi tessellation based on the gas particle distribution inside the bubble. 
    }
    \label{fig:FBmodel:model}
\end{figure*}

A schematic description of the `Spherical superbubble model' developed in this paper is illustrated in Fig.~\ref{fig:FBmodel:model}.
When the SN event occurs, we first calculate the density and metallicity at the SN site. 
An iterative solver is used to find a smoothing length for the stellar particle that satisfies
\begin{equation}
    h_{\mathrm{sml}, i} = \left(\frac{3 N_{\rm ngb}}{4 \pi \sum_j W(r_{ij}, h_{\mathrm{sml}, i})} \right)^{1/3},
    \label{eq:FBmodel:stellarhsml}
\end{equation}
where $r_{ij}$ is the distance from the $i$-th stellar particle to the $j$-th gas particle, $N_{\rm ngb}$ is the number of neighboring SPH particles, and $W$ is the kernel function adopted in the SPH simulation.
We then compute the shock radius $R_{\rm shock}$ using local density and metallicity, using equation (\ref{eq:SB:shockradius}).
% The shock radius is unresolved in most cases, but if the shock radius exceeds SPH smoothing length, we can resolve the spatial scale of feedback.
% In such cases, we limit the shock radius by SPH smoothing length, and calculate the time evolution of superbubble directly in our simulation.
% The smoothing length at the position of the $i$-th stellar particle is estimated as
We search for the gas particles within the shock radius and project them from the position of the star onto a sphere centered at the star with radius $R_{\rm shock}$.
Then, we construct a Voronoi polyhedron using \textsc{STRIPACK}\footnote{\url{https://people.sc.fsu.edu/~jburkardt/f_src/stripack/stripack.html}} \citep{renka1997algorithm}, which %\textsc{STRIPACK} 
is an algorithm that constructs a Voronoi diagram of a set of points on the surface of a sphere.
After constructing the Voronoi polyhedron on the spherical surface, we calculate $\Omega$, which is the solid angle of the corresponding face on the Voronoi polyhedron from the star.
We distribute physical quantities from the SN to neighboring gas particles weighted by $\Omega$.
The mass and metal deposited on the $i$-th gas particle are
\begin{eqnarray}
    \Delta m_i &=& m_{\rm SN}\left(\frac{\Omega_i}{4\pi}\right),\\
    % &\Delta m_i = \Delta m_{\mathrm{kin}, i} + \Delta m_{\mathrm{th}, i}\\
    \label{eq:FBmodel:m}
    \Delta m_{Z, i} &=& m_{Z, {\rm SN}}\left(\frac{\Omega_i}{4\pi}\right),
    % &\Delta m_{Z, i} = \Delta m_{Z, \mathrm{kin}, i} + \Delta m_{Z, \mathrm{th}, i},
    \label{eq:FBmodel:mZ}
    % &\Delta E_{\mathrm{th}, i} = 0.7E_{\rm SN}\frac{\Omega_i}{4\pi}
    % \label{eq:FBmodel:E_th}\\
    % &\Delta E_{\mathrm{kin}, i} = 0.3E_{\rm SN}\frac{\Omega_i}{4\pi}
    % \label{eq:FBmodel:E_kin},
\end{eqnarray}
% where $\Delta m_{\mathrm{kin}, i}$ and $\Delta m_{\mathrm{th}, i}$ are mass input with kinetic and thermal feedback, $\Delta m_{Z, \mathrm{kin}, i}$ and $\Delta m_{Z, \mathrm{th}, i}$ are metal input with kinetic and thermal feedback, $E_{\rm SN}$ is energy from SN calculated using the \textsc{CELib}
where $m_\mathrm{SN}$ and $m_{Z, \mathrm{SN}}$ are the mass and metal inputs from feedback and $\Omega_i$ is the solid angle of the corresponding face on the Voronoi polyhedron from the star. 
If the number of gas particles inside the shock radius is less than four, the Voronoi polyhedron cannot be constructed.
In that case, we search for at least two nearest gas particles and equally assign mass and metals to them.
Note that our method is similar to, but different in detail from, that of \citet{2018MNRAS.477.1578H}.

% The amounts of mass and metal input with each feedback are described respectively in Section \ref{sec:FBmodel:kinetic} and \ref{sec:FBmodel:thermal}. 
% We use $\Delta E_{\mathrm{kin}, i}$ to set an upper limit of momentum input. 
% The thermal energy $\Delta E_{\mathrm{th}, i}$ is injected stochastically as described in Section \ref{sec:FBmodel:thermal}.

% In this spherical superbubble model, we assume a superbubble formed by SN feedback grows spherically. We saw a superbubble grows almost spherically in a turbulent ISM (Figure \ref{fig:SB:slice plot}), which motivates us to develop this model. Superbubbles acquire momentum with sweeping up ISM, and we model this process by projection and Voronoi tessellation on a sphere.
% In a strongly perturbed ISM, the shapes of SN bubbles are no longer spherical, and one should consider the density structure of the ISM \citep{2019MNRAS.485.3887O}. 
% However, in a sub-grid scale, the density structure of the ISM is not resolved well, thus it is difficult to predict the shape of a superbubble. 
% In our model, SN feedback is weighted by a solid angle $\Omega$, and the low-density region is expected to receive stronger feedback.

\subsection{Mechanical feedback}
\label{sec:FBmodel:kinetic}

% from local density and metallicity at the location of the SN explosion
Using the superbubble momentum $p_{\rm SB}$ computed in Eq.~(\ref{eq:SB:terminalmomentum}), we deposit the following momentum on the $i$-th gas particle:
\begin{equation}
    \Delta \bm{p}_i = p_{\rm SB} \frac{\Omega_i}{4\pi}\bm{n}_i,
\end{equation}
where $\bm{n}_i$ is the normal vector of the face on the Voronoi polyhedron.
When the number of neighboring gas particles falls below four (which prevents the construction of a Voronoi polyhedron), we inject the same amount of momentum and determine $\bm{n}_i$ so that the total momentum is conserved.
The total momentum of the surrounding gas should be conserved before and after the SN event. For this, we compute the total momentum input as:
\begin{equation}
    \Delta \bm{p}_{\rm tot} = \sum_i \Delta \bm{p}_i,
\end{equation}
and modify momentum input to $i$-th gas particle to 
\begin{equation}
    \Delta \bm{p}_i' = \Delta \bm{p}_i - \frac{\Omega_i}{4\pi}\Delta \bm{p}_{\rm tot}.
    \label{eq:FBmodel:momentum-conservation}
\end{equation}

% We calculate the shock radius and the terminal momentum self-consistently from local density and metallicity. However, 
The kinetic energy input by momentum kick may exceed the SN energy input due to particle distribution.
Thus, we limit the momentum input to each neighboring particle based on the Sedov--Taylor solution. The resulting momentum input is:
\begin{equation}
    \Delta \bm{p}_i'' = 
    % \begin{cases}
    % \Delta \bm{p}_i' &\left(\left|\Delta \bm{p}_i'\right| > \sqrt{2 (m_i + \Delta m_i) \Delta E_{\mathrm{kin}, i}}\right)\\
    % \frac{\Delta \bm{p}_i'}{\left|\Delta \bm{p}_i'\right|}\sqrt{2(m_i + \Delta m_i) \Delta E_{\mathrm{kin}, i}} & (\mathrm{otherwise})
    % \end{cases},
    \frac{\Delta \bm{p}_i'}{\left|\Delta \bm{p}_i'\right|}\,\min\left(\left|\Delta \bm{p}_i'\right|, \sqrt{2(m_i + \Delta m_i) \Delta E_{\mathrm{kin}, i}}\right),
    \label{eq:FBmodel:energy-conservation}
\end{equation}
where $m_i$ is the mass of the $i$-th gas particle and $\Delta E_{\mathrm{kin}, i}$ is the solid-angle-weighted kinetic energy from SN feedback. $\Delta E_{\mathrm{kin}, i}$ is given as
\begin{equation}
    \Delta E_{\mathrm{kin}, i} = \epsilon_{\rm kin}E_{\rm SN}\frac{\Omega_i}{4\pi},
\end{equation}
where $\epsilon_{\rm kin}$ is a fraction of the SN energy  deposited as kinetic energy, and we adopt $\epsilon_{\rm kin} = 0.3$ as the default value \citepalias{2019MNRAS.484.2632S}.
Equation (\ref{eq:FBmodel:energy-conservation}) essentially corresponds to equation~(A1) in \citet[][]{2014ApJ...788..121K} and equation~(32) in \citet[][]{2018MNRAS.477.1578H}. 

The momentum input above is calculated with respect to the frame moving with the star particle. 
To ensure exact conservation, we require a term accounting for the relative motion between the gas and the star \citep{2018MNRAS.477.1578H}. 
Finally, the momentum input boosted back to the simulation frame is 
\begin{equation}
    \Delta \bm{p}_i''' = \Delta \bm{p}_i'' + \Delta m_i \bm{v}_{\rm star},
\end{equation}
where $\bm{v}_{\rm star}$ is the star velocity.

In summary, we first compute $\Delta \bm{p}_i$ for the momentum that we want to assign the neighboring gas particles with. 
Then $\Delta \bm{p}_i'$ makes it isotropic and we ensure energy conservation by $\Delta \bm{p}''_i$.  
Finally, $\Delta \bm{p}'''_i$ takes care of the motion of the originating stellar particle.  By giving $\Delta \bm{p}'''_i$, the momentum feedback is basically guaranteed to be isotropic, energy-conserving, and momentum-conserving. 

To be more specific, it is possible that momentum conservation could be broken when the second term in the RHS of Eq.~(\ref{eq:FBmodel:energy-conservation}) is chosen.
% we break the momentum conservation ensured in Eq.~(\ref{eq:FBmodel:momentum-conservation}).
However, such a situation does not arise very often because we do not have sufficient to resolve the Sedov--Taylor phase.
In fact, we have checked our isolated galaxy simulations performed in Section~\ref{sec:Isogal}, the second term was chosen 
% of Eq.~(\ref{eq:FBmodel:energy-conservation}) 6 times in 17,879,733 cases 
only $3.4\times10^{-5}$\,\% of the cases for the Fiducial run,
% and 288 in 262,085,790 
and $1.1\times10^{-4}\,\%$ for the High-reso run. 

\subsection{Thermal feedback}
\label{sec:FBmodel:thermal}
Several groups have studied the SN-driven outflow using high-resolution small-box simulations. 
\citet{2019MNRAS.483.3363H} investigated SN-driven outflow of a dwarf galaxy. 
They showed the entropy $S \equiv \kB T n^{1 - \gamma}$ of the hot outflow to be $10^8$ -- $10^9\,\kB\,\kelvin\,\mathrm{cm}^2$, and it is almost constant after being launched from the galaxy. 
Although their result is on a dwarf galaxy, the energy loading factor and specific energy of the hot outflow by \citet{2019MNRAS.483.3363H} are similar to those of a Milky-way-mass galaxy \citep{2017ApJ...841..101L, 2019MNRAS.490.4401A, 2018ApJ...853..173K, 2020ApJ...900...61K}, according to \citet{2020ApJ...890L..30L}. 

It is suggested that the hot outflow is driven by buoyancy and its entropy is its fundamental physical quantity; if the entropy of the hot bubble is higher than that of the surrounding CGM, the hot bubble becomes buoyant and drives the outflow \citep{2017MNRAS.465...32B}. 
\citet{2020MNRAS.493.2149K} analytically calculated the entropy of superbubbles and the virialized Milky-way-mass halos to be about $10^8\,\kB\,\kelvin\,\mathrm{cm}^2$. 
The buoyancy-driven outflow framework is supported by the MUGS2 simulations \citep{2015MNRAS.453.3499K, 2016MNRAS.463.1431K}, and the framework may explain the ineffectiveness of SN feedback in halos more massive than $10^{12}\,\Msun$ \citep{2020MNRAS.493.2149K}.

In this work, we update the stochastic thermal feedback model \citep{2012MNRAS.426..140D} by using the entropy of hot outflow $S_{\rm OF}$ as a free parameter, setting $S_{\rm OF} = 10^8 \kB\,\kelvin\,\mathrm{cm}^2$ as the default value. 
When SN feedback occurs, thermal energy is stochastically injected to heat neighboring particles to target entropy, $S_{\rm OF}$. 
The thermal energy required to heat the $i$-th gas particle is
\begin{equation}
    \Delta E_{\mathrm{req}, i} = \frac{1}{\gamma - 1} \frac{m_i}{\mu m_H} n_i^{2/3} S_{\rm OF},
    \label{eq:FBmodel:E_req}
\end{equation}
where $n_i$ is the number density of the $i$-th gas particle.
The probability of injecting thermal energy is the ratio between the solid-angle-weighted thermal energy from SN feedback (Section~\ref{sec:FBmodel:model}),
\begin{equation}
    \Delta E_{\mathrm{th}, i} = \epsilon_{\rm th} E_{\rm SN}\frac{\Omega_i}{4\pi},
    \label{eq:FBmodel:E_th}
\end{equation}
and the required thermal energy:
\begin{equation}
    P_i = \frac{\Delta E_{\mathrm{th}, i}}{\Delta E_{\mathrm{req},i}},
\end{equation}
where $\epsilon_{\rm th} =1 - \epsilon_{\rm kin}$ is a fraction of the SN energy deposited as thermal energy, and we adopt $\epsilon_{\rm th} = 0.7$ as the default value \citepalias{2019MNRAS.484.2632S}.
We draw a random number $0 < \theta_i < 1$ for each gas particle, and inject thermal energy $\Delta E_{{\rm req}, i}$ to $i$-th gas particle if $\theta_i < P_i$.
When $P_i > 1$, we simply inject a thermal energy of $\Delta E_{\mathrm{th}, i}$ to the $i$-th gas particle.

The original stochastic thermal feedback model by  \citet{2012MNRAS.426..140D} uses temperature increase $\Delta T$ as a free parameter and stochastically heats gas that magnitude. Their default value $\Delta T = 10^{7.5}\,\kelvin$ was set for a numerical reason; %when set to this value, 
the expectation value for the number of heated gas particles per SN is about 1. If this value is much smaller than 1, most SN feedback events do not inject energy into their surroundings, leading to poor sampling of the SN feedback cycle \citep{2015MNRAS.446..521S}.
% On the other hand, however, 
Although they provide a good reason to use $\Delta T$ as a parameter for their stochastic feedback model, the outflow properties depend on $\Delta T$ \citep{2012MNRAS.426..140D} and a better parameter choice is desired. 
%physical motivation to choose a parameter is required. 
In this work, we use outflow entropy as a free parameter, motivated by the high-resolution simulations \citep{2019MNRAS.483.3363H} and buoyancy-driven outflow framework \citep{2017MNRAS.465...32B, 2020MNRAS.493.2149K}.

% Since the stochastic thermal feedback model represents energy injection by clustered Type II SNe, we consider thermal feedback only for Type II SN feedback.
% When clustered SNe inject energy, they should also inject the corresponding amount of mass and metals.
% We parametrize a fraction of metal injected with thermal feedback as a free parameter $f_{\rm th}$.
% When $i$-th particle receive thermal feedback, it also receives following amount of  mass and metals:
% \begin{eqnarray}
%     &\Delta m_i = f_{\rm th} \frac{\Delta E_{{\rm req}, i}}{\Delta E_{{\rm th}, i}} m_{\rm SN}\frac{\Omega_i}{4\pi},\\
%     &\Delta m_{Z, i} = f_{\rm th} \frac{\Delta E_{{\rm req}, i}}{\Delta E_{{\rm th}, i}} m_{Z, {\rm SN}}\frac{\Omega_i}{4\pi}.
% \end{eqnarray}

% \begin{figure*}
%     % \centering
%     % \plotone{Isogal_models_plotmetallicity.pdf}
%     \plotone{Isogal_models_metallicity.png}
%     % \includegraphics[width=\textwidth]{Isogal_models_metallicity.png}
%     % \includegraphics[width=\columnwidth]{Isogal_models_plotmetallicity.pdf}
%     \caption{Projected metallicity (density-weighted) of isolated galaxies given in Table \ref{tab:Isogal:simulationsettings} at $t=1$\,Gyr.
%     Top row shows face-on images, and bottom row shows edge-on images.
%     Metals carried by the outflows are visible below and above the disk.  
%     }
%     \label{fig:Isogal:metallicity20}
% \end{figure*}

\section{Isolated galaxy simulation with {\sc GADGET3-Osaka}}
\label{sec:Isogal}
In this section, we implement the SN feedback model based on the high-resolution \textsc{Athena++} simulations described in the earlier sections, and demonstrate its effect on star formation and galactic outflow in an isolated galaxy simulation. 

\subsection{Simulation Setup}
We use the {\sc GADGET3-Osaka} cosmological smoothed particle hydrodynamics (SPH) code \citep{2017MNRAS.466..105A, 2019MNRAS.484.2632S}, which is a modified version of \textsc{GADGET-3} (originally described in \citealt{2005MNRAS.364.1105S}, as \textsc{GADGET-2}\footnote{\url{https://wwwmpa.mpa-garching.mpg.de/gadget/}}).
% We solved the SPH equation of motion, derived as the Euler--Lagrange equation of the discretized Lagrangian of a perfect fluid \citep{2002MNRAS.333..649S}:
% \begin{equation}
%     L = \frac{1}{2}\sum_i m_i v_i^2 - \sum_i m_i \frac{S_i \rho_i^{\gamma - 1}}{\gamma - 1},
% \end{equation}
% where $m_i$, $v_i$, $S_i$, and $\rho_i$ are the mass, velocity, and entropy (defined as $S \equiv P/\rho^\gamma$), and the density of the $i$-th gas particle, respectively.
We solve the SPH equation of motion, following the entropy-conserving, density-independent SPH formulation \citep{2013MNRAS.428.2840H, 2013ApJ...768...44S}:
\begin{multline}
    \frac{d\bm{v}_i}{dt} = -\sum_j m_j (S_i S_j)^{1/\gamma} \\
    \times \left[ \frac{f_{ij} \bar{P}_i}{\bar{P}_i^{2/\gamma}} \nabla_i W(r_{ij},h_{\mathrm{sml}, i}) + \frac{f_{ji} \bar{P}_j}{\bar{P}_j^{2/\gamma}} \nabla_i W(r_{ij},h_{\mathrm{sml}, j}) \right],
\end{multline}
\begin{equation}
    f_{ij} = 1 - \frac{h_{{\rm  sml},i}}{3S_j^{1/\gamma}\rho_i}\frac{\partial \bar{P}_i^{1/\gamma}}{\partial h_{{\rm sml},i}}\left(1 + \frac{h_{\mathrm{sml}, i}}{3 \rho_i}\frac{\partial \rho_i}{\partial h_{\mathrm{sml}, i}}\right)^{-1},
\end{equation}
where $m_i$, $\bm{v}_i$, $S_i$, $h_{\mathrm{sml}, i}$, and $\bar{P}_i$ denote the mass, velocity, entropy (defined as $S \equiv \bar{P}/\rho^\gamma$), smoothing length, and smoothed pressure defined as
\begin{equation}
    \bar{P}_i = \left[\sum_j m_j S_j^{1/\gamma} W(r_{ij}, h_{\mathrm{sml}, i}) \right],
\end{equation} 
of the $i$-th gas particle, respectively.
In this formulation, the thermal energy of a particle is computed from its entropy and smoothed pressure. 
When thermal feedback injects energy, updating entropy assuming fixed pressure leads to the wrong result because the smoothed pressure field itself depends on the entropy.
Therefore, we use an iterative method to calculate entropy change by energy injection from feedback and energy dissipation by radiative cooling \citep{2015MNRAS.446..521S, 2021MNRAS.505.2316B}.
The self-gravity of SPH and collisionless particles are also considered.
We adopt the quintic B-spline kernel \citep{Schoenberg1946ContributionsTT, 1996PASA...13...97M}, and set the number of neighboring particles for each SPH particle  to $N_{\rm ngb} = 128 \pm 8$. 
Our code includes the time-step limiter \citep{2009ApJ...697L..99S, 2012MNRAS.419..465D}.

Radiative cooling is calculated using the \textsc{grackle-3} chemistry and cooling library\footnote{\url{https://grackle.readthedocs.io}\\The default solar metallicity in \textsc{grackle-3} is $Z_\odot = 0.0134$. This value is smaller than the value of $\Zsun$ that was used in Eq.~(\ref{eq:L_6}) by a factor of 1.45. This difference will affect the calculation of terminal momentum in Eq.~(\ref{eq:psfmulti}), however the power index on $\Lambda_{6,-22}$ term is small so that the impact on the value of $p_{\rm sf, m}$ is negligible.} \citep{2017MNRAS.466.2217S}, which solves non-equilibrium primordial chemistry and cooling for the H, D, and He species, including molecular H$_2$ and HD.
The library also includes tabulated rates of metal cooling calculated with the photoionization code \textsc{Cloudy} \citep[][]{2013RMxAA..49..137F} and photoheating and photoionization from the ultraviolet background (UVB) radiation, and we adopt the UVB value at $z = 0$ by \citet[][]{2012ApJ...746..125H}.
We applied a nonthermal Jeans pressure floor that forces the local Jeans length to be resolved to avoid artificial numerical fragmentation \citep{2011MNRAS.417..950H, 2016ApJ...833..202K}:
\begin{equation}
     P_{\rm Jeans} = \frac{1}{\gamma \pi} N_{\rm Jeans}^2 G \rho_{\rm gas}^2 \Delta x^2,
\end{equation}
where $\gamma = 5/3$ is the adiabatic index, $N_{\rm Jeans} = 4$ is the Jeans number adopted from \citet{1997ApJ...489L.179T}, $G$ is the gravitational constant, and $\rho_{\rm gas}$ is the gas density.
$\Delta x$ is chosen from the larger one of either the gravitational softening length of an SPH particle or the spatial resolution of hydrodynamics $(m_{\rm gas}/\rho_{\rm gas})^{1/3}$, where $m_{\rm gas}$ is the mass of the gas particle.

We used an initial condition taken from the AGORA project\footnote{\url{http://www.AGORAsimulations.org}} \citep{2016ApJ...833..202K}.
The galaxy has properties characteristic of Milky Way-mass galaxies at redshift $z \sim 1$.
% as summarized in table \ref{tab:Isogal:initialcondition}.
The galaxy is composed of the following components: a dark matter halo with $M_{\rm DM} = 1.25 \times 10^{12}\,\Msun$, a stellar disk with $M_{\rm disc} = 4.30 \times 10^9\,\Msun$, a stellar bulge with $M_{\rm bulge} = 3.44 \times 10^{10}\,\Msun$, and a gas disk with $M_{\rm gas} = 8.59 \times 10^9\,\Msun$.
The total mass of the galaxy is $1.3 \times 10^{12}\,\Msun$.
In the fiducial run, we employed $10^5$ dark matter particles, $10^5$ gas (SPH) particles, and $10^5$ and $1.25 \times 10^4$ collisionless particles representing the stars in the disk and the bulge, respectively.
We also added a hot gaseous halo following \citet{2021ApJ...917...12S}. 
We randomly sampled $4\times 10^4$ dark matter particles and added gas particles with the same mass as that of originally existing gas particles at the same position as those sampled dark matter particles.
We adopt a fixed gravitational softening length of $\epsilon_{\rm grav}$ = 80\,pc.
We allowed the minimum gas smoothing length to reach 10 percent of the spline size of gravitational softening $2.8\epsilon_{\rm grav}$.%
\footnote{We use the definitions of the smoothing length and the gravitational softening length in \textsc{GADGET-2} code \citep{2005MNRAS.364.1105S} in this paper; the gravitational softening length is equivalent to the Plummer softening length, while the smoothing length is the kernel size beyond which kernel value vanishes \citep[see also Appendix C in][]{2016ApJ...833..202K}. We define the SPH spatial resolution as the smoothing length widely used in literature, $\Delta x= \eta(m/\rho)^{1/3}$ \citep[see e.g.,][for review]{2009NewAR..53...78R}. The parameter $\eta$ is usually set to $\eta \sim 1.3$, but we set it to $\eta = 1$ for simplicity. In our \textsc{GADGET3-Osaka} simulation, we adopt the quintic spline kernel and $N_{\rm ngb} = 128$. If we consider the quintic spline kernel to truncate at $3h$  as in \citet{2012JCoPh.231..759P} \citep[see also][for another definition of the smoothing length in terms of the smoothing kernel]{2012MNRAS.425.1068D}, we set to $\eta = 1.04$ in our simulation, effectively.}
We first evolve the system to 0.5 Gyr adiabatically for relaxation to set up the initial condition, and then evolve it to 1 Gyr with the sub-grid physics including cooling, UVB heating, star formation, and stellar feedback.

\subsection{Star formation and Stellar feedback}
\label{sec:Isogal:subgridmodel}
\subsubsection{Star formation}
We assume star formation to occur when the gas number density $n > 10\,\cc$ and temperature $T < 10^4$\,K.
We use the same star formation prescription as \citetalias{2019MNRAS.484.2632S}, a Schmidt-type star formation law \citep{1959ApJ...129..243S}. 
The star formation rate (SFR) density is \citep{1992Cen, 1992ApJ...391..502K, 2001ApJ...558..497N, 2003MNRAS.339..289S, 2006MNRAS.373.1074S, 2014ApJS..210...14K, 2016ApJ...833..202K}
\begin{equation}
    \dot{\rho_*} = \epsilon_* \frac{\rho_{\rm gas}}{t_{\rm ff}},
\end{equation}
where $\epsilon_*$ is the star formation efficiency, $\rho_{\rm gas}$ is the gas density, and $t_{\rm ff} = \sqrt{3\pi/(32 G \rho_{\rm gas})}$ is the local free-fall time. We adopt $\epsilon_* = 0.05$.
The star particles are stochastically spawned from gas particles to follow the SFR density. 
Each gas particle can spawn a maximum of $n_{\rm spawn}$ star particles.
The initial mass of the star particle is defined as $m_* = m_{\rm gas}/n_{\rm spawn}$, where $m_{\rm gas}$ is the mass of the gas particle.
We have adopted $n_{\rm spawn} = 2$ throughout this paper, using a simple stellar population (SSP) approximation and assuming each star particle to consist of a cluster of stars whose mass function follows the Chabrier IMF \citep{2003PASP..115..763C} with a mass range of 0.1-100\,$\Msun$.

\subsubsection{Stellar feedback}
\label{sec:Isogal:stellarfeedback}
We consider stellar wind from massive stars, Type II SNe, Type Ia SNe, and stellar wind from the asymptotic giant branch (AGB) stars as stellar feedback, while distributing the physical quantities from them using the spherical superbubble model.
We use \textsc{CELib} to calculate time and metallicity-dependent mass, metal, and energy input from Type II SNe, Type Ia SNe, and AGB stars (see \citetalias{2019MNRAS.484.2632S}, figure 1).
To deposit energy and metals gradually rather than at a single instant, we divide each feedback from the star particle into $n_{\rm fb}$ events and adopt $n_{\rm fb} = 8$ throughout this paper \citepalias{2019MNRAS.484.2632S}.
% \begin{table}
% \caption{Free parameters in our feedback model}
% \label{tab:freeparameters}
% \begin{tabular}{lcc}
% \hline
% Parameter & Symbol & Default value
% Number density threshold of star formation & $n$ & $10\,\pcmq$\\
% Tempereture threshold of star formation & T
% \hline
% \end{tabular}
% \end{table}

\paragraph{Type II SN feedback}
% For Type II SN feedback, we calculate the metallicity-dependent metal yield and mass loss using \textsc{CELib}.
% We also calculate the time-dependent SN rate with the \textsc{CELib}. 
We assume the range of the progenitor mass to be 13 -- 40 $\Msun$ and the hypernova fraction to be $f_{\rm HN} = 0.05$. 
In the case of solar metallicity, the specific energy of Type II SNe is $\epsilon_{\rm SNII} = 7.19\times 10^{48}\,\erg\,\Msun^{-1}$.
We assume the SN energy to be $10^{51}\,\erg$ and estimate shock radius and momentum input from SN feedback using equations (\ref{eq:SB:shockradius}) and (\ref{eq:SB:terminalmomentum}).
When SN feedback occurs at a low-density void formed by previous feedback events, we may overestimate the shock radius.
Thus, we set an upper limit on the mass enclosed inside the shock radius $M(R_{\rm max}) = 2\times 10^3 m_* / n_{\rm fb}$.
This limit comes from a rough estimate of the SNR mass at fadeaway under an assumption that the terminal momentum per SN is $p \sim 2\times 10^5\,\Msun\,\kms$, the sound speed of star-forming cloud is $c_s \sim 1\,\kms$, and the SN rate is $N_{\rm SN} \sim m_*/(100\,\Msun)$.
% We limit the temperature of the heated particles to $10^9$\,K to prevent an unphysically huge explosion which occurs with a very small probability by stochastic feedback. 

\paragraph{Type Ia SN feedback}
% We calculate the metallicity-dependent metal yield and mass loss using \textsc{CELib} also for Type Ia SN feedback.
We adopted the delay-time distribution function with a power law of $t^{-1}$ for Type Ia SNe event rate \citep[e.g.,][]{2008PASJ...60.1327T, 2012PASA...29..447M}, using equations (\ref{eq:SB:shockradius}) and (\ref{eq:SB:terminalmomentum}) to estimate shock radius and momentum input, although Type Ia SN explosions are intermittent. 
One can estimate the momentum input from a single Type Ia SN using equation (\ref{eq:SB:psfsingle}), assuming the SNR formed by a single SN to acquire about 77 \% of its terminal momentum by the shell-formation time \citep{2015ApJ...802...99K}.
For Type Ia SN feedback, we ignore thermal feedback because the formation of a hot bubble by stochastic thermal feedback represents superbubble formation by clustered Type II SNe.
We used the same upper limit as Type II SN feedback on the shock radius.

\paragraph{Stellar wind from OB stars}
For stellar wind feedback from OB stars, we consider only mechanical feedback. 
We calculated the energy of the stellar wind from OB star using \textsc{Starburst99}\footnote{\url{https://www.stsci.edu/science/starburst99/docs/default.htm}} \citep{1999ApJS..123....3L, 2005ApJ...621..695V, 2010ApJS..189..309L, 2014ApJS..212...14L} and set the specific energy of stellar wind to $\epsilon_{\rm SW} = 1.5\times10^{48}\,\mathrm{erg}\,\Msun^{-1}$. 
Assuming 30 \% of the energy turns into kinetic energy, we estimate the shock radius and momentum input using the equation (\ref{eq:SB:shockradius}) and (\ref{eq:SB:terminalmomentum}) by setting $N_{\rm SN} = m_* \epsilon_{\rm SW} / (10^{51} \mathrm{erg})$.
% It is valid 
Here we choose to use the equation of superbubble momentum because the specific luminosity of stellar wind and Type II SNe are similar \citep[e.g.,][]{2013ApJ...770...25A}.
We set an upper limit on the mass enclosed inside the shock radius $M(R_{\rm max}) = 2\times 10^2 m_* / n_{\rm fb}$. 
This limit is one order smaller than that of SN feedback because the amount of energy injected by stellar winds is about one order smaller than that injected by SNe. 
The adopted numbers above translates to a specific momentum injection rate of $\dot{p}/M \sim 100$ km/s/Myr.  However, \citet[][]{2021ApJ...914...90L} have shown that the wind specific momentum injection rate from O star winds in a cluster is $\dot{p}/M \sim 10$\,km/s/Myr due to fractal mixing boundary layer.  Our model here can be regarded as an upper limit of O star winds.

We do not use our stochastic thermal feedback model for OB stars because ionization heating is limited to $2\times10^4$ K by hydrogen recombination, and heating to higher temperature is unphysical. Just heating to $2\times10^4$ K is ineffective due to overcooling.
In reality, radiation pressure and ionization heating form low-density \ion{H}{2} bubbles, but we don't have a reasonable subgrid model for the bubble formation process.

\paragraph{AGB feedback}
% We use \textsc{CELib} to calculate mass loss and metal yield from AGB stars. 
For stellar wind from AGB stars, we only consider mass and metal input.
We cannot resolve %the radius of 
the circumstellar envelope of an AGB star at $ R \lesssim 1\,\mathrm{pc}$ \citep[e.g.,][]{2018A&ARv..26....1H}.
In our simulations, 
%we treat stars as star particles consisting of a cluster of stars, 
a star particle represents a star cluster; however, in reality, stars are dispersed.
To account for this, we consider the star particle to be representative of stars in the surrounding gas with a mass of $100 m_*$ assuming the star formation efficiency to be $\epsilon_* \sim 0.01$.
Thus, we set the mass enclosed inside the shock radius of the AGB feedback to $M(R_{\rm shock}) = 100 m_*$ and distribute the mass and metals to gas particles inside the shock radius.
We do not consider the mechanical feedback from stellar winds because it is slower ($\sim 10\,\kms$) than (and negligible in comparison to) SN feedback. 
However, we add a boost-back momentum for momentum conservation. 
If the relative velocity between the star and gas particles is large, the boost-back momentum can have a significant impact \citep{2018MNRAS.477.1578H, 2019MNRAS.487.4393S}.

\subsection{Results from {\sc GADGET3-Osaka} simulations}

% \begin{table*}
% \caption{List of isolated galaxy simulations used in this paper.}
% \label{tab:Isogal:simulationsettings}
% \begin{tabular}{\textwidth}{lccccX}
% \hline
%  & Type II SN & Type II SN & Stochastic & Type Ia \& OB star & \\
% Run Name & mechanical feedback & thermal feedback & thermal feedback  & mechanical feedback & Notes\\
% \hline
% Fiducial & \checkmark & \checkmark & \checkmark & \checkmark & Fiducial run with all feedback models\\
% Non-stochastic & \checkmark & \checkmark &  & \checkmark  & Fiducial but thermal energy is distributed without stochastic treatment \\
% SNII-kinetic & \checkmark &  &  &   & Type II SN mechanical feedback only\\
% SNII-thermal &  & \checkmark & \checkmark &   & Type II SN thermal feedback only\\
% No-FB &  &  &  &  & No feedback\\
% \hline
% \end{tabular}
% \end{table*}

\begin{deluxetable*}{lccccl}
\label{tab:Isogal:simulationsettings}
% \tablenum{2}
\tablecaption{List of isolated galaxy simulations used in this paper.}
% \tablewidth{0pt}
\tablehead{
\colhead{Run Name} & \multicolumn{4}{c}{Feedback Type} & \colhead{Note}\\
\cline{2-5}
\colhead{} & \colhead{Type II SN} & \colhead{Type II SN} & \colhead{Stochastic} & \colhead{Type Ia \& OB star} & \colhead{}\\
\colhead{} & \colhead{mechanical FB} & \colhead{thermal FB} & \colhead{thermal FB} & \colhead{mechanical FB} & \colhead{}
}
% \decimalcolnumbers
\startdata
Fiducial & \checkmark & \checkmark & \checkmark & \checkmark & Fiducial run with all feedback models\\
Non-stochastic & \checkmark & \checkmark &  & \checkmark  & Fiducial but thermal energy is \\
    &   &   &   &   & distributed without stochastic treatment \\
SNII-kinetic & \checkmark &  &  &   & Type II SN mechanical feedback only\\
SNII-thermal &  & \checkmark & \checkmark &   & Type II SN thermal feedback only\\
No-FB &  &  &  &  & No feedback\\
\enddata
\end{deluxetable*}

To explore the thermal and mechanical feedback effects, we ran five simulations with different feedback (FB) settings, as summarized in Table~\ref{tab:Isogal:simulationsettings}.
We retained the amount of mass and metals ejected by SNe and AGB stars and changed only the energy and momentum injection models.
When a FB is turned off, the energy for the FB is reduced. All simulations have AGB feedback.
 
\subsubsection{Projection Maps}

\begin{figure*}[t]
    % \centering
    \begin{interactive}{animation}{Isogal_models_density.mp4}
    \plotone{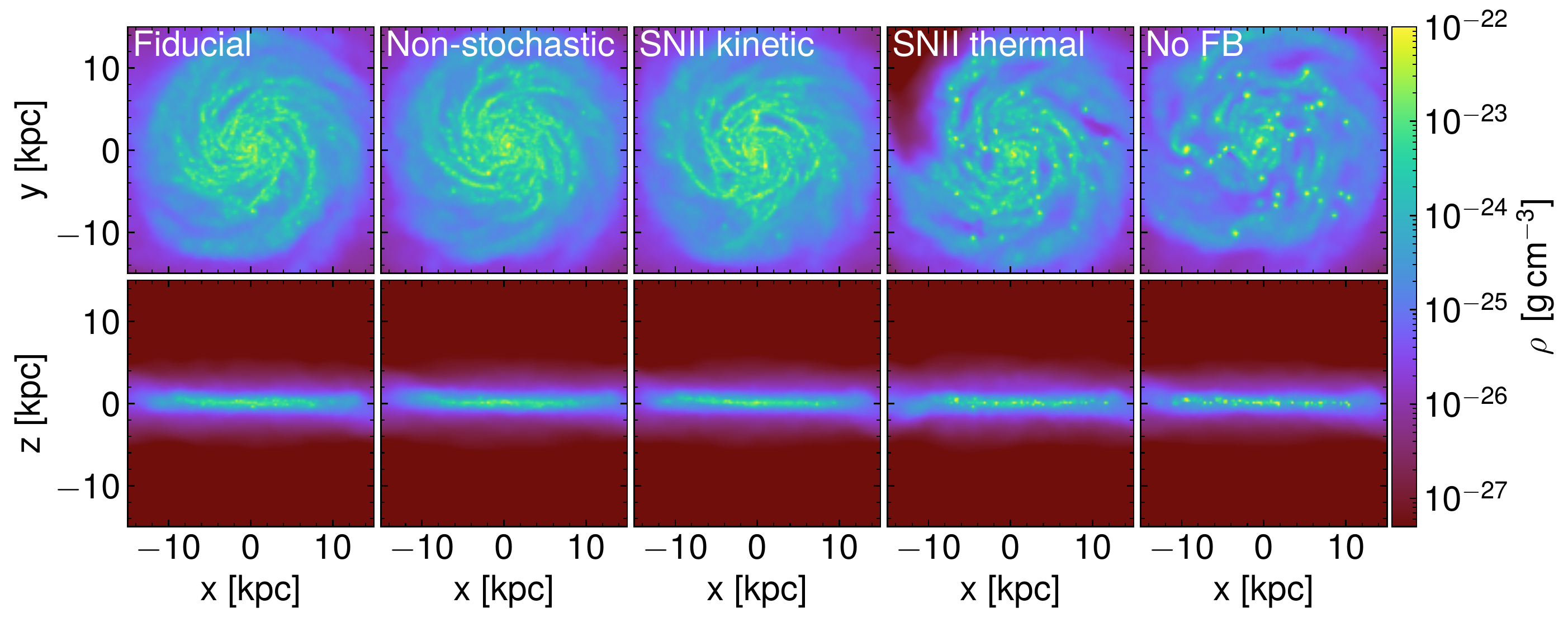}
    \end{interactive}
    \caption{Projected gas density plots of the isolated galaxies in given Table~\ref{tab:Isogal:simulationsettings} at $t\simeq 1$\,Gyr in 30\,kpc$\times$30\,kpc images. The top and bottom rows depict face-on and edge-on images, respectively.
    Here, we plot $\sum \rho^2/\sum \rho$ to enhance the contrast and to use the same density-weighted method for both temperature and metallicity. 
    An animated version of this figure, showing the evolution of these panels during $t$ = 0–1\,Gyr, is available in the HTML version of this article.
    }
    \label{fig:Isogal:density20}
\end{figure*}

\begin{figure*}[t]
    % \centering
    \begin{interactive}{animation}{Isogal_models_temperature.mp4}
    \plotone{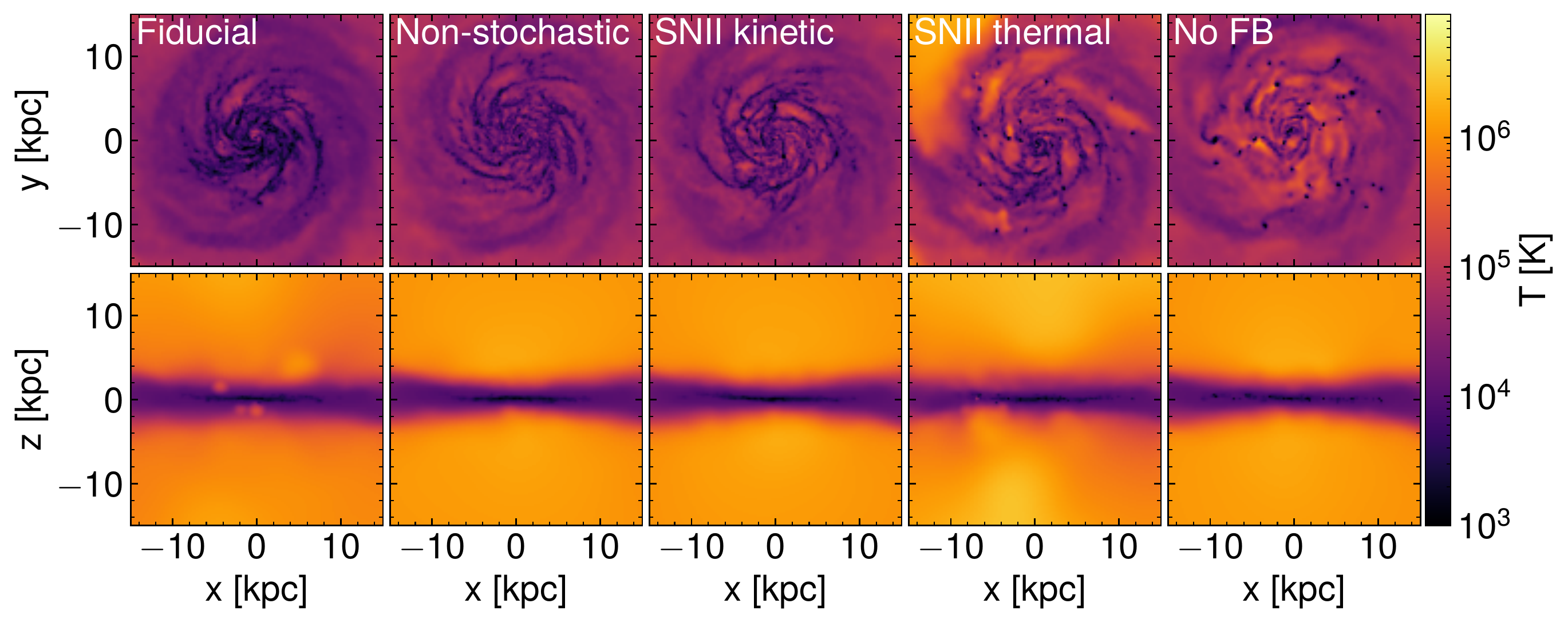}
    \end{interactive}
    \caption{Projected temperature (density-weighted) of the isolated galaxies given in Table \ref{tab:Isogal:simulationsettings} at $t\simeq 1$\,Gyr.
    The top and bottom rows depict face-on and edge-on images.
    The hot gas above the disk in the No-FB run is due to the galactic halo gas with $T=10^6$\,K implemented in the initial condition. 
    An animated version of this figure, showing the evolution of these panels during $t$ = 0–1\,Gyr, is available in the HTML version of this article.}
    \label{fig:Isogal:temperature20}
\end{figure*}

\begin{figure*}[t]
    % \centering
    \begin{interactive}{animation}{Isogal_models_metallicity.mp4}
    \plotone{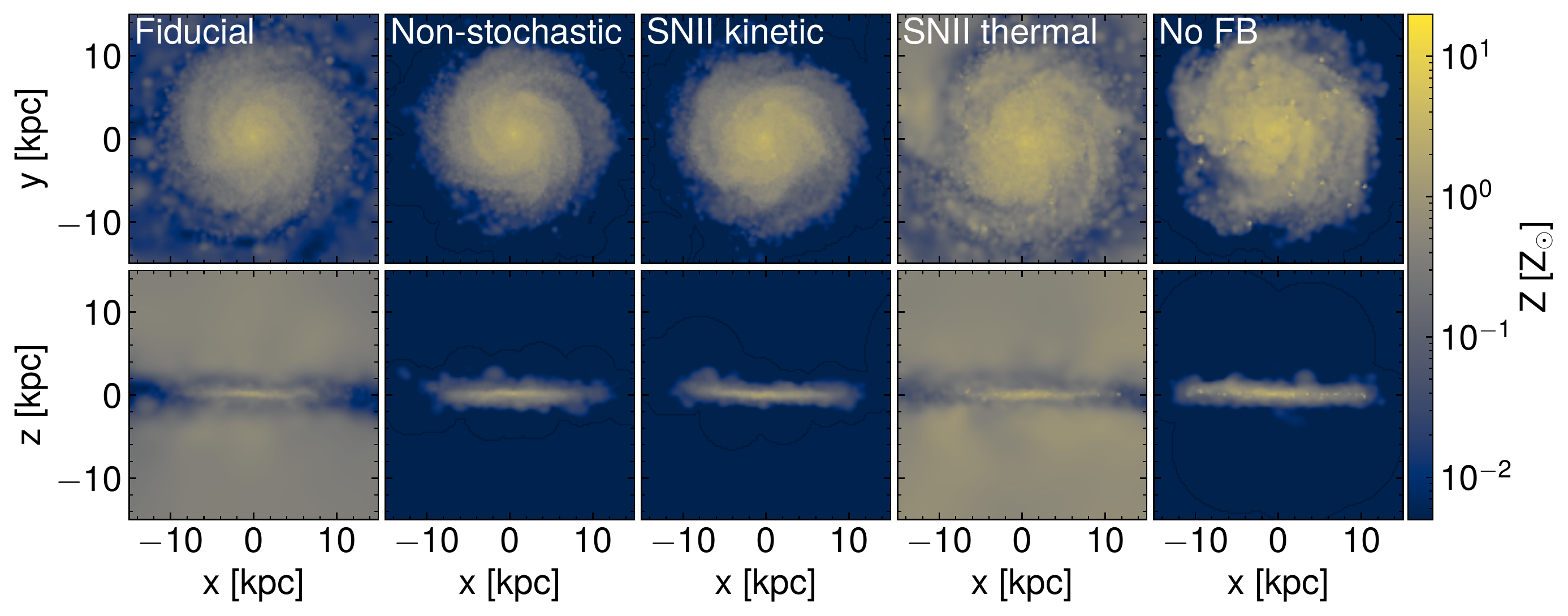}
    \end{interactive}
    \caption{Projected metallicity (density-weighted) of the isolated galaxies given in Table \ref{tab:Isogal:simulationsettings} at $t\simeq 1$\,Gyr.
    The top and bottom rows depict face-on and edge-on images.
    Metals carried by the outflows are visible above and below the disk.  
    An animated version of this figure, showing the evolution of these panels during $t$ = 0–1\,Gyr, is available in the HTML version of this article.
    }
    \label{fig:Isogal:metallicity20}
\end{figure*}

Figures \ref{fig:Isogal:density20} ,\ref{fig:Isogal:temperature20}, and \ref{fig:Isogal:metallicity20} show the density-weighted projection plots of density, temperature, and metallicity of the simulated galaxies at $t \simeq 1\,\mathrm{Gyr}$.
Galaxies are centered at the density-weighted center of gas mass, 
\begin{equation}
    \bm{r}_{\rm center} = \frac{\sum_i \rho_i m_i \bm{r}_i}{\sum_i \rho_i m_i}.
\end{equation}
In Figure\ref{fig:Isogal:density20}, one can see that the density structure in the presence of mechanical feedback differs from that in its absence. 
The gas disk is maintained for 1 Gyr in three runs with Type II SN mechanical feedback (Fiducial, Non-stochastic, and SNII-kinetic) while the gas disks become clumpy in the other two runs.
This result suggests that the galactic disk is supported against gravity by the turbulence driven by mechanical feedback.
Comparing the Fiducial and Non-stochastic runs, the effect of stochastic thermal feedback is not clearly seen in the density distribution. 

Figure \ref{fig:Isogal:temperature20} shows the density-weighted temperature.
One can see the hot bubbles formed by the stochastic thermal feedback inside the gas disks in the Fiducial and SNII-thermal runs, which escape from the galactic disk to produce the hot outflow.
This feature is observed in X-ray by \citet{2018ApJ...862...34N}, who suggested that the hot, gaseous halo of the Milky Way is formed by the hot outflow resulting from stellar feedback.
The low-density regions created by mechanical feedback or gas depletion are shown as $\sim10^6$ K because we set the initial temperature of the halo to $10^6$ K.

The density-weighted metallicity is depicted in Figure~\ref{fig:Isogal:metallicity20}.
It should be noted that we did not change the metal injection model, and that these five runs are different at the models of energy and momentum injection.
The metal distribution above the galactic disks in the presence of stochastic thermal feedback differs from that in its absence.
\begin{figure*}
    % \centering
    \begin{interactive}{animation}{Isogal_models_metal_density.mp4}
    \plotone{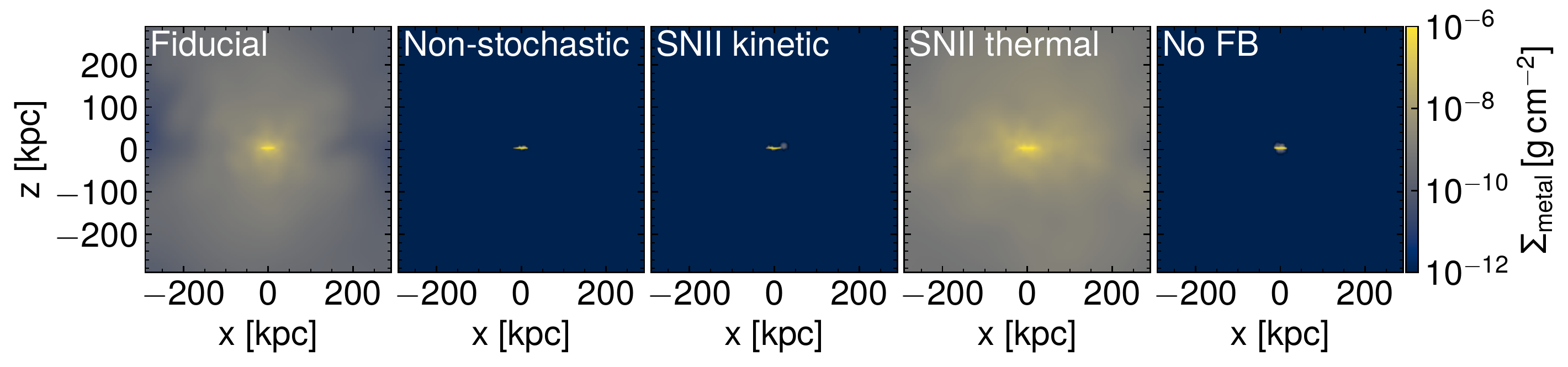}
    \end{interactive}
    \caption{Large-scale, edge-on view of metal column density at $t\simeq 1$\,Gyr.
    It is clear that the metals are carried over large distances in the Fiducial and SNII-thermal runs. 
    An animated version of this figure, showing the evolution of these panels during $t$ = 0–1\,Gyr, is available in the HTML version of this article.
    }
    \label{fig:Isogal:metaldensity200}
\end{figure*}
We also indicate the distribution of metals on the virial radius\,(205\,kpc) scale in Figure~\ref{fig:Isogal:metaldensity200}.
Metal outflows were not observed in the Non-stochastic run.
This result is consistent with those of \citet{2021ApJ...917...12S}, who demonstrated that metal outflow is suppressed by ram pressure from the hot gaseous halo using \textsc{GADGET2} code with a simple thermal dump SN feedback model.
On the other hand, the Fiducial run shows the metal outflows beyond the virial radius.
We will further compare our work with that of \citet[][]{2021ApJ...917...12S} in Section~\ref{sec:Isogal:outflowprofile}.
% No-FB run shows more powerful outflow than Non-stochastic or SNII-kinetic. This could be due to the diffusion because we do not see such an outflow in the case without the gaseous halo.

\subsubsection{Phase Diagrams}
% phase diagram
\begin{figure*}
    \centering
    \includegraphics[width=\textwidth]{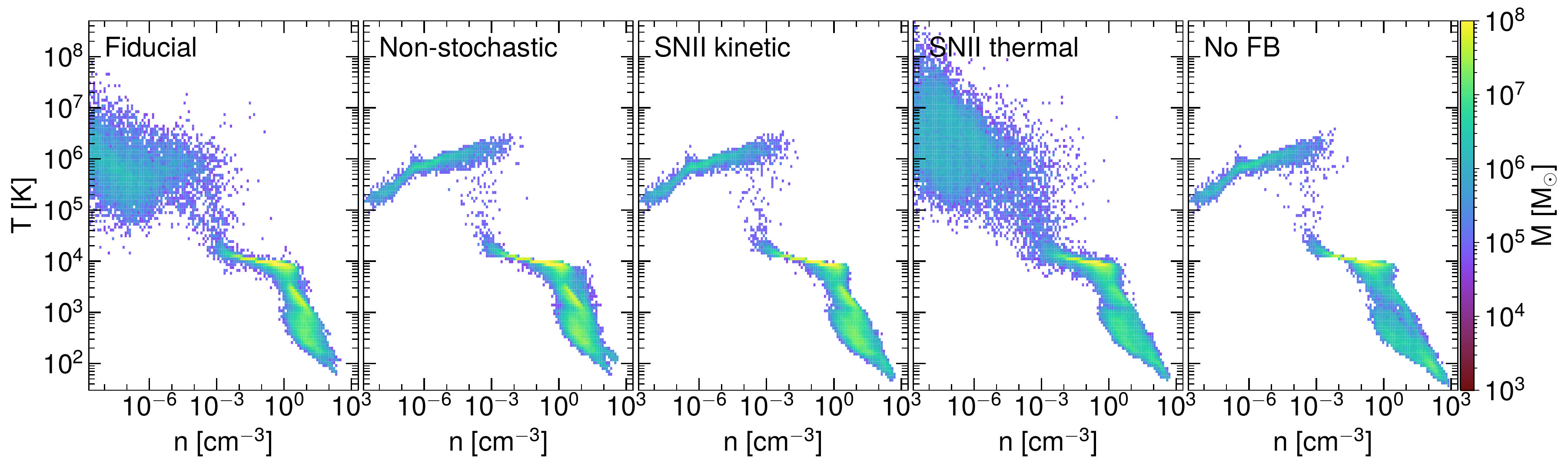}
    \caption{Phase diagram of the gas in the isolated galaxies given in Table \ref{tab:Isogal:simulationsettings} at $t\simeq 1$\,Gyr.  The dense, cold gas ($T<10^3$\,K) in the galactic disk is visible on the lower right corner. The hot, tenuous gas heated by feedback adiabatically cools and then joins the hot gaseous halo in the upper left region with $T\sim 10^6$\,K. The hot component is most visible in the SNII-thermal run due to the highest star formation rate and subsequent strong thermal feedback. 
    The hot gas seen in the No-FB run is due to the hot halo implemented in the initial condition. 
    }
    \label{fig:Isogal:phase}
\end{figure*}
Fig.~\ref{fig:Isogal:phase} shows the phase diagrams of simulated galaxies at $t=1$ Gyr.
The dense, cold gas ($T < 5\times10^3$\,K) in the galactic disk is visible on the lower right corner.
In three runs with mechanical SN feedback (Fiducial, Non-stochastic, and SNII-kinetic), two-phase ISM of the warm ($T \sim 2\times10^3$\,K) and cold ($T \sim 3\times10^2$\,K) phases were observed.
Compared with them, the warm phase is less visible and the cold gas is more condensed at a lower temperature ($T < 10^2$\,K) in SNII-thermal and No-FB runs.
The hot, tenuous gas heated by thermal feedback adiabatically cools and joins the hot gaseous halo in the upper left region with $T \sim 10^6$\,K.
We do not see a difference in the upper left gas distributions of the Non-stochastic and No-FB runs, which indicates ineffective thermal feedback in the Non-stochastic run.
The hot component is most visible in the SNII-thermal run due to its higher star formation rate and subsequent thermal feedback.

\subsubsection{Star Formation Histories}
% star formation rate
\begin{figure}
    \centering
    \includegraphics[width=\columnwidth]{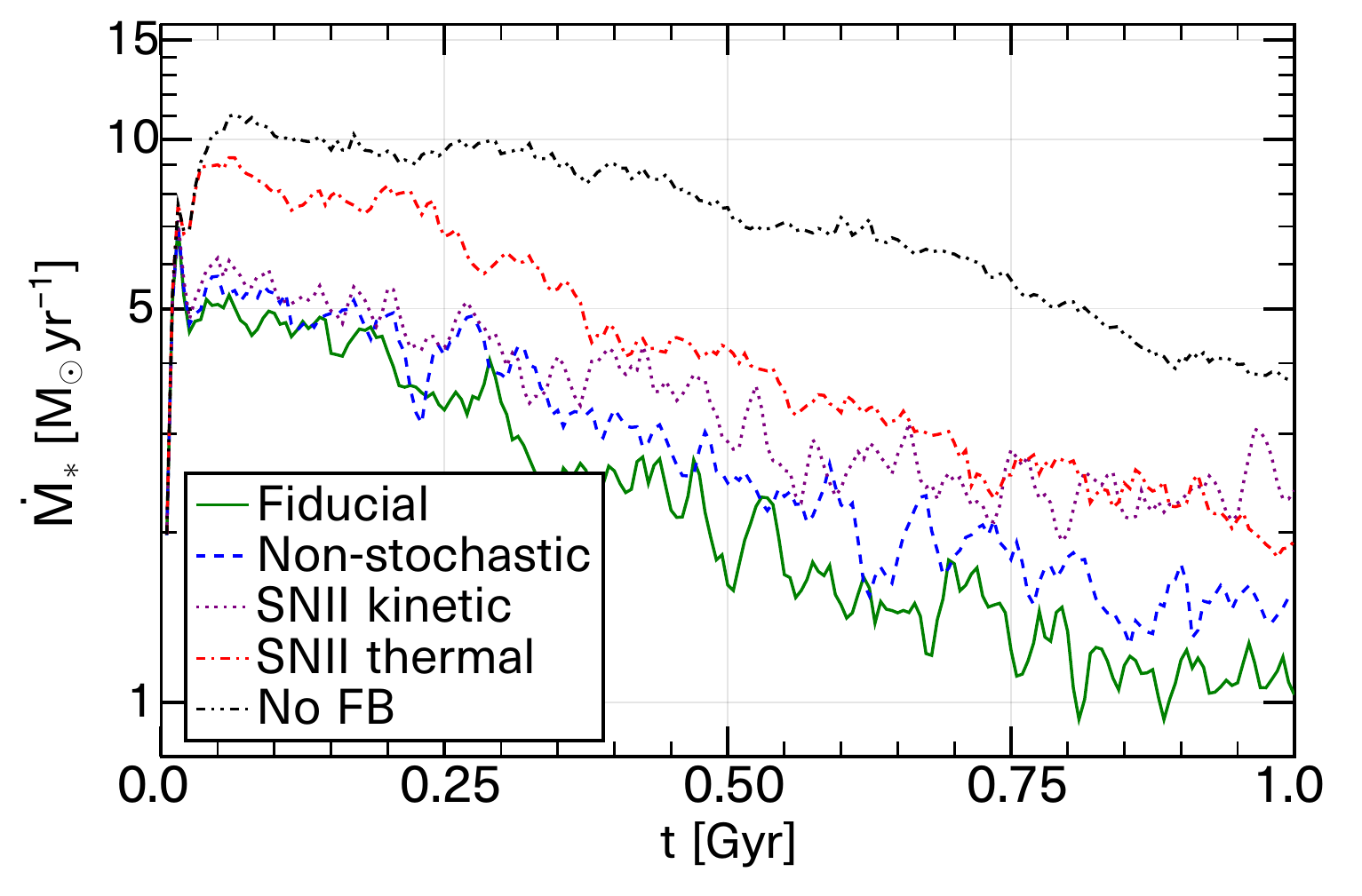}
    \caption{Star formation rate as a function of time for the isolated galaxy simulations given in Table \ref{tab:Isogal:simulationsettings}.
    The No-FB run has the highest SFR, and the SNII-thermal run shows weak suppression of SFR. In other runs, SFR is suppressed by the mechanical SN feedback efficiently.}
    \label{fig:Isogal:sfr}
\end{figure}
Figure \ref{fig:Isogal:sfr} depicts the star formation history of the simulated galaxy.
Star formation is suppressed in all the runs with feedback compared to the No-FB run.
% the Fiducial, Non-stochastic, and SNII-kinetic runs than in the No-FB run.
% On the other hand, the star formation history of the SNII-thermal run is similar to that of the No-FB run.
% This indicates that mechanical feedback suppresses star formation while thermal feedback does not.
The thermal feedback supresses star formation, and the mechanical feedback supresses it further.
Star formation rate in the Fiducial and Non-stochastic runs are slightly lower than that in the SNII-kinetic run because the mechanical feedback from Type Ia SNe and OB stars are not considered in the SNII-kinetic run.
In our previous work \citepalias{2019MNRAS.484.2632S}, we observed an `initial star burst' which occurs just after the beginning of the simulations due to the collapse of gas without feedback.
This phenomenon is weakened in this work because we first adiabatically evolve the initial condition to 0.5\,Gyr for relaxation and then plot the star formation history with an interval of 5\,Myr, which is approximately equal to the lifespan of a massive star.

\subsubsection{Kennicutt--Schmidt Relation}
% Kennicutt-Schmidt relation
\begin{figure}
    \centering
    \includegraphics[width=\columnwidth]{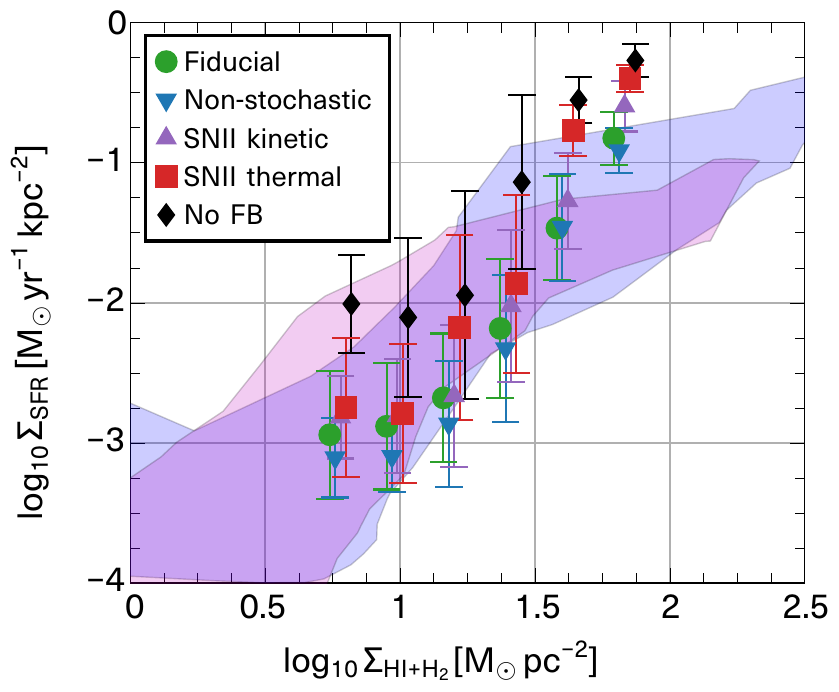}
    \caption{The Kennicutt--Schmidt relation of the simulated galaxies given in Table \ref{tab:Isogal:simulationsettings} at $t\simeq1$\,Gyr, calculated with 500\,pc $\times$ 500\,pc patches.
    %Following the observations, the surface density is averaged over 500\,pc $\times$ 500\,pc patches. 
    Coloured regions depict the observational data of nearby spiral galaxies (blue: \citealt{2011AJ....142...37S}; magenta: \citealt{2008AJ....136.2846B}). 
    }
    \label{fig:Isogal:KS}
\end{figure}
Figure~\ref{fig:Isogal:KS} shows the Kennicutt--Schmidt relations of the simulated galaxies given in Table~\ref{tab:Isogal:simulationsettings} at $t = 1$\,Gyr calculated with 500\,pc $\times$ 500\,pc patches up to the galactic radius of $R=10$\,kpc. 
The results from our simulations agree with the observed range by \citet[][]{2008AJ....136.2846B, 2011AJ....142...37S} with a similar slope and normalization, which is encouraging. 
The three runs with mechanical feedback (Fiducial, Non-stochastic, and SNII-kinetic) show similar results, and their star formation rate density is lower by a factor of 2--3 than that of No-FB run.
The SNII-thermal run is inbetween No-FB and other runs consistently with Fig.~\ref{fig:Isogal:sfr}.
The simulated data fluctuates as a function of time but on average stays within the observed range. 
The Fiducial and Non-stochastic runs have the lowest $\Sigma_{\rm SFR}$ among all runs due to low SFR, as shown in Fig.~\ref{fig:Isogal:sfr}.

The simulation data point corresponding to the highest gas column density is above the observed data; at this point, it is unclear if our results will turn over at higher column densities due to limitations in our resolution. 
The cut-off at the lower end of the simulation data is determined by mass resolution and by how well we can track star formation in a low-density region, which will be explored in future works. 
% *** CHECK LOWER DENSITY REGION *** 

\subsubsection{Outflow Profiles of Mass, Energy, Velocity, and Metals}
\label{sec:Isogal:outflowprofile}
% outflow rate
\begin{figure*}
    \centering
    \includegraphics[width=\textwidth]{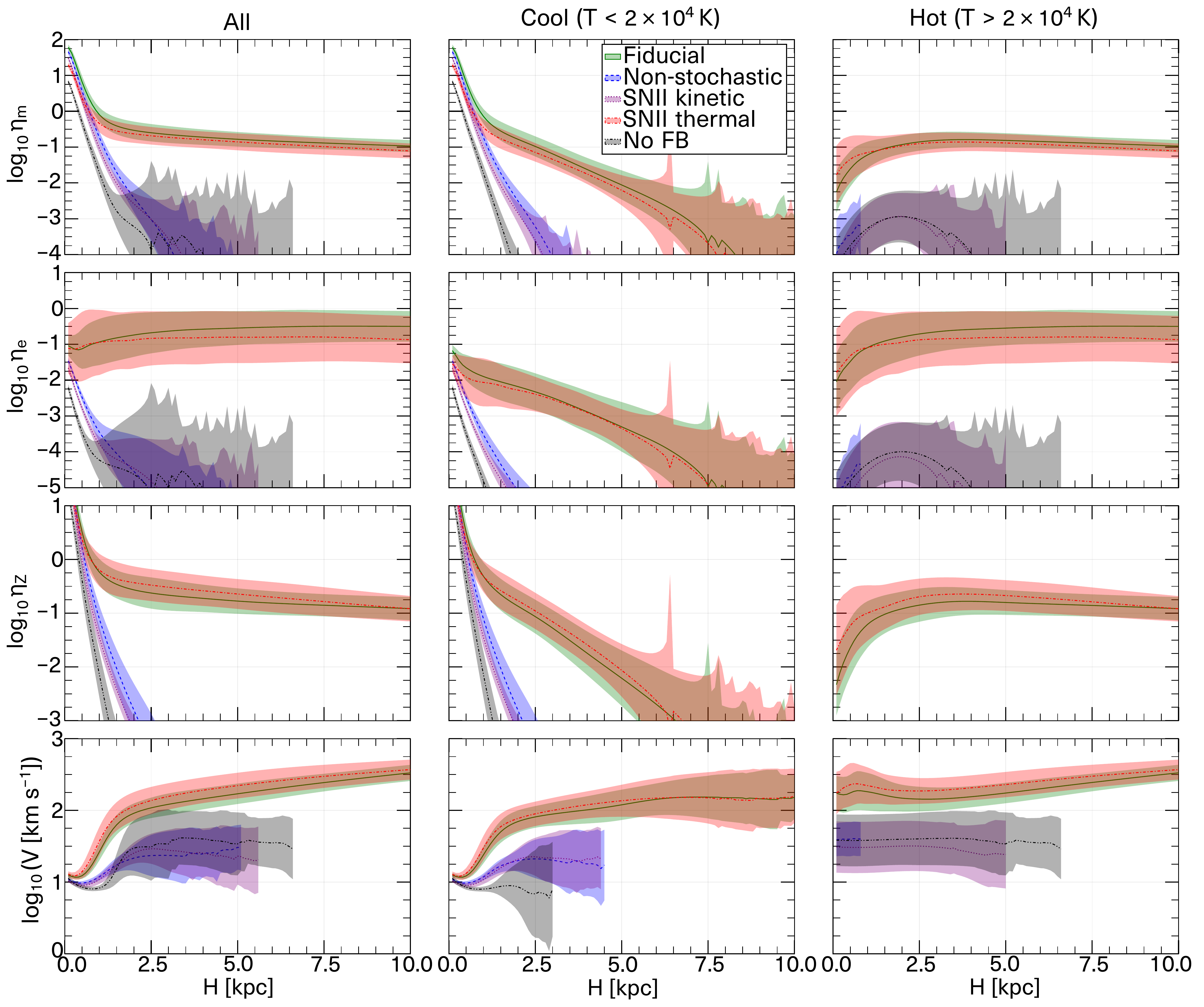}
    \caption{Mass (top row), energy (second row), and metal (third row) loading factors vs. the height above the galactic plane.
    The fourth row depicts outflow velocity vs. the height above the galactic plane.
    All quantities are measured using cylinders of different heights and a 10 kpc radius and averaged over $t=0.5-1.0$\,Gyr.
    The lines indicate the average values during this time period, with $\pm 1 \sigma$ shade for time variation.}
    \label{fig:Isogal:outflowprofile_10}
\end{figure*}

The outflow rate of physical value $X$ at a height $H$ above the galactic plane is %the flux of $X$ integrated over the plane,
\begin{multline}
    \dot{X}_{\rm out}(H) = \int \rho_X v_z \,dS_H\\
    = \int_{S_H} \sum_i X_i v_{i, z} W(r_{ij}, h_{{\rm sml},i}) \,d \bm{r}_j,
\end{multline}
where $S_H$, $\rho_X$, and $v_z$ denote the plane at height $H$, the density of $X$, and velocity in the $z$-direction, while $W$ indicates the kernel function used in the simulation.
Other symbols, such as $r_{ij}$, $v_{i, z}$, and $h_{ \mathrm{sml},i}$ denote the distance from the $i$-th gas particle to $\bm{r}_j$, the $z$-direction outflow velocity, and the smoothing length of the $i$-th gas particle, respectively.
% We compute the outflow rate of physical value $X$ at a height of $H$ from the galactic plane using following equation 
We rewrite this equation as follows, to compute the outflow rate from simulation snapshots \citep{2020MNRAS.494..598S}:
\begin{multline}
    \dot{X}_{\rm out} (H) =\\
    \sum_i X_i v_{i, z} \int_0^{\sqrt{h_{\mathrm{sml},i}^2 - \zeta_i^2}} 2 \pi \xi W\left(\sqrt{\zeta_i^2 + \xi^2}, h_{{\rm sml},i}\right) d\xi,
\end{multline}
where $\zeta_i = |z_i - H|$ and $z_i$ is the $z$ coordinate of $i$-th particle.\footnote{We used the quintic B-spline kernel function in our simulations. In this case, the integral accounting for the cross-section of the SPH particle and the plane of height $H$ can be analytically calculated as 
\begin{multline}
    \int_0^{\sqrt{h^2 - \zeta^2}} 2 \pi \xi W\left(\sqrt{\zeta^2 + \xi^2}, h\right) d\xi \\
    = f_{\frac{1}{3}, 1}(\zeta, h) + f_{-2, \frac{2}{3}}(\zeta, h) + f_{5, \frac{1}{3}}(\zeta, h),
\end{multline}
where
\begin{equation}
    f_{a, b}(\zeta, h) = \frac{3^9 a}{359 h^2} \left(b - \frac{\zeta}{h}\right)_+^6  \left(\zeta + \frac{h}{7}\left(b - \frac{\zeta}{h}\right)_+\right).
\end{equation}
Here, $(\cdot)_+ \equiv \max (0, \cdot)$.
}
To evaluate the intrinsic outflow, we take the summation over those outflowing gas particles that have non-zero metallicity.
Since the metallicity is initially zero and metal mixing between gas particles is not considered, such particles are considered as metal-enriched outflows.
%The parameter $\xi$ in the integral is the radial coordinate of the cylinder. 

We depict the outflow profiles of mass, energy, and metal loading factor
\begin{eqnarray}
    \eta_m &= \frac{\dot{M}_{\rm out}}{\dot{M}_*},\\
    \eta_e &= \frac{\dot{E}_{\rm out}}{\epsilon_{\rm SNII}\dot{M}_*},\\
    \eta_Z &= \frac{\dot{M}_{\rm Z, out}}{\mu_{\rm Z, SN}\dot{M}_*},
\end{eqnarray}
averaged over $t = 0.5$ -- $1.0$ Gyr, in Fig.~ \ref{fig:Isogal:outflowprofile_10}. 
Here, $\dot{M}_{\rm out}$, $\dot{E}_{\rm out}$, and $\dot{M}_{\rm Z, out}$ are the outflow rates of mass, energy, and metals, while $\dot{M}_*$ is the star formation rate and $\mu_{\rm Z, SN} = 0.01$ is the specific metal mass released by Type II SNe.
We also %show the outflow velocity computed as
compute density-weighted, average outflow velocity,
\begin{multline}
    V_{\rm out} (H) = \frac{\int \rho v_z \,dS_H}{\int \rho \,dS_H} = \\
    \frac{\sum_i m_i v_{i, z} \int_0^{\sqrt{h_{ \mathrm{sml},i}^2 - \zeta_i^2}} 2 \pi \xi W\left(\sqrt{\zeta_i^2 + \xi^2}, h_{{\rm sml},i}\right) d\xi}{\sum_i m_i \int_0^{\sqrt{h_{ \mathrm{sml},i}^2 - \zeta_i^2}} 2 \pi \xi W\left(\sqrt{\zeta_i^2 + \xi^2}, h_{{\rm sml},i}\right) d\xi},
\end{multline}
where $m_i$ is the mass of the $i$-th gas particle.
The profiles of the total mass and the metal loading factor are high at low altitudes, decrease rapidly to $H \sim 1.0$ kpc, and become flatter at $H \gtrsim 1.0$ kpc.

At $H \lesssim 1.0$ kpc, the cool ($T < 2\times10^4$\,K) component dominates mass outflow (higher $\eta_m$ in the top middle panel of Fig.~\ref{fig:Isogal:outflowprofile_10}), which is slow ($V \lesssim 30\,\kms$) and falls back to the galaxy.
We see that the mass loading factor of cool outflow
in the runs with feedback is greater by $\sim 0.8$ dex over that of No-FB run. 
%in runs with mechanical feedback (Fiducial, Non-stochastic, and SNII-kinetic) is larger by 0.75 dex over that of others.
% The higher mass loading factor in runs with mechanical feedback indicates that it launches cool outflows.

At $H \gtrsim 2.5$ kpc, the hot ($T > 2\times10^4$\,K) component dominates outflow.
% Compared to the cool component, the hot component has higher specific energy and lower mass.
We see that the mass loading factor of hot outflow in runs with stochastic thermal feedback (Fiducial and SNII-thermal) is much higher %by 1.5 dex at $H = 10$\,kpc 
than that of others, which indicates that stochastic thermal feedback launches hot outflow.
% No difference is observed in the profiles of hot outflow between Non-stochastic and SNII-kinetic runs,
Non-stochastic and SNII-kinetic runs shows much weaker hot outflow,
demonstrating that simply distributing thermal energy from SN feedback is ineffective in launching hot outflow.
The outflow profile of Fiducial run is qualitatively consistent with high-resolution shearing box simulations by \citet[][]{2020ApJ...900...61K}.

\begin{figure}
    \centering
    \includegraphics[width=.8\columnwidth]{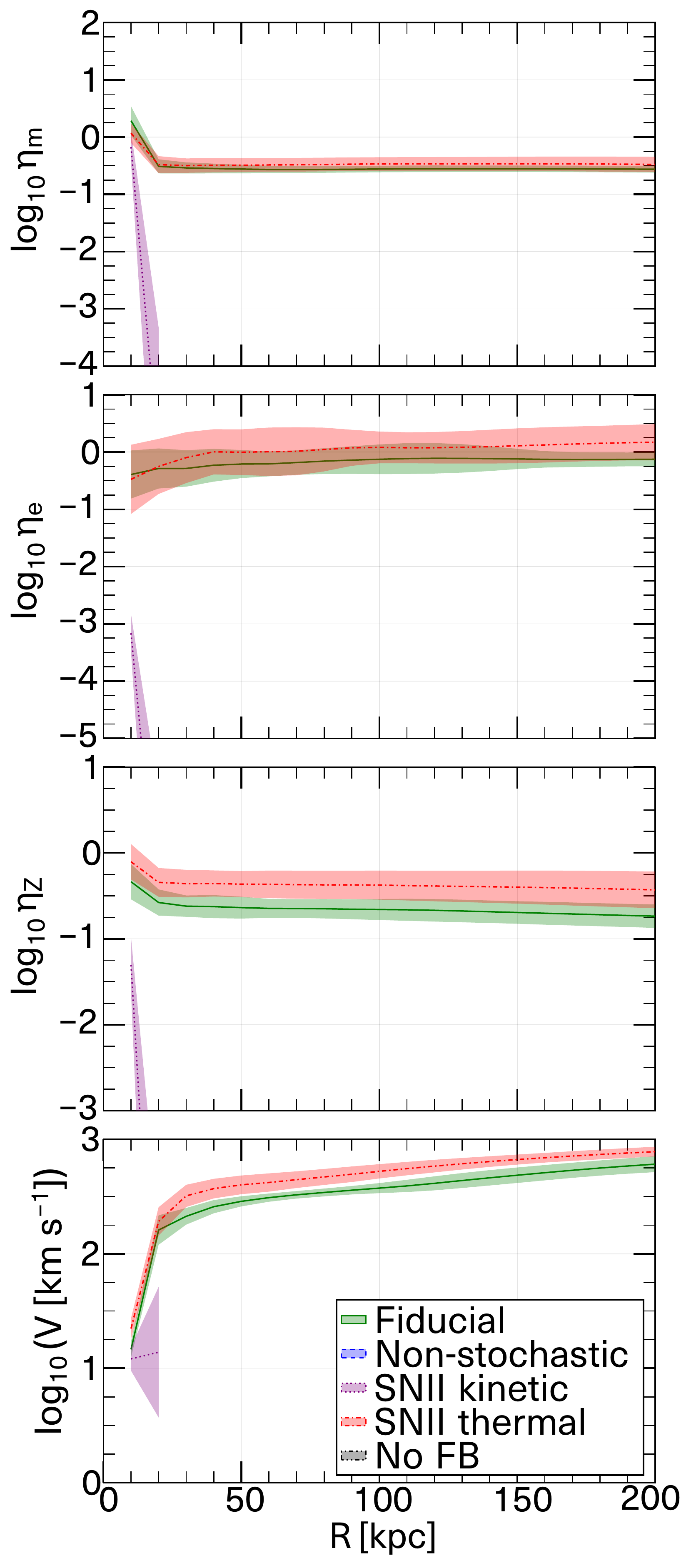}
    \caption{Same as Figure~\ref{fig:Isogal:outflowprofile_10} but measured with spheres of different radii up to 200 kpc, centered at the galactic center and averaged between $t=0.5-1.0$\,Gyr.
    The lines show the average during this time period with $\pm 1 \sigma$ shade for the time variation.}
    \label{fig:Isogal:outflowprofile_200}
\end{figure}
Figure \ref{fig:Isogal:outflowprofile_200} shows the outflow profiles up to 200 kpc. 
We calculate the outflow rates at the surface of a sphere with a radius $R$ centered at the galactic center, and show the loading factors and outflow velocity of all components.
The hot component dominates at $R > 10$\,kpc with cool component disappearing at large $R$.
We see the outflow is accelareted towards large $R$ and reaches $\gtrsim 560\,\kms$ at $R=200$\,kpc.
The mass loading factor of Fiducial run at $R = 200$\,kpc is $\eta_m \sim 0.3$.
% , which shows that the outflow is not strong enough to expel gas from CGM to IGM.
In all the panels the SNII-thermal model shows somewhat higher loading factors and outflow velocity than the Fiducial run, which is due to higher SFR in the former model as was shown in Fig.~\ref{fig:Isogal:sfr}.

For the presented isolated galaxy test, we followed \citet{2021ApJ...917...12S} in adding a gaseous halo to the initial condition. 
As shown in Fig.~\ref{fig:Isogal:outflowprofile_10}, metals are transported via the hot outflow driven by thermal feedback, while the cool outflow cannot transport metals beyond $H = 5$\,kpc.
This is because the hot outflow driven by thermal feedback is not impeded by the ram pressure of the gaseous halo $P_{\rm ram} = \rho v^2$ due to its higher thermal pressure.
On the other hand, the cool outflow is decelerated by the ram pressure of the gaseous halo because it is not accelerated after its initial launch by momentum feedback. 
%, which is comparable to the kinetic energy density of the outflow.
% does not have sufficient kinetic energy density compared to 
% The hot outflow driven by thermal feedback is accelerated %to $R=200$\,kpc 
% while the cool outflow driven by momentum is decelerated because the ram pressure of the gaseous halo $P_{\rm ram} = \rho v^2$, which is comparable to the kinetic energy density of the outflow, is not effective in impeding hot outflow dominated by thermal energy but effective in cool outflow dominated by kinetic energy.

% The cool outflow driven by momentum is decelerated because the ram pressure of the gaseous halo $P_{\rm ram} = \rho v^2$ is effective in impeding it.
% The hot outflow driven by thermal feedback is accelerated because the ram pressure of the gaseous halo $P_{\rm ram} = \rho v^2$
% %, which is comparable to the kinetic energy density of the outflow, 
% is not effective in impeding it.
% %hot outflow dominated by thermal energy 
% %cool outflow dominated by kinetic energy.
% %is comparable to the kinetic energy density of the outflow

Our stochastic thermal feedback model heats the gas particles up to a higher temperature than simple thermal dump models do, driving hot outflow without increasing the total SN feedback energy by hand. 
The discrepancy in metal outflow between mesh-based (\textsc{Enzo}) and particle-based (\textsc{GADGET-2} and \textsc{GIZMO}) simulations reported by \citet[][]{2021ApJ...917...12S} could be due to the difference in mass resolution;
particle-based simulations fix the mass resolution, but mesh-based simulations do not fix it and can effectively heat gas cells at low-density voids created by previous SN feedback events.
% The gaseous halo of \citet{2021ApJ...917...12S} has a typical number density of $n = 10^{-6}\,{\rm cm}^{-3}$ and a temperature of $T = 10^6$\,K, so the typical entropy is $S = 10^{10} k_{\rm B}\,\kelvin\,{\rm cm}^2$. 
% This is larger than the entropy given by the stochastic feedback, $10^8 k_{\rm B}\,\kelvin\,{\rm cm}^2$, but our simulations show an outflow. 
% This result suggests that the outflow seen in the isolated galaxy test is driven by a different mechanism than the buoyancy-driven outflow model (see Section~\ref{sec:FBmodel:thermal} for relevant discussion). The outflows seen in this study are likely to be driven by pressure gradient forces, as in the \citet{1985Natur.317...44C} model.
% *** TO BE CONFIRMED *** 

\section{Discussion}
\label{sec:Discussion}

\subsection{On the momentum of superbubble}

\paragraph{Pressure-driven versus Momentum-driven}
When $\dtsn < 0.1t_{\rm PDS}$, the gas inside the shell is expected to be hot even after shell formation; thus, we expect a pressure-driven evolution,
% In the pressure-driven model, the momentum evolves as 
$p\propto t^{7/5}$ (Eq.~\ref{eq:pressure-driven momentum}), but we observe $p \propto t$.
There could be three reasons for this.
% This could be because 

First is that the deceleration due to ram pressure is ignored in the pressure-driven model, as explained below. 
The ram pressure and the pressure of the hot gas immediately after shell formation are

\begin{multline}
    P_{\rm ram} = \rho V_{\rm sf, m}^2\\
    = 5.5\times10^6\,\kB\,\kelvin \,\pcmq~E_{51}^{0.18}\,\Delta t_{\rm SN, 6}^{-0.18}\,n_0^{1.2}\,\Lambda_{6, -22}^{0.34}
\end{multline}
\begin{multline}
    P_{\rm hot} = (\gamma - 1)\frac{(1/2) f_{\rm th} E_{\rm SN}(t_{\rm sf, m}/\dtsn)}{(4/3)\pi R_{\rm sf, m}^3}\\
    = 1.7\times10^6\,\kB\,\kelvin\,\pcmq~E_{51}^{0.18}\,\Delta t_{\rm SN, 6}^{-0.18}\,n_0^{1.2}\,\Lambda_{6, -22}^{0.34},
\end{multline}
% \begin{eqnarray}
%     P_{\rm ram} &=& \rho V_{\rm sf, m}^2 \nonumber \\
%     &=& 5.5\times10^6\,\kB\,\kelvin \,\pcmq~E_{51}^{0.18}\,\Delta t_{\rm SN, 6}^{-0.18}\,n_0^{1.2}\,\Lambda_{6, -22}^{0.34}\\
%     P_{\rm hot} &=& (\gamma - 1)\frac{(1/2) E_{\rm SNe, th}(t_{\rm sf, m}/\dtsn)}{(4/3)\pi R_{\rm sf, m}^3} \nonumber \\
%     &=& 1.7\times10^6\,\kB\,\kelvin\,\pcmq~E_{51}^{0.18}\,\Delta t_{\rm SN, 6}^{-0.18}\,n_0^{1.2}\,\Lambda_{6, -22}^{0.34},
% \end{eqnarray}
where $R_{\rm sf, m}$ and $V_{\rm sf, m}$ are the radius and velocity at shell formation from Eqs.~(\ref{eq:Rsfmulti}) and (\ref{eq:Vsfmulti}), respectively, $E_{\rm SN} = 1\times10^{51}\,\erg$ is the SN energy, $f_{\rm th} = 0.78$ is the fraction of thermal energy in superbubble energy \citep[][]{1977ApJ...218..377W}, and factor $1/2$ is for energy dissipation at shell formation (Eq.~\ref{eq:residualthermalenergy}).
Since $P_{\rm ram} > P_{\rm hot}$, the ram pressure cannot be neglected immediately after shell formation.
Therefore, even if the gas inside is hot, the superbubble is not expected to obey the pressure-driven model.

The second reason is that the timescale of our simulations is not long enough, therefore continuous luminosity approximation does not apply. %, and each SN deposits momentum impulsively.
\citet[][]{2019MNRAS.490.1961E} discussed the timescale $t_{\rm subsonic}$, at which an SN blast becomes subsonic before reaching the outer shell.  
Our simulation timescale is shorter than $t_{\rm subsonic}$.
In supersonic case ($t < t_{\rm subsonic}$), each SN deposits momentum impulsively, and superbubble is driven by the momentum injection. 
\citet[][]{2019MNRAS.490.1961E} performed long-term 1D simulations, and showed that the superbubble makes a transition from momentum-driven to pressure-driven after $t_{\rm subsonic}$.
It will be our future work to run long-term 3D 
simulations to investigate whether the transition is observed also in 3D simulations.

%Even after the ram pressure decreases, the time evolution of momentum is approximated as $p \propto t$ and does not follow the pressure-driven model.
%This may be due to 
The third reason is that the pressure of the hot gas decreases as the heat is transported to the shell at the mixing layer.
\citet{2017MNRAS.465.1720Y} performed 3D simulations of the superbubble and showed that about 90\% of the injected energy is lost by radiation %after the shell formation.
after the energy is transported into the shell.
% Also, \citet{2017ApJ...834...25K} performed 3D simulations without considering heat conduction and obtained a time evolution of momentum consistent with the results of this study.
% On the other hand, 
%
% \citet{2019MNRAS.490.1961E} performed 1D simulations modeling gas mixing at the rear surface of the shell and showed that the superbubble radius and momentum evolve according to a modified, pressure-driven model that considers energy loss, which is inconsistent with our result. % following momentum-driven model.
% One difference between the 1D and 3D calculations is that the shell deformation is not considered in the 1D calculation.
%
\citet[][]{2021ApJ...914...89L} pointed out that the energy losses are greater in 3D than in 1D simulations due to the fractal area of the interface originating from inhomogeneous ISM, and it would be necessary to use a turbulent diffusion coefficient depending on the bubble size as in Eq.~(37) of \citet[][]{2021ApJ...914...89L} to capture this.
% The fractal interface is because of the clumpy nature of the inhomogeneous ISM.

%Other sources of turbulence includes KHI, RM, Vishniac... 
%An isothermal decelerating shock wave
% while being pushed by hot gas, the shell 
%is deformed by the Vishniac instability \citep{1983ApJ...274..152V, 2013A&A...550A..49K, 2018A&A...617A.133M}, which occurs due to the difference in inertia caused by mass flow in the shell.
As the turbulence develops through various instabilities such as Kelvin--Helmholtz \citep{thompson1871hydrokinetic, helmholtz1868discontinuous,2006ApJ...648.1052H, 2020ApJ...894L..24F}, Rayleigh--Taylor \citep{rayleigh, taylor,2004A&A...423..253B,2016MNRAS.459.2188B}, Richtmyer--Meshkov \citep{richtmyer1960commun, 1972FlDy....4..101M, 2013ApJ...772L..20I}, or Vishniac instabilities 
\citep{1983ApJ...274..152V, 2013A&A...550A..49K, 2018A&A...617A.133M}, the area of turbulent contact surface with the hot gas increases, which in turn increases the energy loss.
Therefore, the time dependence becomes shallower than the pressure-driven model, $p \propto t^{7/5}$.

\paragraph{Multiple SN explosions and mixing}
In our \textsc{Athena++} calculations, the momentum per SN explosion is about a few $\times 10^5\,\Msun\,\kms$. 
This is comparable to the results of previous studies of the momentum per isolated SN explosion \citep{2015ApJ...802...99K, 2015MNRAS.450..504M}, and does not show an increase in momentum due to multiple SN explosions, contrary to \citet{2017MNRAS.465.2471G}. 
The results for the superbubble momentum depend on whether the mixing of the shell and hot gas is considered \citep{2019MNRAS.483.3647G, 2019MNRAS.490.1961E}. 
When the mixing of the hot bubble-shell boundary is considered, the hot bubble is efficiently cooled at the bubble-shell boundary. 
In our 3D Eulerian \textsc{Athena++} simulations, the mixing between the shell and the hot gas is considered, and thus the momentum enhancement is not observed, which is consistent with the findings of \citet[]{2017ApJ...834...25K} for multiple SNe.

\paragraph{Thermal conduction and turbulence}
In this study, we also added the setup of \citet{2017ApJ...834...25K} to account for thermal conduction and turbulence, and the superbubble momentum was comparable to their results.
\citet{2019MNRAS.490.1961E} showed that this is because mixing is more dominant than heat conduction in energy transport at the bubble--shell boundary, and the energy conducted to the shell is returned to the hot bubble by evaporation. 
%In other words, heat conduction increases the mass of the hot bubble but does not affect its pressure. 
Therefore, we do not see significant differences in momentum when heat conduction is accounted for, as was shown by \citet{2017ApJ...834...25K}. 
The resolution of our simulation is not as high as that of \citet{2019MNRAS.490.1961E} and conduction is probably not captured so well, but we do observe the transfer of mass from cold shell to the hot interior as we saw in Fig.~\ref{fig:radius-time}.

The turbulence considered in this study corresponds to a Mach number of ${\cal M} \sim 1$.
If the turbulence is much stronger with ${\cal M} > 10$, the momentum could be reduced by about 30 \% \citep{2015MNRAS.450..504M} due to increased radiative cooling caused by mixing between cold gas clouds and hot bubbles in the ISM.

\paragraph{Comments on $\dtsn$, $t_{sf}$, and caveat on non-equilibrium cooling effect}
The time evolution of the superbubble momentum can be modeled in a unified manner using Eq.~(\ref{eq:SB:fit_momentum}) under different metallicity, density, and $\dtsn$ by normalizing the time and momentum by those at the shell-formation time\,(Fig.~\ref{fig:SB:momentum}), which is useful to estimate the momentum of superbubble for the application to the subgrid model of galaxy simulations.
Therefore, the accurate determination of the shell-formation time is important.% for estimating the superbubble momentum. 

When the superbubble momentum is translated to momentum per single SN, it sensitively depends on $\dtsn$ as was also shown by \citet[][]{2017ApJ...834...25K}. 
Because the time interval of SN explosions depends on the mass of the stellar cluster, the initial stellar cluster mass function will be important for the discussion of SN feedback in the context of galaxy formation.
%Therefore, we argue that the accurate determination of the shell-formation time is important for estimating the superbubble momentum. 

We assumed collisional ionization equilibrium in this study because the shell-formation time remains unchanged even after accounting for non-equilibrium cooling, radiation transport, and heat conduction \citep{2021MNRAS.504..583S} in the case of an isolated supernova explosion.
As for the temperature, the timescale where the ion and electron temperatures balance in an isolated supernova explosion is $10^4$\,yr \citep{1978PASJ...30..489I}, which is shorter than the shell-formation time, so the different temperatures between ions and electrons do not affect the shell-formation time significantly.  
However, these assumptions need to be checked for intermittent SN explosions in the future. 

%To calculate the shell-formation time more precisely, it is necessary to solve the non-equilibrium cooling.  %using a two-temperature model.
%However, in this study, the ion and electron temperatures were assumed to be identical, and the 
%cooling function was based on the assumption of collisional ionization equilibrium was assumed. 
%In the case of an isolated supernova explosion, the shell-formation time is almost the same as that of the case assuming collisional ionization equilibrium, even when non-equilibrium cooling, radiation transport, and heat conduction are taken into account \citep{2021MNRAS.504..583S}. %, but it is necessary to verify what happens in the case of an intermittent supernova explosion. 
%As for the temperature, the timescale where the ion and electron temperatures balance in an isolated supernova explosion is $10^4$ yr \citep{1978PASJ...30..489I}, which is shorter than the shell-formation time, so it is not expected to affect the shell-formation time. 
%, but the case of intermittent energy injection needs to be verified. 
%Therefore, the future work is to calculate the shell-formation time more precisely by solving the non-equilibrium cooling using a two-temperature model.

\subsection{On the SN feedback model}
In our model, we assume that the superbubbles spread isotropically. 
However, when the ISM is highly non-uniform, the superbubbles spread selectively towards the low-density channel \citep{2019MNRAS.485.3887O}. 
In particular, when the superbubbles break out, the kinetic feedback to the local star formation may be weaker, as the energy flows out of the galaxy before the superbubbles gain momentum \citep{2018MNRAS.481.3325F}.
Although our feedback model considers the non-uniformity of the ISM via anisotropic particle distribution, the impact of non-uniform ISM needs to be examined in the future with a more realistic setup and a higher resolution. 
For example, an analytical model considering the effect of superbubble breakout could be incorporated to construct a more realistic feedback model \citep{2021arXiv210914656O}.
%that estimates the evolution (breakout or not) of the superbubble could be incorporated to construct a more realistic feedback model.

There are several different weighting schemes to distribute mechanical feedback in particle simulation such as \citet{2018MNRAS.477.1578H, 2019MNRAS.483.3363H, 2019MNRAS.489.4233M}.
However, the detailed distribution of particles on sub-resolution scales is somewhat random, and the exact weighting scheme does not matter very much in the final outcome as long as it is isotropic. 

% \citet{2019MNRAS.489.4233M} 
The models of \citet{2014ApJ...788..121K} and \citet{2018MNRAS.477.1578H} estimated the momentum by scaling that of isolated supernova explosions. 
However, the superbubble momentum also depends on the time interval of supernova explosions as we show in this work. 
Here we develop a more realistic model by using a universal scaling relation for the superbubble terminal momentum and radius, and this is where our model is different from \citet{2014ApJ...788..121K} and \citet{2018MNRAS.477.1578H}.  

In addition to momentum feedback, 
thermal feedback was also considered by \citet{2014ApJ...788..121K} and \citet{2018MNRAS.477.1578H}, where they simply injected thermal energy. However, thermal feedback will not work if the model is not constructed to avoid overcooling \citep[][]{2019MNRAS.483.3363H}. 
Therefore, in our model, we adopt a stochastic thermal feedback model. In the isolated galaxy test, we show that the thermal outflow is driven by the stochastic model.

\citet{2012MNRAS.426..140D} constructed a stochastic thermal feedback model with temperature rise, $\Delta T$, as a free parameter. 
However, the temperature is not constant during the evolution of high-temperature bubbles formed by supernova explosions. 
Therefore, a parameter survey is necessary to determine $\Delta T$, and using temperature as a model parameter is not so ideal
%appropriate for constructing a model based on physical evidence. Therefore, 
as we discuss in Section~\ref{sec:FBmodel:thermal}. 
Instead, we use entropy as a parameter in our model. 
%Since the gas evolves adiabatically during the outflow of high-temperature gas out of the galaxy to form a high-temperature outflow, 
Since the high-temperature outflow gas expands adiabatically,  
the entropy is constant and its value can be determined based on high-resolution simulations. 

In our model, we also provide momentum feedback in addition to thermal feedback. 
As shown in the isolated galaxy test, thermal feedback can only suppress star formation weakly.
This is because the spatial resolution required to solve for the time evolution of SNR is more stringent than the mass resolution required to resolve for hot SN bubbles. 

%On the other hand, however, star formation is suppressed in the calculations of \citet{2012MNRAS.426..140D}. 
%There are three possible reasons for this. 
%First, their SN energy is  $1.7 \times 10^{49} \erg/\Msun$ which is 3.4 times higher than our value $5.0\times 10^{48} \erg/\Msun$. 
%
%Second, they used effective EoS for unresolved multiphase gas. 
%%By using the effective EoS, the effect of thermal feedback on the sub-resolution scale may be taken into account. 
%However, the multiphase gas model with effective EoS does not explicitly distinguish between the high and low-temperature phases \citep{2014MNRAS.442.3013K}, but gives effective EoS as a temperature floor and considers the gas heating and cooling as a single phase. 
%
%\citet{2019MNRAS.488.4400I} used the AREPO code with the wind feedback model of \citet{2003MNRAS.339..289S}, and the ISM model does not show any difference in the star formation rate of galaxies. 
%
%The third is the use of a star formation model based on the Kennicutt-Schmidt relation \citep{2008MNRAS.383.1210S}, 
%where the KS-relation is the relationship between the gas areal density and the star formation rate areal density on the galaxy scale, 
%and it has been suggested that this relation may be formed as a result of self-regulation of star formation by feedback \citep[e.g.,][]{2019MNRAS.488.4753D}. 
%Therefore, one might expect the KS-based SF recipe to incorporate the effect of feedback-induced star formation suppression.

Our thermal feedback model heats gas particles up to a fixed entropy, which means denser gas particles are heated to higher temperatures. 
In high-$z$ galaxies where the gas density is higher than in the local ones, our thermal feedback would become stronger.
\citet[][]{2020MNRAS.498.5541A} performed cosmological galaxy simulations using the stochastic thermal feedback model by \citet{2012MNRAS.426..140D}, with $\Delta T = 10^{7.5}$\,K
%, thermal feedback model to heat a randomly selected gas particle by $10^{7.5}$\,K, 
and studied metal-emission lines from high-$z$ galaxies. % at $z \sim 6$.
They showed that the radial profile of [\ion{C}{2}] surface brightness of simulated galaxies is lower at outer radius than observation.
This discrepancy could be resolved %may be reduced 
by our stronger thermal feedback. 
% On the other hand, \citet{2021arXiv210914626O} suggest that superbubbles are less effective in driving hot outflow in high-$z$ galaxies due to high gas surface density? (***CHECK***)
%because superbubbles stall before break-out due to high gas surface density and highly turbulent ISM at high redshift. 
In future works, we will investigate the impact of our SN feedback model in a cosmological context. 
%on cosmological galaxy formation process.

Our SN feedback model injects both terminal momentum and thermal energy from SNe, which may overestimate the feedback effect.
In reality, some SN energy is used to drive a superbubble and the other is used to accelerate outflows after the break-out of the superbubble.
\citet[][]{2021arXiv210914656O} developed an analytic model for clustered SNe in a galactic disk, and their model predicts that the superbubble is more likely to break out in an environment with rich gas and short dynamical times, which is consistent with observations and simulations \citep[][]{2021arXiv210914626O}.
Their model could be incorporated into our SN feedback model in the future.

\section{Conclusions}
\label{sec:Conclusions}

In the first part of this paper, we performed idealized superbubble simulations using \textsc{Athena++} to study the dependency of its momentum on the density, metallicity, and interval of SN explosions. 
Our \textsc{Athena++} simulations included cooling, heating, thermal conduction, and turbulence.
We injected thermal SN energy $E_{\rm SN} = 10^{51}$\,erg into the turbulent medium 10 times with fixed time intervals of $\dtsn$ = 0.01, 0.1, 1\,Myr.
We find that superbubble momentum at metallicity $Z = 10^{-3}\,\Zsun$ is larger by a factor of two than that at $Z = 1\,\Zsun$ owing to a long Sedov-Taylor phase caused by later shell-formation time.

To interpret the simulation results, we extend the analytic solution of the SNR shell-formation time by \citet{2015ApJ...802...99K} to include the effect of metallicity (Eq.~\ref{eq:SB:tsfsingle}), and obtained the analytic solution of the superbubble shell-formation time in Eq.~(\ref{eq:tsfmulti}).
We normalized the time evolution of superbubble momentum in the simulation by the analytic time and momentum at shell formation and find that the normalized time evolution is universal (Fig.~\ref{fig:SB:momentum}) as previously shown and described as ``congruent" by \citet[][]{2015ApJ...802...99K}. 
We note that \citet[][]{2017ApJ...834...25K} normalized only the time variable by $t_{\rm sf,m}$ for the multiple SN explosion simulation, however we also do the same for the momentum and derived the fitting function.
From the universal time evolution model of superbubble momentum (Eq.~\ref{eq:SB:fit_momentum}), we derived a momentum per SN, averaged over the initial stellar cluster mass function (Eq.~\ref{eq:SB:terminalmomentum}), which is useful for the application to the subgrid model of galaxy simulations.

In Appendix~\ref{sec:ResolutionRequirement}, we outline the physical conditions for effective thermal/kinetic feedback and the necessity of subgrid feedback models.
Therefore, in the second part of this paper (Sec~\ref{sec:FBmodel}), we developed mechanical and thermal SN feedback models.
% based on 
% our \textsc{Athena++} simulation results and 
% previous high-resolution simulation works on SN-driven outflow.
We distribute ejecta mass, metal, momentum, and energy from the SN feedback to neighboring gas particles based on the solid angle of a face on a Voronoi polyhedron, obtained by utilizing the STRIPACK algorithm with projected particle distribution on a sphere (Fig.~\ref{fig:FBmodel:model}), guaranteeing isotropy, energy- and momentum-conservation. 
%We perform Voronoi tesselation on a sphere to model the formation of superbubble on sub-grid scale and to distribute momentum isotropically (Fig.~\ref{fig:FBmodel:model}).
Note that our method is similar to, but different in detail from that of \citet{2018MNRAS.477.1578H} (See Section~\ref{sec:FBmodel:model}).
We estimate the momentum input from SN feedback using Eq.~(\ref{eq:SB:terminalmomentum}) considering the effect of multiple SN explosions rather than just using a formula for a single SN. 
%
%In Appendix~\ref{sec:resolutionrequirement}, we outline the physical conditions for thermal/kinetic feedback to be effective and the reason why a subgrid feedback models are necessary. 
%Therefore 
We also developed a stochastic thermal feedback model to reproduce hot outflow driven by SNe, as described in Section~\ref{sec:FBmodel:thermal}.
% Our thermal feedback model heats gas particle up to target entropy.
Our thermal feedback model uses entropy as a controlling parameter
% to control outflow 
rather than temperature.

In the third part of this paper, we implemented the above models into isolated galaxy simulation using \textsc{GADGET3-Osaka} code.
We find that mechanical feedback maintains the galactic disk against gravitational collapse (Fig.~\ref{fig:Isogal:density20}) and suppresses star formation, while stochastic thermal feedback can only suppress star formation weakly. (Fig.~\ref{fig:Isogal:sfr}).
The mechanical feedback is more efficient in driving cool outflow than the thermal feedback model by about a factor of two as shown in the middle column of Fig.~\ref{fig:Isogal:outflowprofile_10}.
We also find that the stochastic thermal feedback drives hot outflow while non-stochastic thermal feedback does not (Figs.~\ref{fig:Isogal:outflowprofile_10}, \ref{fig:Isogal:outflowprofile_200}).
Our result is consistent with that of \citet[][]{2019MNRAS.483.3363H}.

%In summary, there are three basic methods to follow the thermal feedback effect by supernovae in galaxy simulations: multiphase ISM, delayed cooling, and stochastic injection. 
As we discussed in Section \ref{sec:introduction}, there are various methods to overcome the overcooling problem. 
In this work, we adopted stochastic injection and avoided the delayed cooling model since it is unphysical. 
The multiphase ISM model becomes effectively similar to the delayed cooling model when thermal energy is stored in the hot phase, and here we did not consider this model. 
In the case of our stochastic thermal feedback model, we adopted a sufficiently high value of target entropy ($S_{\rm OF}$) as the controlling parameter, which allowed us to alleviate the overcooling problem and eject metals out of the galactic plane. 
However, thermal feedback can only suppress star formation weakly (Figs.~\ref{fig:Isogal:sfr}, \ref{fig:Isogal:KS}), and mechanical feedback is also necessary as we formulated in Section~\ref{sec:FBmodel:kinetic}.% in terms of terminal momentum. 

Our next step will be to perform zoom-in and full-box cosmological simulations using our SN feedback models developed in this paper and compare them with observations such as star formation histories and mass--metallicity relations of galaxies. 
% and metal distribution in the CGM/IGM.
% We hope to differentiate our feedback models through such comparisons.
% in the star formation history and metal distribution in the CGM between the models as seen in the isolated galaxy test.
% Additional measures of galaxy observables, such as mass-metallicity relation, could be more discriminatory among the models, and comprehensive testing will be needed.
Further comparisons against recent CGM metal line observations (both absorption and emission) such as the Keck Baryonic Structure Survey \citep[KBSS;][]{2012ApJ...750...67R, 2014ApJ...795..165S} and the MusE GAs FLOw and Wind (MEGAFLOW) survey \citep[][]{2016ApJ...833...39S, 2019MNRAS.490.4368S, 2021MNRAS.506.1355S, 2019MNRAS.485.1961Z, 2020MNRAS.492.4576Z, 2021MNRAS.507.4294Z, 2021MNRAS.502.3733W, 2021MNRAS.501.1900F}, 
can be more discriminatory tests of different models \citep[][]{2017MNRAS.471..690T, 2019ApJ...887..107F, 2020MNRAS.499.1721C, 2021ApJ...914...66N, 2021ApJ...923...56D}.
% will allow us to constrain the details of our feedback models 
% similary to Fujimoto19, Nagamine21, Chen19, Illustris, Eagle.
% more about gas dynamics around galaxies 

\section*{Acknowledgements}

We are grateful to Volker Springel for providing the original version of {\sc GADGET-3}, on which the {\sc GADGET3-Osaka} code is based. 
Our numerical simulations and analyses were carried out on the XC50 systems at the Center for Computational Astrophysics (CfCA) of the National Astronomical Observatory of Japan (NAOJ), {\sc Octopus} at the Cybermedia Center, Osaka University, and {\small Oakforest-PACS} at the University of Tokyo as part of the HPCI System Research Project (hp200041, hp210090).
This work is supported by the JSPS KAKENHI Grant Number 21J20930 (YO), 16H05998, 18H05440, 21H04487 (KT), JP17H01111, 19H05810, 20H00180 (KN). 
KN acknowledges the travel support from the Kavli IPMU, World Premier Research Center Initiative (WPI), where part of this work was conducted.

%%%%%%%%%%%%%%%%% APPENDICES %%%%%%%%%%%%%%%%%%%%%

\appendix
\section{Analytic theory of single SNR}
\label{sec:SNRtheory}

In the following section, the number density ratio of hydrogen to helium atoms in the interstellar medium (ISM) is assumed to be 10:1 unless otherwise specified. Other heavy elements are not considered. 
We also assume the ISM to be unionized. In this case, the mean molecular weight is $\mu = \rho/(m_{\rm H}n) = 1.4$ and the specific heat ratio is $\gamma = 5/3$.

In the initial stage of SNR time evolution, the mass ejected by the SN explosion expands with velocity 
\begin{equation}
    v_{\rm ej} = \sqrt{\frac{2E_{\rm SN}}{M_{\rm ej}}} = 7\times10^3\,\kms~\left(\frac{E_{\rm SN}}{10^{51}\,\erg}\right)^{1/2}\left(\frac{M_{\rm ej}}{2\,\Msun}\right)^{-1/2}
\end{equation}
while conserving kinetic energy and momentum, where $E_{\rm SN}$ is the kinetic energy of the SN ejecta and $M_{\rm ej}$ is the ejecta mass. 
This is called the free expansion phase.

When the SNR sweeps up the surrounding gas and the collected mass increases to approximately that of the ejecta, the mass gain of the swept-up gas can no longer be ignored, and the SNR enters the Sedov--Taylor phase.  The radius and time at this point are: 
\begin{eqnarray}
    &R_{\rm free} = \left( \frac{2\,M_{\rm ej}}{(4/3)\pi \rho}\right)^{1/3} = 3\,{\rm pc}~\left(\frac{M_\mathrm{ej}}{2\,\Msun} \right)^{1/3}\,n_0^{-1/3}, \\
    &t_{\rm free} = \frac{R_{\rm free}}{v_{\rm ej}} = 4\,{\rm kyr}~\left(\frac{M_\mathrm{ej}}{2\,\Msun} \right)^{5/6}\,n_0^{1/3}.
\end{eqnarray}
In the Sedov--Taylor phase, the SNR is hot, and the cooling effect is negligible until the cooling time. 
Therefore, the SNR grows while conserving energy, and its momentum increases as it sweeps up the surrounding gas and increases its mass. 
Its time evolution is represented by the Sedov--Taylor self-similar solution. The dimensionless quantity $\xi$ of the self-similar solution is created from the radius $r$, energy $E$, density $\rho$ of the ISM, and time $t$;
\begin{equation}
    \xi = \left(\frac{\rho}{E}\right)^{1/5}\frac{r}{t^{2/5}}.
    \label{eq:sedovtaylor}
\end{equation}
The value of the dimensionless quantity at the shock surface is $\xi_0 \sim 1.15$ \citep[][]{2011piim.book.....D}.

Here, we estimate the thermal energy of the hot gas inside the shell at the shell-formation to calculate the SNR evolution after shell formation.
We have the Rankine--Hugoniot relations for a strong shock\,($\mathcal{M} \gg 1$, where $\mathcal{M}$ is the Mach number):
\begin{eqnarray}
    &\rho_{\rm shell} = 4\rho_{\rm ISM}, \label{eq:rankinehugoniotdensity}\\
    &v_{\rm shell} = \frac{3}{4}\dot{R}, \label{eq:rankinehugoniotvelocity}\\
    &P_{\rm shell} = \frac{3}{4}\rho_{\rm ISM} \dot{R}^2,
    \label{eq:rankinehugoniotpressure}
\end{eqnarray}
where $\rho_{\rm shell}$, $v_{\rm shell}$, and $P_{\rm shell}$ are the shell density, velocity, and pressure, respectively.
We assume that the shell has a uniform density at the Sedov--Taylor phase. 
From the Rankine--Hugoniot relation and the mass conservation, 
\begin{equation}
   \rho_{\rm ISM} \frac{4}{3}\pi R^3  = \rho_{\rm shell} 4\pi R^2\Delta R,
\end{equation}
we obtain the thickness of the shell $\Delta R = (1/12)R$. 

From the equation of motion,
\begin{equation}
    \frac{d}{dt}\left(\rho_{\rm ISM}\frac{4}{3}\pi R^3 v_{\rm shell}\right) = 4\pi R^2 P_{\rm in},
\end{equation}
we obtain the averaged pressure of the SNR,
\begin{equation}
    P_{\rm in} = (3/8)\rho_{\rm ISM} \dot{R}^2.
    \label{eq:pin}
\end{equation}
% Here, we use the Sedov--Taylor solution (\ref{eq:sedovtaylor}) for the derivation. Also,
% Note that the velocity of the shell is $v_{\rm shell} = (3/4)\dot{R}$ from the Rankine--Hugoniot relation.
% The pressure determined here, $P_{\rm in}$, is the pressure driving the shell and is considered to be the average pressure inside the SNR. 
The ratio of the thermal energy of the entire SNR, $E_{\rm tot}$, to that of the shell, $E_{\rm shell}$, is 
\begin{equation} \label{eq:residualthermalenergy}
    \frac{E_{\rm shell}}{E_{\rm tot}} = \frac{\frac{3}{2}P_{\rm shell}4\pi R^2\Delta R}{\frac{3}{2}P_{\rm in}\frac{4}{3}\pi R^3} = \frac{1}{2}.
\end{equation}
This means that one half of the thermal energy of the SNR is in the shell, while the other is in the gas inside the shell. 
When a low-temperature, high-density shell is formed, its thermal energy is radiated out. 
In the Sedov--Taylor phase, 72\% of the total energy is thermal energy \citep{1988ApJ...334..252C}. During the formation of the low-temperature high-density shell, half of the thermal energy, or 36\% of the total energy, will be radiated.

In the PDS phase, we assume that there is hot gas inside the low-temperature, high-density shell, and that the shell is driven by the adiabatic expansion of the hot gas.
The SNR time evolution is obtained from Poisson's equation and the equation of motion of the shell, 
\begin{eqnarray}
    &P\left(\frac{4}{3}\pi R^3\right)^{5/3} = const.\\
    &\frac{d}{dt}\left(\frac{4}{3}\pi R^3 \rho \dot{R}\right) = 4\pi r^2 P,
\end{eqnarray}
and by assuming a solution of the form $R\propto t^{\alpha}$, we obtain $R\propto t^{2/7}$. Thus, $Pt^{10/7} = const.$. Since the pressure inside the shell after shell formation is 
\begin{multline}
    P_{\rm sf} = \frac{(\gamma - 1)0.36E_{51}}{(4/3)\pi R_{\rm sf}^3} \\
    = 1.2\times10^6 k_{\rm B}\,\kelvin\,\pcmq~E_{51}^{0.14}\,n_0^{1.3}\,\Lambda_{6,-22}^{0.39},
\end{multline}
the time $t_{\rm PDS}$ at which the pressure inside the shell and the pressure in the ISM $P_{\rm ISM}$ are equal is estimated to be 
\begin{equation}
    t_{\rm PDS} = 5.0\times10^6\,{\rm yr}~T_{\mathrm{ISM},3}^{-0.7}\,E_{51}^{0.32}\,n_0^{-0.36}\,\Lambda_{6,-22}^{-0.053},
    \label{eq:t_PDS}
\end{equation}
where $T_{\mathrm{ISM},3} = T_\mathrm{ISM}/(10^3\,\kelvin)$ is the temperature of the ISM. The radius and velocity at this time are
\begin{gather}
    R_{\rm PDS} = 89\,{\rm pc}~T_{\mathrm{ISM},3}^{-0.2}\,E_{51}^{0.32}\,n_0^{-0.37}\,\Lambda_{6,-22}^{-0.053},\\
    V_{\rm PDS} = 5.0\,\kms~\,T_{\mathrm{ISM},3}^{0.5}.
\end{gather}
Since the speed of sound in the ISM is 
\begin{equation}
    c_s = \sqrt{\gamma\frac{P}{\rho}} = 3.1\,T_3^{0.5}[\,\kms],
\end{equation}
$V_{\rm PDS}$ is greater than the sound speed. In reality, the duration of the pressure-driven snowplow phase is shorter than the value of $t_\mathrm{PDS}$ derived here because of the loss of thermal energy of the hot gas inside the shell, which in turn is a result of the mixing between the shell and the gas, which was left unaccounted for here \citep{2015ApJ...802...99K}.

When the pressure inside the shell drops to about the pressure of the ISM, the pressure-driven snowplow phase ends and transitions to the momentum-conserving snowplow phase. The time evolution is expressed in terms of momentum conservation,
\begin{equation}
    \frac{d}{dt}\left(\frac{4}{3}\pi R^3 \rho \dot{R}\right) = 0,
\end{equation}
and $R\propto t^{1/4}$. As the velocity of the SNR decreases to about the speed of sound, the SNR mixes with the ISM.

\section{Analytic theory of superbubbles after shell formation}
\label{sec:SBtheory}

Here, we briefly review two classical analytic models to describe the time evolution of a superbubble after shell formation: the pressure-driven and momentum-driven models.
These models are distinguished by whether or not the pressure inside the shell is high enough to accelerate the shell.
%In this section we briefly explain these two models.

\subsection{Pressure-driven model} \label{sec:SB:pressuredrivenmodel}
The shell is considered to be driven by gas pressure while the gas interior to the shell is overpressured against ISM \citep[][]{1977ApJ...218..377W}. 
Its time evolution is described by the equation of motion,
\begin{equation}
    \frac{d}{dt}\left( \frac{4}{3}\pi R^3 \rho \dot{R}\right) = 4\pi R^2 P,
\end{equation}
and the equation of the time evolution of energy inside the shell
\begin{equation}
    \frac{d}{dt}\left( \frac{3}{2}P\left(\frac{4}{3}\pi R^3\right)\right) = \frac{E_{\rm SN}}{\dtsn} - 4\pi R^2 P \dot{R},
\end{equation}
where the first term in the rhs is the energy injection rate from SN and the second term is the energy loss rate due to $PdV$ work.
Assuming a solution of the form $R = A t^{\alpha}$, we obtain 
\begin{equation} 
    R = \left (\frac{125}{154\pi}\right)^{1/5}\,E^{1/5}\,\dtsn^{-1/5}\,\rho^{-1/5}\,t^{3/5}.
    \label{eq:pressure-driven radius}
\end{equation} 
The momentum evolves in time with the following equation:
\begin{equation} 
    p = \frac{4}{3}\pi R^3 \dot{R}\propto t^{7/5}.
    \label{eq:pressure-driven momentum}
\end{equation}

\subsection{Momentum-driven model} \label{sec:SB:momentumdrivenmodel} 
After the pressure of the gas interior to the shell becomes comparable to that of ISM, the SNR 
% formed by the SN explosion inside the shell 
gives momentum to drive the expanding shell \citep[][]{1972SvA....15..708A}. 
The time evolution of the shell is 
\begin{equation}
    \frac{d}{dt}\left(\frac{4}{3}\pi R^3 \rho \dot{R}\right) = \frac{p_{\rm SNR}}{\dtsn}\label{eq:momentum-driven radius},
\end{equation} 
where $p_{\rm SNR}$ is the momentum given by the SNR of SN explosion inside the shell.
Assuming that $p_{\rm SNR}$ is constant, the radius and momentum evolve in time with $R \propto t^{1/2},\,p \propto t$, respectively.

\section{Resolution requirement for thermal and kinetic feedback}
\label{sec:ResolutionRequirement}

Here, we analytically calculate the mass resolution requirement for thermal and kinetic feedback.
\citet{2012MNRAS.426..140D} estimated the resolution required for effective thermal feedback by comparing the sound-crossing time, $t_s$, with the cooling time, $t_c$.
The sound-crossing time across a scale of spatial resolution $\Delta x = (m_{\rm gas}/\rho)^{1/3}$ is 
\begin{multline}
    t_s = \frac{\Delta x}{c_s} 
    = \left(\frac{\mu m_{\rm H}}{\gamma k_{\rm B}T}\right)^{\frac{1}{2}}\left(\frac{m_{\rm gas}}{\rho}\right)^{\frac{1}{3}}
    = 1.0 \times 10^5\,\mathrm{yr} \\
    \times \left(\frac{\mu}{0.59}\right)^{\frac{1}{6}} \left(\frac{T}{10^{7.5}\,\kelvin}\right)^{-\frac{1}{2}}
    \left(\frac{m_{\rm gas}}{10^4\,\Msun}\right)^{\frac{1}{3}} \left(\frac{\nh}{1\,\pcmq}\right)^{-\frac{1}{3}},
\end{multline}
where $c_s$ is the local sound speed. 
The cooling time is 
\begin{equation}
    t_c = \frac{n \kB T}{(\gamma - 1)\Lambda_{ff}},
    \label{eq:coolingtime}
\end{equation}
where $\Lambda_{ff}$ is the cooling rate by bremsstrahlung.
Here, we consider the cooling time of gas with $T > 10^7$\,K, where bremsstrahlung is the dominant cooling process.
Note that the cooling time we consider here is not that corresponding to a point explosion, i.e., the shell-formation time for SNR.
We compare the cooling time for a gas particle with fixed density to the sound-crossing time to see if the heated gas particle is cooled down within the simulation time step or not.
We can avoid overcooling if the cooling time is longer than the simulation time step, however, avoiding overcooling does not ensure that we resolve the Sedov--Taylor phase as we will discuss later in this section.
The cooling rate by bremsstrahlung is \citep[][]{2011piim.book.....D}
\begin{equation}
    \Lambda_{ff} = 1.14 \times 10^{-23}\,\erg\,\pcmq\,\psec\,\left(\frac{\nh}{1\,\pcmq}\right)^{2}\left(\frac{T}{10^{7.5}\,\kelvin}\right)^{1/2},
    \label{eq:coolingrate_ff}
\end{equation}
where we have assumed that the ratio of number density of hydrogen to helium is 10:1, the plasma is fully ionized, and the gaunt factor is 1.19.
Combining Eq.~(\ref{eq:coolingtime}) and (\ref{eq:coolingrate_ff}), we obtain the cooling time as:
\begin{equation}
    t_c = 4.2 \times 10^7\,{\rm yr}\,\left(\frac{\nh}{1\,\pcmq}\right)^{-1}\left(\frac{T}{10^{7.5}\,\kelvin}\right)^{1/2}.
\end{equation}
The ratio between the sound-crossing and cooling times is:
\begin{multline}
    \frac{t_c}{t_s} = 4.2 \times 10^{2} \\
    \times \left(\frac{\mu}{0.59}\right)^{-\frac{1}{6}} \left(\frac{\nh}{1\,\pcmq}\right)^{-\frac{2}{3}} \left(\frac{T}{10^{7.5}\,\kelvin}\right)  \left(\frac{m_{\rm gas}}{10^4\,\Msun}\right)^{-\frac{1}{3}}.
\end{multline}
The thermal feedback is effective when the cooling time is sufficiently longer than to the sound-crossing time, $t_c/t_s > 10$, 
which translates to 
%The temperature at which thermal feedback becomes effective is 
\begin{equation}
    T > 7.5 \times 10^5\,\kelvin\,\left(\frac{\mu}{0.59}\right)^{1/6}\left(\frac{\nh}{1\,\pcmq}\right)^{2/3}\left(\frac{m_{\rm gas}}{10^4\,\Msun}\right)^{1/3}.
    \label{eq:Tthermal}
\end{equation}
The criterion temperature $T = 7.5 \times 10^5$\,K is lower than the temperature where metal line radiative cooling dominates over bremsstrahlung (Fig.~\ref{fig:SB:coolingcurve}), 
%Since the gas rapidly cools down in the temperature range where metal line radiative cooling dominates, 
therefore thermal feedback is effective when a gas particle is heated up to $T > 10^{7.5}\,\kelvin$.

The conditions for kinetic feedback to be effective are more challenging than those for thermal feedback.
\citet{2015ApJ...802...99K} studied the resolution dependence of the terminal momentum of SNRs in 3D simulations using the \textsc{Athena} code, and showed that the criterion for convergence within 25\% of the terminal momentum is $\Delta x/ R_{\rm sf} < 1/3$, where $\Delta x$ is the spatial resolution, and $R_{\rm sf}$ is the shell-formation radius (Eq.~\ref{eq:SB:Rsfsingle}).
In order to satisfy this requirement in SPH, the following is necessary:
\begin{equation}
    \frac{(m_{\rm gas}/\rho)^{1/3}}{23 E_{51}^{0.29} n_0^{-0.42} \Lambda_{6, -22}^{-0.13} {\rm pc}} < \frac{1}{3}, 
    \label{eq:FBmodel:KO15_criterion}
\end{equation}
where $m_{\rm gas}$ is the SPH particle mass. 
Solving this inequation for $m_{\rm gas}$, we obtain
\begin{equation}
    m_{\rm gas} < 6.6\,\Msun\,\left(\frac{\mu}{0.59}\right) n_0^{-0.26} E_{51}^{0.87} \Lambda_{6, -22}^{-0.39}.
    \label{eq:FBmodel:mass_criterion}
\end{equation}
\citet{2019MNRAS.483.3363H} investigated the resolution dependence of SNR terminal momentum with $E_{\rm SN} = 10^{51}\,\erg$ in a uniform medium of $n_{\rm H} = 1\,\cc$ at four mass resolutions, $m_{\rm gas} =$ 0.1, 1, 10, and 100\,$\Msun$ using their modified version of \textsc{GADGET-3} SPH code. 
They show that the terminal momentum converges in the cases of $m_{\rm gas} =$ 0.1, and 1\,$\Msun$, and that the terminal momentum in the case of $m_{\rm gas} = 10\,\Msun$ is about 15\% smaller than that for $m_{\rm gas} =$ 0.1, and 1\,$\Msun$, which is roughly consistent with the criterion in Eq.~(\ref{eq:FBmodel:mass_criterion}).
The reason why the terminal momentum converges within 15\% even in the case of $m_{\rm gas} = 10\,\Msun$ might be because the spatial resolution of SPH becomes better in a dense shell.
The terminal momentum in the case of $m_{\rm gas} = 100\,\Msun$ is underestimated by a factor of 3 due to overcooling \citep{2019MNRAS.483.3363H}.

% In SPH simulations of galaxy formation, the mass of stellar particle is equal to that of gas particle, and there is a one-to-one correspondence between the energy of a SN feedback and the mass of gas particle, $E_{\rm SN} = \epsilon_{\rm SN} m_* = \epsilon_{\rm SN} m_{\rm gas}$.
We can also solve Eq.~(\ref{eq:FBmodel:KO15_criterion}) for $E\,(=E_{51}\times10^{51}\,\erg$) to find the injection energy when the shell-formation radius becomes large enough to be resolved well with the given mass resolution $m_{\rm gas}$:
\begin{equation}
    E > 4.5\times10^{54}\,\erg\,\left(\frac{\mu}{0.59}\right)^{-1.2} \left(\frac{m_{\rm gas}}{10^4\,\Msun}\right)^{1.2} n_0^{0.30} \Lambda_{6, -22}^{0.45},
\end{equation}
and obtain the temperature at which the criterion, $\Delta x/ R_{\rm sf} < 1/3$, is satisfied:
\begin{multline}
    T = \frac{\mu m_{\rm H}}{(\gamma - 1)}\frac{E}{m_{\rm gas}k_{\rm B}} \\
    > 1.6 \times 10^9\,\kelvin\,\left(\frac{\mu}{0.59}\right)^{-0.2} \left(\frac{m_{\rm gas}}{10^4\,\Msun}\right)^{0.2} n_0^{0.3} \Lambda_{6, -22}^{0.45}.
    \label{eq:Tkinetic}
\end{multline}
This temperature is much higher than the one for the thermal feedback to become effective (Eq.~\ref{eq:Tthermal}), which means that the condition for kinetic feedback is more stringent and difficult to be realized in actual simulations. 
Therefore, kinetic feedback needs to be implemented as a subgrid model in current hydrodynamic simulations.

\section{Resolution Test of SN Feedback Model in an Isolated Galaxy Simulation}
\label{sec:ResolutionTest}

In this section, we perform a higher-resolution isolated galaxy simulation to test its dependence on resolution.
In the `High-reso' run, ten times more particles are employed, so the mass resolution is ten times better.
We adopt a fixed gravitational softening length of $\epsilon_{\rm grav}$ = 40\,pc for the High-reso run.
We use the same input parameters with the Fiducial run.

\begin{figure}[h]
    \centering
    \includegraphics[width=\columnwidth]{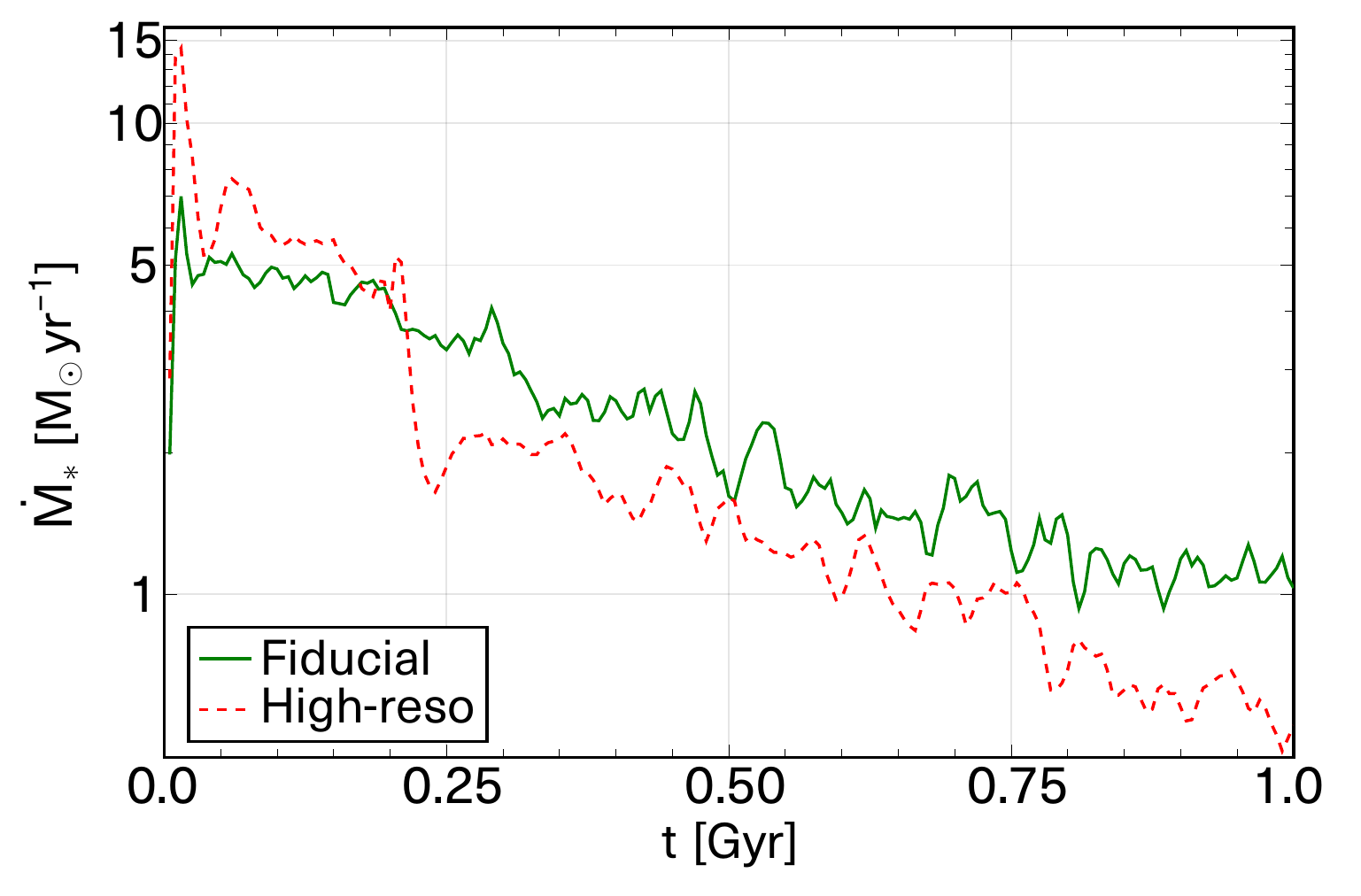}
    \caption{Star formation rate as a function of time for the Fiducial and High-reso runs.}
    \label{fig:Isogal:SFR_reso}
\end{figure}

Fig.~\ref{fig:Isogal:SFR_reso} shows the SFR history of the Fiducial and High-reso runs.
The SFR in the High-reso run is higher than that in the Fiducial run up to $t \sim0.2$\,Gyr, after which the SFR declines rapidly due to local instability and a bursty star formation event.
The discrepancy in SF history was also observed by \citet[][]{2018MNRAS.477.1578H}, who performed cosmological zoom-in simulations using \textsc{GIZMO} with their mechanical feedback model with different mass resolutions, and they demonstrate that the SF history converges at $m_{\rm gas} < 100\,\Msun$. %, which convergence is faster than other sub-grid models.
\citet[][]{2019MNRAS.484.1687L} performed isolated galaxy simulations using \textsc{GIZMO} with a similar but slightly modified SN feedback model by \citet[][]{2018MNRAS.477.1578H}, and an SF recipe considering log-normal density distribution in sub-grid turbulent ISM \citep[][]{2012ApJ...761..156F, 2017ApJ...845..133S}.  They show that the SF history converges at the same resolution with our Fiducial and High-reso runs.
Thus, the discrepancy we see in Fig.~\ref{fig:Isogal:SFR_reso} could be reduced with different SF recipes.

\begin{figure*}
    \centering
    \includegraphics[width=\textwidth]{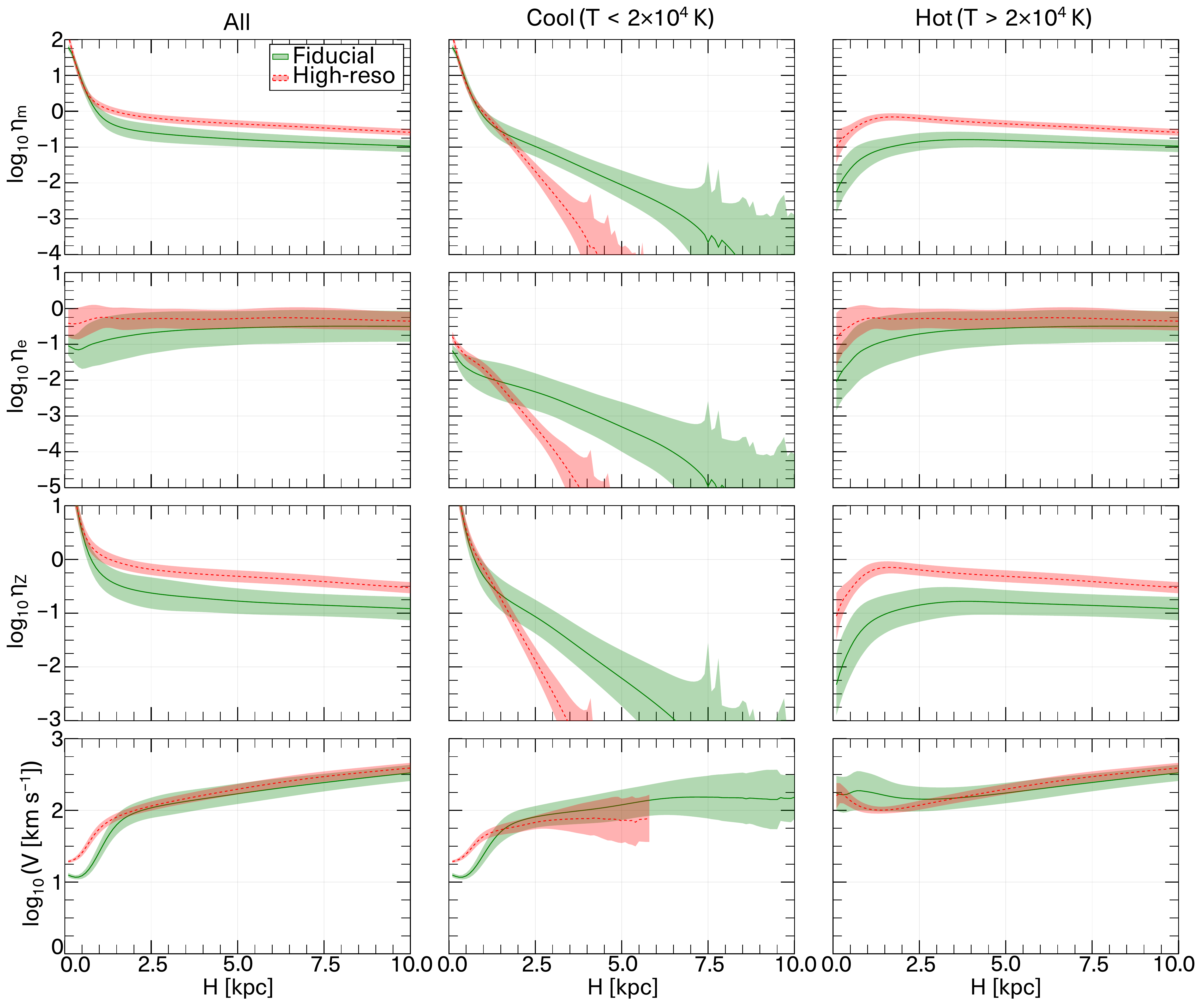}
    \caption{Same as Figure~\ref{fig:Isogal:outflowprofile_10} but for the Fiducial and High-reso runs.}
    \label{fig:Isogal:outflowprofile_10_reso}
\end{figure*}
Fig.~\ref{fig:Isogal:outflowprofile_10_reso} shows the outflow profiles of the Fiducial and High-reso runs up to 10\,kpc above the galactic plane. 
% They show good agreement overall. 
The mass, energy, and metal loading factors of the hot component for the High-reso run is about a factor of two higher than those of the Fiducial run.
% which is likely due to the lower SN feedback energy per feedback event in High-reso run % due to smaller stellar mass, 
% and smaller shock radius. % of SN feedback is smaller.
% We see that the outflow profiles of hot components converge well thanks to stochastic feedback treatment.}

\begin{figure}
    \centering
    \includegraphics[width=.8\columnwidth]{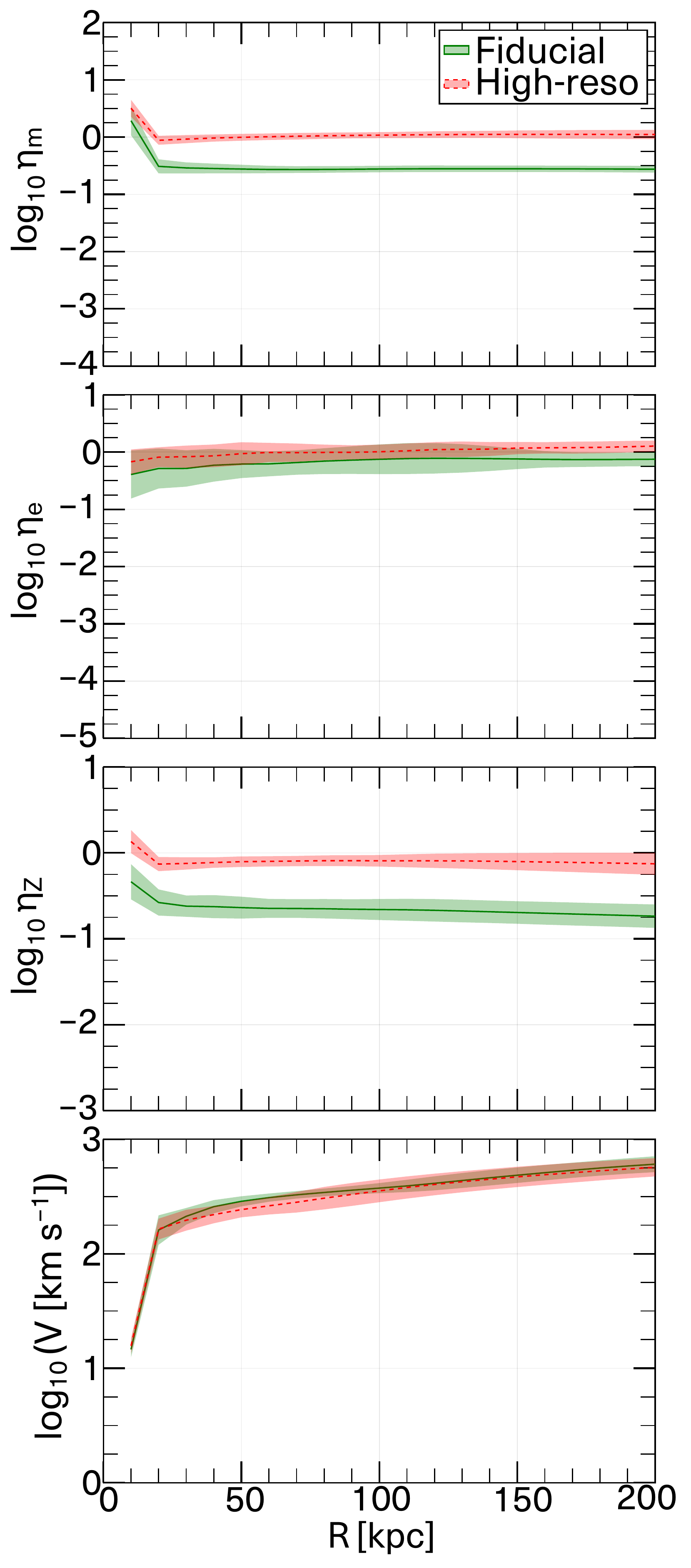}
    \caption{Same as Figure~\ref{fig:Isogal:outflowprofile_200} but for the Fiducial and High-reso runs.}
    \label{fig:Isogal:outflowprofile_200_reso}
\end{figure}
Fig.~\ref{fig:Isogal:outflowprofile_200_reso} shows the outflow profiles of the Fiducial and High-reso runs up to 200 kpc in spherical radius.  
We see that the High-reso run has higher mass, energy, and metal loading factors than the Fiducial run due to its higher SFR during the first $\sim 0.2$\,Gyr. 
%, which is because the outflow profile reflects the SF history, and the High-reso run have higher SFR at first 0.5 Gyr.

These two figures (Figs. \ref{fig:Isogal:outflowprofile_10_reso} \& \ref{fig:Isogal:outflowprofile_200_reso}) show that the perfect convergence is still difficult to achieve even with the implementation of subgrid kinetic feedback model, given the very stringent resolution requirement described in Appendix~\ref{sec:ResolutionRequirement} as well as the impact of fluctuating star formation rate when the resolution is changed. 
In particular, the initial starburst at $t < 0.1$\,Gyr has a significant impact in the later SF history as we see in Fig.~\ref{fig:Isogal:SFR_reso}, such as the sharp drop in SFR of High-reso run at $t\simeq0.2$\,Gyr.
In order to understand the differences we see in Fig.~\ref{fig:Isogal:outflowprofile_10_reso} we would need to understand the origin of hot and cold outflow component in more detail, but this is beyond the scope of current paper and we will study these issues in our future work.

\bibliography{ref}
\bibliographystyle{aasjournal}

%% This command is needed to show the entire author+affiliation list when
%% the collaboration and author truncation commands are used.  It has to
%% go at the end of the manuscript.
%\allauthors

%% Include this line if you are using the \added, \replaced, \deleted
%% commands to see a summary list of all changes at the end of the article.
\listofchanges

\end{document}